\documentclass[12pt]{article}
\usepackage{graphicx}
\usepackage{amsmath}
\usepackage{amsfonts}
\usepackage{amssymb}
\usepackage{cite}

\newcommand{\lsim}{\raisebox{0.1mm}{$\, <$} \hspace{-3.0mm}\raisebox{-1.8mm}{$\sim \,$}}
\newcommand{\gsim}{\raisebox{0.1mm}{$\, >$} \hspace{-3.0mm}\raisebox{-1.8mm}{$\sim \,$}}

\newcommand{\beq}{\begin{eqnarray}}
\newcommand{\eeq}{\end{eqnarray}}

\newcommand{\lsp}{ \left ( }
\newcommand{\rsp}{ \right ) }

\newcommand{\msusy}{ m_{\rm SUSY} }
\newcommand{\mgut}{ M_{\rm GUT} }
\newcommand{\mpl}{ M_{Pl} }

\newcommand{\eps}{ \epsilon }
\newcommand{\epsb}{ \overline{\epsilon} }
\newcommand{\epsY}{ {\epsilon_Y} }
\newcommand{\epsLY}{ {\epsilon_{LY}} }
\newcommand{\epsg}{ {\epsilon_g} }
\newcommand{\epsL}{ {\epsilon_L} }
\newcommand{\epsR}{ {\epsilon_R} }
\newcommand{\epsLR}{ {\epsilon_{LR}} }

\newcommand{\epsp}{ \epsilon' }
\newcommand{\epsbp}{ \overline{\epsilon}' }
\newcommand{\epsYp}{ {\epsilon'_Y} }
\newcommand{\epsLYp}{ {\epsilon'_{LY}} }
\newcommand{\epsgp}{ {\epsilon'_g} }
\newcommand{\epsLp}{ {\epsilon'_L} }
\newcommand{\epsRp}{ {\epsilon'_R} }
\newcommand{\epsLRp}{ {\epsilon'_{LR}} }

\newcommand{\tgb}{ t_\beta }
\newcommand{\ttgb}{ t_\beta^2 }

\newcommand{\sinb}{ s_\beta }
\newcommand{\cosb}{ c_\beta }
\newcommand{\cotb}{ t_\beta^{-1} }

\newcommand{\Vb}{ \overline{V} }
\newcommand{\yb}{ \overline{y} }
\newcommand{\mb}{ \overline{m} }

\newcommand{\RE}{ {\rm Re} }
\newcommand{\IM}{ {\rm Im} }

\newcommand{\uL}{ {u_L} }
\newcommand{\uR}{ {u_R} }
\newcommand{\dL}{ {d_L} }
\newcommand{\dR}{ {d_R} }
\newcommand{\eL}{ {e_L} }
\newcommand{\eR}{ {e_R} }
\newcommand{\nL}{ {\nu_L} }
\newcommand{\uLb}{ {{\overline u}_L} }
\newcommand{\uRb}{ {{\overline u}_R} }

\newcommand{\nLb}{ {{\overline \nu}_L} }

\newcommand{\tWLm}{  { \tilde{W}_L^-} }

\newcommand{\tHdLm}{  { \tilde{H}_{1L}^-} }

\newcommand{\tWRpb}{  { \overline{\tilde{W}} \!\!\ _R^+} }

\newcommand{\tWRmb}{  { \overline{\tilde{W}} \!\!\ _R^-} }

\newcommand{\tWRzb}{  { \overline{\tilde{W}} \!\!\ _R^0} }

\newcommand{\tBRb}{  { \overline{\tilde{B}} \!\!\ _R} }
\newcommand{\tBLb}{  { \overline{\tilde{B}} \!\!\ _L} }
\newcommand{\tchiRpb}{  { \overline{\tilde{\chi}} \!\!\ _R^+} }
\newcommand{\tchiLpb}{  { \overline{\tilde{\chi}} \!\!\ _L^+} }
\newcommand{\tchiRmb}{  { \overline{\tilde{\chi}} \!\!\ _R^-} }
\newcommand{\tchiLmb}{  { \overline{\tilde{\chi}} \!\!\ _L^-} }
\newcommand{\tchiRzb}{  { \overline{\tilde{\chi}} \!\!\ _R^0} }
\newcommand{\tchiLzb}{  { \overline{\tilde{\chi}} \!\!\ _L^0} }
\newcommand{\tHdRpb}{  { \overline{\tilde{H}} \!\!\ _{1R}^+} }
\newcommand{\tHdLmb}{  { \overline{\tilde{H}} \!\!\ _{1L}^-} }
\newcommand{\tHdRzb}{  { \overline{\tilde{H}} \!\!\ _{1R}^0} }
\newcommand{\tHdLzb}{  { \overline{\tilde{H}} \!\!\ _{1L}^0} }
\newcommand{\tHuLpb}{  { \overline{\tilde{H}} \!\!\ _{2L}^+} }
\newcommand{\tHuRmb}{  { \overline{\tilde{H}} \!\!\ _{2R}^-} }
\newcommand{\tHuRzb}{  { \overline{\tilde{H}} \!\!\ _{2R}^0} }
\newcommand{\tHuLzb}{  { \overline{\tilde{H}} \!\!\ _{2L}^0} }

\def\slashchar#1{\setbox0=\hbox{$#1$}           
   \dimen0=\wd0                                 
   \setbox1=\hbox{/} \dimen1=\wd1               
   \ifdim\dimen0>\dimen1                        
      \rlap{\hbox to \dimen0{\hfil/\hfil}}      
      #1                                        
   \else                                        
      \rlap{\hbox to \dimen1{\hfil$#1$\hfil}}   
      /                                         
   \fi}                                         %

\begin{document}

\begin{flushright}
TUM-HEP-706/08\\
\hfill \today \\
\end{flushright}
\vskip   1 true cm 
\begin{center}
{\Large \textbf{
A Complete Analysis of \\
``Flavored" Electric Dipole Moments \\
in Supersymmetric Theories}}
\\ [20 pt]
\textsc{Junji Hisano}${}^{(a,b)}$, \textsc{Minoru Nagai}${}^{c}$ and 
\textsc{Paride Paradisi}${}^{d}$ \\ [20 pt]
${}^{a}~$\textsl{ICRR, University of Tokyo, Kashiwa 277-8582, Japan } \\ 
[5pt]
${}^{b}~$\textsl{Institute for the Physics and Mathematics of the Universe (IPMU), \\
University of Tokyo, Kashiwa, Chiba 277-8568, Japan} \\ 
[5pt]
${}^{c}~$\textsl{Excellence Cluster Universe, Technische Universit\"at M\"unchen,
D-85748 Garching, Germany} \\
[5pt]
${}^{d}~$\textsl{Physik-Department, Technische Universit\"at M\"unchen,
D-85748 Garching, Germany}

\vskip   1 true cm

\textbf{Abstract\\}

\end{center}
\noindent
The Standard Model predictions for the hadronic and leptonic electric dipole
moments (EDMs) are well far from the present experimental resolutions, thus,
the EDMs represent very clean probes of New Physics effects. Especially,
within supersymmetric frameworks with flavor-violating soft terms large and
potentially visible effects to the EDMs are typically expected. In this work,
we systematically evaluate the predictions for the EDMs at the beyond-leading-order
(BLO). In fact, we show that BLO contributions to the EDMs dominate over the
leading-order (LO) effects in large regions of the supersymmetric parameter
space. Hence, their inclusion in the evaluation of the EDMs is unavoidable.
As an example, we show the relevance of BLO effects to the EDMs for a SUSY
$SU(5)$ model with right-handed neutrinos.

\section{Introduction} 
\label{Chap:introduction}

The Standard Model (SM) of elementary particles has successfully passed
all the electroweak tests at the LEP and also all the low-energy flavor
physics tests.  On the other hand, there is a general agreement that the
SM has to be considered as an effective field theory, valid up to some
unknown scale of New Physics (NP). A natural solution of the hierarchy
problem would point towards a scale of new physics close to the TeV, an
energy scale that will be explored by the LHC.

Besides the direct NP search at the colliders (the so-called {\em high-energy
frontier}), a complementary tool to shed light on NP is provided by high-precision
low-energy experiments (the so-called {\em high-intensity frontier}) specially 
to determine the symmetry properties of the underlying NP theory. The hadronic
uncertainties and the overall good agreement of flavor-changing neutral current
(FCNC) data with the SM predictions prevent any conclusive evidence of NP effects 
in low-energy precision tests of the quark sector. However, a closer look at several
CP-violating observables indicates that the CKM phase might not be sufficient to 
describe simultaneously CP violation in $K$, $B_d$ and $B_s$ meson decays. 
In particular:
\begin{itemize}
\item The values of $\sin 2\beta$ extracted from Penguin-dominated
  modes, {\it i.e.} $B_d\to \phi K_{S},~\eta^{\prime}K_{S}$,
  $\pi^{0}K_{S}$, $\omega K_{S}$, $K_{S}K_{S} K_{S}$ {\it etc.}, are
  significantly lower than the corresponding value from the tree-level
  process $B_d\to \psi K_S$, {\it i.e.}  $(\sin 2\beta)_{\psi K_S}=
  0.680\pm 0.025$~\cite{hfag}.

\item CP violation in the $B_d-\overline{B}_d$ system, {\it i.e.}
  $(\sin 2\beta)_{\psi K_S}$, combined with the $\Delta M_d/\Delta M_s$
  constraint, appears insufficient to describe the experimental value
  of $\epsilon_K$ within the SM, after the new SM value for $\epsilon_K$
  is taken into account~\cite{BG}. Similarly, a simultaneous description
  of $\epsilon_K$ and $\Delta M_d/\Delta M_s$ within the SM implies
  $\sin 2\beta= 0.88\pm 0.11$~\cite{SL}~\cite{BG}, well far from the
  measured $(\sin 2\beta)_{\psi K_S}$.

\item There are some hints for the asymmetry in $B_s\rightarrow
  {\psi\phi}$, $S_{\psi\phi}$, to be significantly larger than the SM
  value $S_{\psi\phi}\approx 0.04$ \cite{first_evidence}.

\item There are also other interesting tensions observed in the data,
  as the rather large difference in the direct CP asymmetries
  $A_{CP}(B_d^{-}\to K^{-}\pi^{0})$ and $A_{CP}(\overline{B}_d^{0}\to
  K^{-}\pi^{+})$ and certain puzzles in $B_d\to\pi K$ decays.
\end{itemize}
All the above tensions might be accommodated through the introduction
of new sources of CP violation in a given NP scenario. 

We know that some sources of CP violation, in addition to the unique
CKM phase, are required to reproduce the observed baryon asymmetry of
the universe.  In this context, it seems of particular interest the
study of those low-energy observables that are quite sensitive
to new sources of CP violation, as the Electric Dipole Moments (EDMs).
In fact, the EDMs offer a unique possibility to shed light in NP,
given their strong suppression within the SM and their high
sensitivities to NP effects.

The minimal supersymmetric SM (MSSM), that is probably the most
motivated model beyond the SM, exhibits plenty of CP-violating phases
\cite{pospelov} able to generate the EDMs at an experimentally visible
level \cite{expedm,Regan:2002ta,Romalis:2000mg}. After SUSY-breaking 
terms are introduced, new CP-violating sources may naturally appear 
through {\it i)} flavor-conserving terms such as the $B\mu$ parameter 
in the Higgs potential or the $A$ terms for trilinear scalar couplings
and {\it ii)} flavor-violating terms such as the squark and slepton mass
terms. It seems quite likely that the two categories {\it i)} and {\it ii)}
of CP violation are controlled by different physical mechanisms, thus,
they should be distinguished and discussed independently.

In the case {\it i)}, non-vanishing quark EDMs and lepton EDMs are generated 
already at one-loop level. It is well-known that the flavor-conserving phases 
$\phi_A$ and $\phi_\mu$ have to be very small, at the level of ${\cal O}(10^{-(2-3)})$, 
if a low-energy SUSY is realized around the electroweak scale. Obviously,
some mechanisms are required to suppress them in a natural way. The SUSY
CP problem can be avoided in a number of SUSY scenarios. The gaugino/gauge
mediation models for SUSY breaking are candidates when the $A$ and $B\mu$
parameters are derived by the renormalization group (RG) effects. We also
remind the ``effective SUSY scenarios''~\cite{Cohen:1996vb}, the ``Split SUSY limit''~\cite{ArkaniHamed:2004fb}, or the ``Dirac gaugino models''~\cite{Hisano:2006mv}.

If the flavor-conserving terms do not have CP-phases, CP might be violated only by 
those soft terms that additionally change the flavors, {\it i.e.} the case {\it ii)}.
The {\it flavored} EDMs, induced by the flavor-violation, turn out to be naturally
suppressed because of the smallness of the mixing angles regulating the flavor
transitions. In such a case, the EDMs and flavor-changing processes are intimately
related as they both come from the same sources and their correlated study can
represent a useful tool to test or to falsify this scenario.

In this paper, we analyze in detail the predictions for the {\it flavored} EDMs and
Chromo-EDMs (CEDM), at the beyond-leading-order (BLO) level within SUSY theories.
As already stressed in Refs.~\cite{Hisano:2006mj,Hisano:2007cz}, the inclusion of BLO
effects is essential for a correct evaluation of the {\it flavored} EDMs. In fact,
BLO effects do not just represent a correction to the leading-order (LO) contributions,
but, they can dominate over the LO ones in a broad region of the SUSY parameter space.

At the LO, the hadronic EDMs are generated by the one-loop exchange of gluinos
($\tilde{g}$) and charginos ($\tilde{\chi}^\pm$) with squarks. The dominant BLO 
contributions are computed by including all the one-loop induced ($\tan\beta$-enhanced) 
non-holomorphic corrections for the charged Higgs ($H^\pm$) couplings with 
fermions and for the $\tilde{\chi}^\pm$/$\tilde{g}$ couplings with fermions-sfermions.
The above effective couplings lead to the generation of $H^\pm$ effects to the (C)EDMs,
absent at the LO, via the one-loop $H^\pm$/top-quark exchange. Moreover, the chargino
contributions, suppressed at the LO by the light quark masses, are strongly enhanced 
at the BLO by the heaviest-quark Yukawa couplings.
Finally, also the gluino effects receive large BLO contributions that are competitive,
in many cases, with the LO ones. As a result, BLO effects dominate over the LO effects
in large regions of the SUSY parameters space and their inclusion in the evaluation of
the {\it flavored} (C)EDMs is mandatory.

The organization of the paper is as follows. In Section~2, we review the LO contributions
to the flavored EDMs. Here, we also discuss the Jarlskog invariants, to which the (C)EDMs
are proportional to. The invariants are a useful tool to classify the CP violation in the
flavor-changing SUSY-breaking terms. In Section~3, we show the BLO contributions to the
flavored EDMs and we compare them with the LO contributions. For completeness, the BLO
contributions to the electron EDM and the CP-violating four-Fermi operators are also
given in the section. In Section~4, the EDMs in the SUSY GUTs are discussed. Section~5 is
devoted to the conclusions.

In Appendix~A, we show the derivation of various effective couplings induced by the $\tan\beta$-enhanced radiative corrections with explicit CP phase dependence. The effective vertices in the mass insertion approximation is shown there. The loop factors used in Appendix~A are given in Appendix~B. In Appendix~C
we present the effective vertices in the mass eigenstate basis necessary for the numerical calculation. Appendix~D contains the formulae for the (C)EDMs in the mass eigenstate basis, using the effective vertices.
In Appendix~E and F the loop functions used in this paper are defined.

%
\section{Jarlskog Invariants and Flavored EDMs at the LO}
\label{Sec:Jarlskog}

The effective CP-odd Lagrangian, which contributes to the (C)EDMs, is given as
\beq
 {\cal L}_{\rm eff}
   \!\!\! & = & \!\!\!
   \frac{g_s^2}{32\pi^2} \bar{\theta}\, G^a_{\mu\nu} {\tilde G}^{\mu\nu,a}
   -
   \sum_{i=u,d,s,e,\mu} i \frac{d_f}{2} \bar{\psi}_i (F\cdot \sigma)\gamma_5 \psi_i
   -
   \sum_{i=u,d,s} i \frac{d^c_f}{2} g_s \bar{\psi}_i (G\cdot \sigma)\gamma_5 \psi_i
   \nonumber\\
   && \!\!\!
   +
   \frac{1}{3} w\, f^{abc}G^a_{\mu\nu} {\tilde G}^{\nu\rho,b}G_{\rho}^{\mu,c}
   +
   \sum_{i,j} C_{ij}\, (\bar{\psi}_i \psi_i)(\bar{\psi}_j i\gamma_5 \psi_j)
   + \cdots,
\label{Eq:CPodd}
\eeq
where $F_{\mu\nu}$ and $G^a_{\mu\nu}$ are the electromagnetic and the $SU(3)_{C}$
gauge field strengths, respectively. The first term of Eq.~(\ref{Eq:CPodd}) is the
well-known QCD theta term. The second and third terms of Eq.~(\ref{Eq:CPodd}) are the
fermion EDMs and CEDMs, respectively, while the second line of Eq.~(\ref{Eq:CPodd})
contains dimension-six CP-odd operators, such as the Weinberg operator and the CP-odd
four-Fermi interactions.

The QCD theta parameter is tightly constrained by the neutron EDM at the level of
$\bar{\theta}\lsim 10^{-(9-10)}$; this naturalness problem is commonly referred to
as the strong CP problem. One natural way to achieve the required suppression for
$\bar{\theta}$ is to impose the Peccei-Quinn symmetry, since the axion field makes
$\bar{\theta}$ dynamically vanishing.
Under the above assumptions, the quark (C)EDMs are very suppressed in the SM as they
are generated only at the three-loop level by the phase of the CKM matrix. Long-range
effects to the neutron EDM, arising at the two-loop level, are still far below the
current bound. In this paper we assume the Peccei-Quinn symmetry, for simplicity.

Among the various atomic and hadronic EDMs, a particularly important role is played by
the thallium EDM ($d_{\rm{Tl}}$) and by the neutron EDM ($d_n$) that can be estimated as~\cite{LiuKelly,MP1,MP2}
\beq
 d_{\rm Tl} = -585\, d_e - e\, 43\, {\rm GeV}\, C_S^{(0)},
\label{Eq:dTl}
\eeq
where $C_s^{(0)}$ has been evaluated as~\cite{Demir:2003js}
\beq
 C_S^{(0)} =
 C_{de} \frac{29\, {\rm MeV}}{m_d}
  +
  C_{se} \frac{\kappa \times 220\, {\rm MeV}}{m_s}
  +
  C_{be} \frac{66\, {\rm MeV} (1-0.25 \kappa)}{m_b},
\eeq
with $\kappa \simeq 0.5$ \cite{KBM}. Moreover, $d_n$ can be estimated as~\cite{Pospelov:1999ha,Pospelov:2000bw}
\beq
 d_n = (1\pm 0.5)\Big[ 1.4\,(d_d-0.25\, d_u) + 1.1\, e\,(d^c_d+0.5\,d^c_u)\Big]\,.
\label{Eq:dn_odd}
\eeq
In our analysis, we will use the above formulae for the evaluation of the (C)EDMs~\footnote{
  It has been argued in Ref.~\cite{Narison:2008jp} that the evaluation of the neutron
  EDM still suffers from sizable uncertainties even when evaluated within the QCD sum 
  rules approach.}.

The strong suppression of the quark (C)EDMs in the SM could be understood by considering
Jarlskog invariants (JIs) which represent a basis-independent measure of CP violation.
In the SM, we have the following JIs,
\begin{eqnarray}
 &&
 J_{\rm SM}^{d_i} = {\rm Im} \left\{ y_d Y_u [Y_d, Y_u] Y_d \right\}_{ii},
 \nonumber\\
 &&
 J_{\rm SM}^{u_i} = {\rm Im} \left\{ y_u Y_u [Y_d, Y_u] Y_d \right\}_{ii}
\end{eqnarray}
for down- and up-type quarks, respectively. Here, $(y_{d(u)})_{ij}$ is
the down(up)-type quark Yukawa couplings and $Y_{d(u)}\equiv
y_{d(u)}^\dagger y_{d(u)}$.  Since the EDMs and CEDMs of the ${\it
  i}$-th down- and up-type quarks are proportional to the above
Jarlskog invariants, which are of the ninth order in the Yukawa
coupling, the quark (C)EDMs are highly suppressed in the SM.

In the MSSM, the EDMs are not necessarily suppressed at the same level
as in the SM.  As long as one remains within the class of
models belonging to the so-called {\it minimal flavor violation} (MFV)
principle \cite{Hall:1990ac,D'Ambrosio:2002ex,Paradisi:2008qh}, where the
flavor-violating terms in the MSSM arise only from the CKM matrix, the
resultant EDMs are highly suppressed, at the same level of the SM. In
contrast, within an MSSM framework with general flavor violation, the
invariants introduced by the new flavor-violating interactions are
less suppressed by the Yukawa couplings and large EDMs can be
generated.

Let us first parameterize the effects of flavor violation as usual by
means of the mass insertion (MI) parameters
\cite{Hall:1985dx,Gabbiani:1996hi}, \beq && (\delta^q_{LL})_{ij}
\equiv \frac{(m^2_{\tilde q})_{ij}} {{m}^2_{\tilde q}},\quad
 (\delta^d_{RR})_{ij} 
 \equiv
  \frac{(m^2_{\tilde d})_{ij}} {{m}^2_{\tilde d}}, \quad 
 (\delta^u_{RR})_{ij} 
 \equiv
  \frac{(m^2_{\tilde u})_{ij}} {{m}^2_{\tilde u}},
 \nonumber\\
 &&
 (\delta^l_{LL})_{ij} 
 \equiv
  \frac{(m^2_{\tilde l})_{ij}} {{m}^2_{\tilde l}},\quad 
 (\delta^e_{RR})_{ij} 
 \equiv
  \frac{(m^2_{\tilde e})_{ij}} {{m}^2_{\tilde e}}, \quad 
\eeq
for $i\neq j$ and ${m}^2_{\tilde f}~(f=q,u,d,l,e)$ are the
corresponding average sfermion masses. Notice that the above MIs are
basis-dependent quantities.  In this paper, we define the MIs in the
{\it bare} SCKM basis, where the fermions and sfermions are rotated in
the same way so that the corresponding tree-level Yukawa couplings are
flavor-diagonal; in contrast, the threshold corrections induced by
non-holomorphic Yukawa couplings, will be off-diagonal in the flavor
space. We also take $\delta^d_{LL}=\delta^q_{LL}$ and the MI
parameters for left-handed up- and down-type quarks are related by the
$SU(2)_L$ symmetry as $\delta^u_{LL}=\Vb\delta^d_{LL}\Vb^\dagger$
where $\Vb$ is the {\it bare} CKM matrix appearing in tree-level
Yukawa couplings.

The size and the pattern of the MIs are unknown, unless we assume
specific models.  They are regulated by the SUSY-breaking mechanism
and by the interactions of the high-energy theories beyond the
MSSM. In this way, it makes sense to study the individual impact of
different kinds of MIs on the low-energy observables.

When only $(\delta^q_{LL})_{ij}\neq 0$, the following JIs appear,
\begin{eqnarray}
 &&
 J^{(d_i)}_{LL}= {\rm Im} \left\{y_d [Y_u,\delta_{LL}^q] \right\}_{ii},
 \nonumber\\
 &&
 J^{(u_i)}_{LL}= {\rm Im} \left\{y_u [Y_d,\delta_{LL}^q] \right\}_{ii}.
 \label{inv_LL}
\end{eqnarray}
When $(\delta_{RR}^d)_{ij}\neq 0$ or
$(\delta_{RR}^u)_{ij} \neq 0$, the new JIs are
\begin{eqnarray}
 &&
 J^{(d_i)}_{RR}= {\rm Im} \left\{\delta_{RR}^d y_d Y_u \right\}_{ii},
 \nonumber\\
 &&
 J^{(u_i)}_{RR}= {\rm Im} \left\{\delta_{RR}^u y_u Y_d \right\}_{ii}.
 \label{inv_RR}
\end{eqnarray}

Notice that both $J^{(q_i)}_{LL}$ and $J^{(q_i)}_{RR}$ are of the third order
in the Yukawa coupling constants and they arise from the relative phases between
the corresponding MIs and CKM mixings.

When both left- and right-handed squarks have mixings, the new JIs are proportional 
to only one Yukawa coupling constant as
\begin{eqnarray}
 J^{(d_i)}_{LR}= {\rm Im} \left\{\delta_{RR}^d y_d\delta_{LL}^q
 \right\}_{ii},
 \nonumber\\
 J^{(u_i)}_{LR}= {\rm Im} \left\{\delta_{RR}^u y_u\delta_{LL}^q
 \right\}_{ii}\,,
 \label{inv_LR}
\end{eqnarray}
and the relative phases between the left- and right-handed squark mixings
contribute to them. Notice that if only the left-handed squarks have
mixing, the JIs are proportional to the external light quark masses.
However, when the right-handed squarks have mixing, the JIs, and then the
quark (C)EDMs, are enhanced by the Yukawa coupling constants of heaviest quarks.

Finally, for the lepton EDMs, only the following JI is relevant,
\begin{eqnarray}
 J^{(e_i)}_{LR}= {\rm Im} \left\{\delta_{RR}^e y_e\delta_{LL}^l\right\}_{ii}
 \label{inv_LR_e}.
\end{eqnarray}
This is because the leptonic sector does not have a counterpart of the CKM mixing
within the MSSM.

\begin{figure}[t]
\begin{center}
\begin{tabular}{cc}
\includegraphics[scale=0.4]{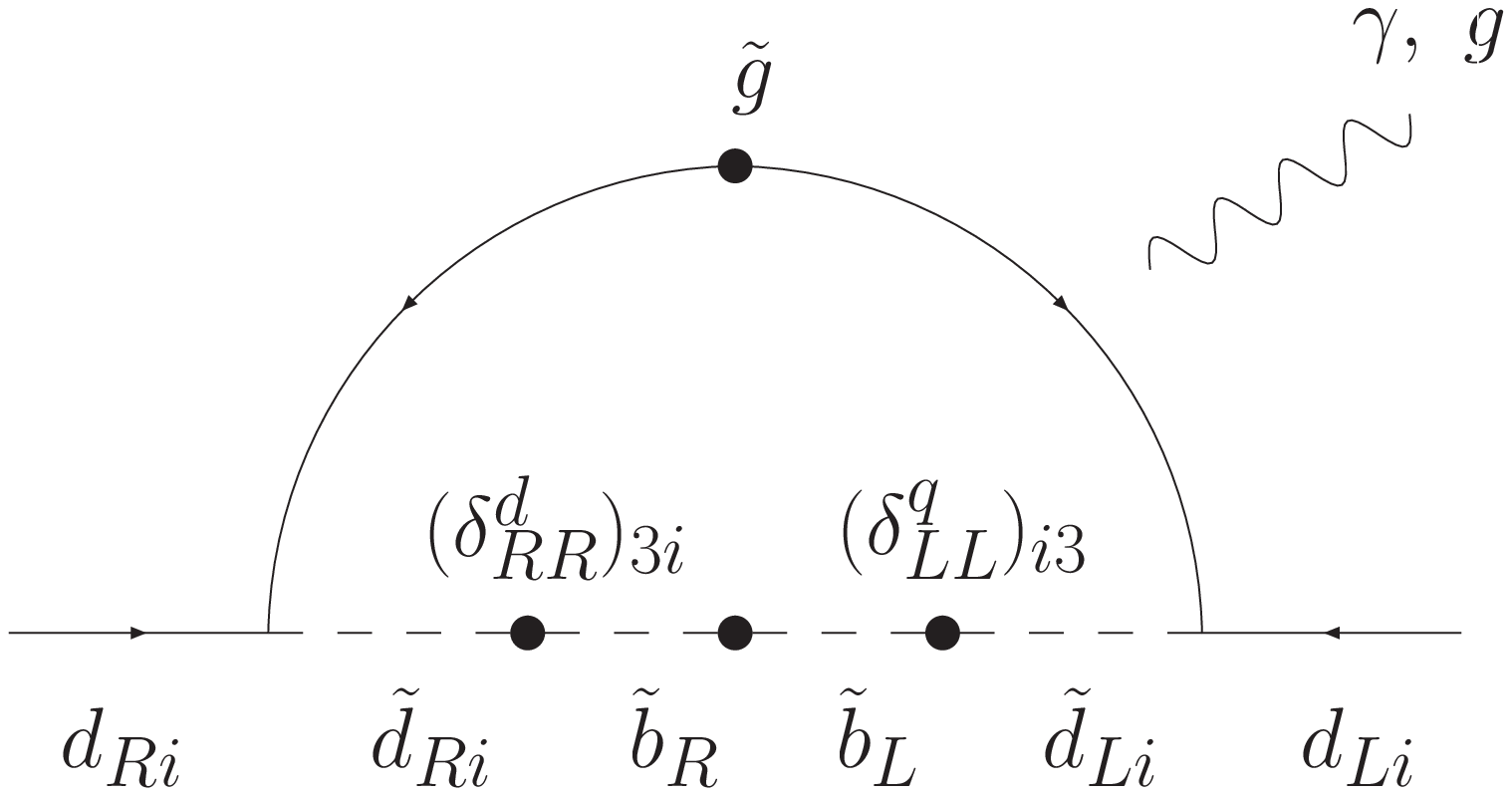} &
\includegraphics[scale=0.4]{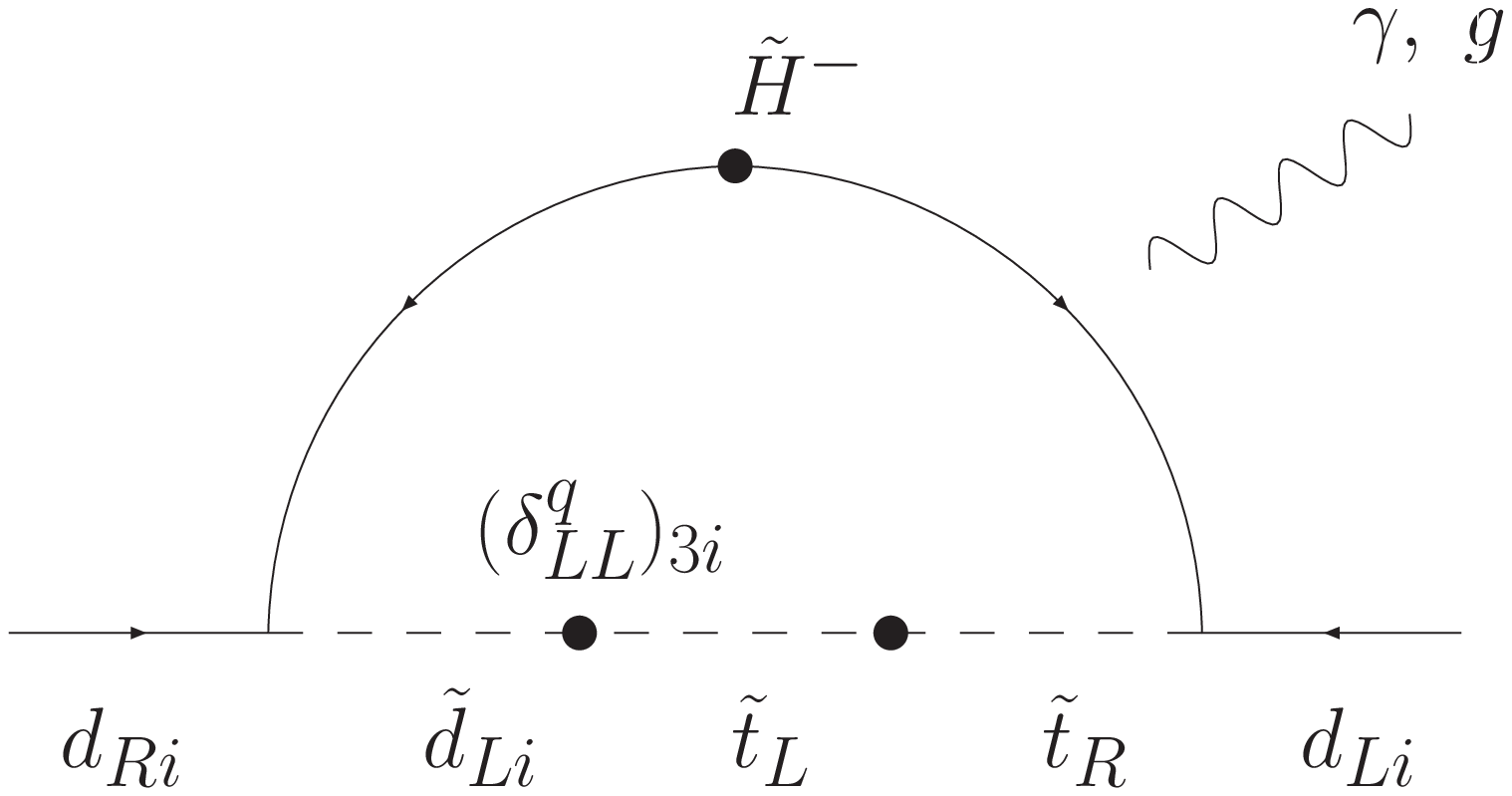}
\end{tabular}
\caption{\label{Fig:leading} Leading contributions to the down-type quark (C)EDMs
proportional to $J^{(d_i)}_{LR}$ (left) and $J^{(d_i)}_{LL}$ (right).}
\end{center}
\end{figure}

Here, we summarize the LO contributions to the flavored EDMs leaving the
derivation of the BLO ones to the next section. The SUSY contributions to
the quark (C)EDMs can arise at the one-loop level, and they are generated
by the JIs of Eqs.~(\ref{inv_LL}) and (\ref{inv_LR}).
In particular, when the left-handed squarks have mixing, the Higgsino diagram
of Fig.~(\ref{Fig:leading}) (on the right) contributes to the down-type quark
(C)EDM through $J^{(d_i)}_{LL}$ and the resultant (C)EDM is given by
\begin{eqnarray}
 \left\{ \frac{d_{d_i}}{e},~d^c_{d_i} \right\}_{{\tilde H}^\pm} =
 \frac{\alpha_2}{16\pi}
 \frac{m_{d_i}}{m^2}
 \frac{m_t^2}{m_W^2}
 \frac{A_t \mu}{m^2}
 \tgb~
  {\rm Im}\left[ V^*_{3i}(\delta^d_{LL})_{3i} \right]
  \left\{
   f_5^{(2)}(y),~
   f_0^{(2)}(y) 
  \right\},
\label{Eq:LO_Higgsino}
\end{eqnarray}
where $i=1,2$, $\tgb=\tan\beta$, $y=\mu^2/m^2$ ($m$ is the average squark mass)
and $\{f_5^{(2)}(1),~f_0^{(2)}(1)\}=\{2/15,~1/10\}$, with the complete expressions
for $\{f_5^{(2)}(y),~f_0^{(2)}(y)\}$ given in Appendix~\ref{App:loop_func}.
Eq.~(\ref{Eq:LO_Higgsino}) clearly shows that the (C)EDMs are suppressed by the
external light quark masses, as expected from the invariants of Eq.~(\ref{inv_LL}).
Assuming CKM-like MIs, {\it i.e.} $|(\delta^q_{LL})_{13}| \sim (0.2)^3$,
a typical SUSY mass scale $m_{\rm SUSY}=500\,{\rm GeV}$ and $\tan\beta=10$,
$d_d/e$ and $d^c_d$ are of order $\sim 10^{-28}~{\rm cm}$. For up-type quarks, there 
is also an analogous expression for $d^{(c)}_{u_i}$ proportional to $J^{(u_i)}_{LL}$; 
however, it turns out that $d^{(c)}_{d_i}/d^{(c)}_{u_i}\sim [m_t^2/(m_b^2\tgb^{2})]\times
[A_t\mu/\mu^2]\times\tgb$, thus, $d^{(c)}_{u_i}$ is always negligible.

At one-loop level, the are also gluino contributions to the quark (C)EDMs
through the invariants $J^{(q_i)}_{LR}$ (see Fig.~(\ref{Fig:leading})). The
induced down-type quark (C)EDMs are given as
\begin{eqnarray}
 \left\{ \frac{d_{d_i}}{e},~d^c_{d_i} \right\}_{\tilde g}
 =
 -
 \frac{\alpha_s}{4\pi}
 \frac{m_b}{m^2}
 \frac{M_3 \mu}{m^2}
 \tgb~ 
 {\rm Im}\left[(\delta^d_{LL})_{i3}(\delta^d_{RR})_{3i} \right]
 \left\{
   f_3^{(3)}(x), \,
   f_2^{(3)}(x)
 \right\}\,,
 \label{Eq:LO_gluino}
\end{eqnarray}
where $i=1,2$, $x=M_3^2/m^2$ and $\{f_3^{(3)}(1),~f_2^{(3)}(1)\}=\{4/135,~11/180\}$.
Eq.~(\ref{Eq:LO_gluino}) shows that $d_{d_i}/e$ and $d^c_{d_i}$ are enhanced by the
Yukawa coupling of the heaviest quark and they can reach the level of
$\sim 10^{-(25-26)}\,{\rm cm}$ for $\msusy=500\,{\rm GeV}$,
$|(\delta^d_{LL})_{13}|\sim|(\delta^d_{RR})_{31}|\sim(0.2)^3$ and $\tan\beta=10$.
Thus, the current experimental bounds on the hadronic EDMs already set constraints
on the SUSY parameter space. Similarly, also the up-type quark (C)EDM is generated
at the LO through a gluino mediated diagram and the corresponding prediction is
\begin{eqnarray} 
\label{Eq:up_gluino}
 \left\{ \frac{d_u}{e},~d^c_u \right\}_{\tilde g}
 \!\!\!&=&\!\!\!
 -
 \frac{\alpha_s}{4\pi}\frac{m_t}{m^2}\frac{M_3 A_t}{m^2}
 {\rm Im}\left[(\delta^u_{LL})_{i3}(\delta^u_{RR})_{3i} \right]
 \left\{
   f_4^{(3)}(x), \,
   f_2^{(3)}(x)
 \right\}\,,
 \label{Eq:LO_gluino_up}
\end{eqnarray}
where $x=M_3^2/m^2$ and $\{f_4^{(3)}(1),~f_2^{(3)}(1)\}=\{-8/135,~11/180\}$.

When only the right-handed squarks have mixing, the (C)EDMs are proportional 
to $J^{(q_i)}_{RR}$ and they are generated only at the two-loop level. The most 
relevant effects are provided by non-holomorphic Yukawa coupling corrections,
and they can still lead to sizable contributions, because {\it i)} they are 
enhanced by the heaviest Yukawa coupling via the CKM and the right-handed squark
mixings, and {\it ii)} the non-holomorphic Yukawa couplings entering the (C)EDMs
provide an additional $\tan\beta$ factor that can partially compensate the loop
suppression $\epsilon \sim \alpha_s/(4\pi)$ if $\epsilon\tan\beta \sim 1$.

Passing to the leptonic sector, the dominant contribution to the lepton EDMs arises
from the one-loop exchange of binos/sleptons, and the corresponding EDM, proportional
to $J^{(e_i)}_{LR}$, is given as
\begin{eqnarray}
\label{Eq:lEDM_LO}
 \frac{d_{e_i}}{e}
 \!\!\!&=&\!\!\!
  \frac{\alpha_Y}{8\pi}
  \frac{m_\tau}{m^2}
  \frac{\mu M_1}{m^2}
  \tgb~ 
  {\rm Im}\left[ (\delta^l_{LL})_{i3}(\delta^e_{RR})_{3i} \right]
     f_0^{(3)}(x) \,,
\end{eqnarray}
where $i=1,2$, $x=M_1^2/m^2$ ($m$ is the average slepton mass) and $f_0^{(3)}(1)=-1/15$.
The electron EDM approaches the level of $d_e\sim 10^{-26}\,e\,{\rm cm}$ for
$\msusy=200\,{\rm GeV}$, $|(\delta^l_{LL})_{13}|\sim|(\delta^e_{RR})|\sim (0.2)^3$ and
$\tan\beta=10$.

As stressed above, one peculiar feature of the {\it flavored} EDMs
is that they can result to be proportional to the heaviest Yukawa
couplings. This huge enhancement factor can bring the (C)EDMs 
close to the current and future experimental sensitivities, providing
a splendid opportunity to probe the flavor structure of the MSSM.

On the other hand, the predictions for the {\it flavored} EDMs are not enough
accurate yet. In fact, in order to take into account the contributions arising
from $J^{(q_i)}_{RR}$, two-loop calculations are unavoidable. Moreover, for
large values of $\tan\beta$, the LO predictions for the (C)EDMs discussed in this
section receive significant corrections that have not been accounted for so far.
In the following sections, we discuss the {\it flavored} EDMs at the BLO, stressing
the importance of these contributions to get a correct prediction for the EDMs.

\section{Flavored EDMs at the BLO}
\label{Sec:FEDM}

The effective Lagrangian necessary to evaluate all the relevant BLO effects to the (C)EDMs
includes effective Higgs couplings with fermions and effective fermion-sfermion couplings
with charginos and gluinos. The dominant BLO contributions are computed by including all the
one-loop induced ($\tan\beta$-enhanced) non-holomorphic corrections for the charged Higgs
($H^\pm$) couplings with fermions and for the $\tilde{\chi}^\pm$/$\tilde{g}$ couplings with
fermions-sfermions. The above effective couplings lead to the generation of $H^\pm$ effects
to the (C)EDMs, absent at the LO, via the one-loop $H^\pm$/top-quark exchange. Moreover, the
chargino contributions, suppressed at the LO by the light quark masses, are strongly enhanced
at the BLO by the heaviest-quark Yukawa couplings. Finally, also the gluino effects receive
large BLO contributions that are comparable, in many cases, to the LO ones. As a result, BLO
effects dominate over the LO effects in large regions of the SUSY parameters space and their
inclusion in the evaluation of the flavored EDMs is mandatory. The detailed implementation of
the procedure leading to the above effective Lagrangian and the result in the mass insertion approximation are reported in Appendix~\ref{Sec:effective_vertex}.

As stated before, the BLO expressions for the EDMs are obtained by inserting effective
(one-loop induced) vertices into the one-loop expressions for the EDMs. We would like to
note that such an approach accounts for all the non-decoupling ($\tan\beta$-enhanced)
contributions to the EDMs but it cannot provide the full set of two-loop effects. The
latter requires a full diagrammatic calculation, which is outside the scope of this work.

For instance, the expressions we find for the charged Higgs contributions will be valid as
long as the typical supersymmetric scale $\msusy$ is sufficiently larger than the electroweak
scale $m_{\rm weak}~(\sim m_W,~m_t)$ and the mass of the charged Higgs boson $M_{H^\pm}$.
Therefore, our results can be regarded as the zeroth-order expansion in the parameters 
$(m_{\rm weak}^2,M_{H^\pm}^2)/\msusy^2$ of the full computation. However, it has been shown in Ref.~\cite{borzumati} that this zeroth-order approximation works very well, at least in the
$b\to s\gamma$ case, even for $M_{H^\pm}\geq \msusy$, provided $\msusy$ is sufficiently 
heavier than $m_{\rm weak}$. This finding should also apply to our case, given that both 
$b\to s\gamma$ and the EDMs arise from a similar dipole transition.

In addition, there exist potentially relevant two-loop effects,
proportional to large logarithms of the ratio $\msusy/m_{\rm weak}$, stemming from: {\it i)}
the renormalizations of Yukawa couplings in the Higgs/Higgsino vertices, and {\it ii)} the
anomalous dimensions of the magnetic and chromo-magnetic effective operators. These two classes
of terms become important when the scale of the supersymmetric colored particles is significantly
higher than the $W$ boson and top quark masses.
We account for these effects in the numerical analysis of Section~\ref{Sec:GUT}.

However, we emphasize that the new two-loop effects we are dealing with in the present work
are comparable to and often larger than the leading one-loop effects to the EDMs induced by
gluino/squarks loops. Thus, a full inclusion of all the two-loop effects to the EDMs, although
desirable, is not compulsory.

Keeping the above considerations in mind, in the following sections, we derive the BLO formulae
for the (C)EDM at the SUSY scale. The values attained by the (C)EDMs at the hadron scale 
($\sim 1\,{\rm GeV}$) are obtained according to Ref.~\cite{Degrassi:2005zd}.

For simplicity, we assume degenerate squark masses and we work in the mass insertion
approximation, to make the dependences of the (C)EDMs on the SUSY parameters more 
transparent. 

Since the BLO effects to the quark (C)EDMs mostly affect the predictions for the down-type quark
(C)EDMs, hereafter we do not discuss the subdominant BLO effects to the up-type quark (C)EDMs.

\subsection{Gluino Contribution to Down-type Quark (C)EDMs}
\begin{figure}[t]
\begin{center}
\begin{tabular}{cc}
\includegraphics[scale=0.4]{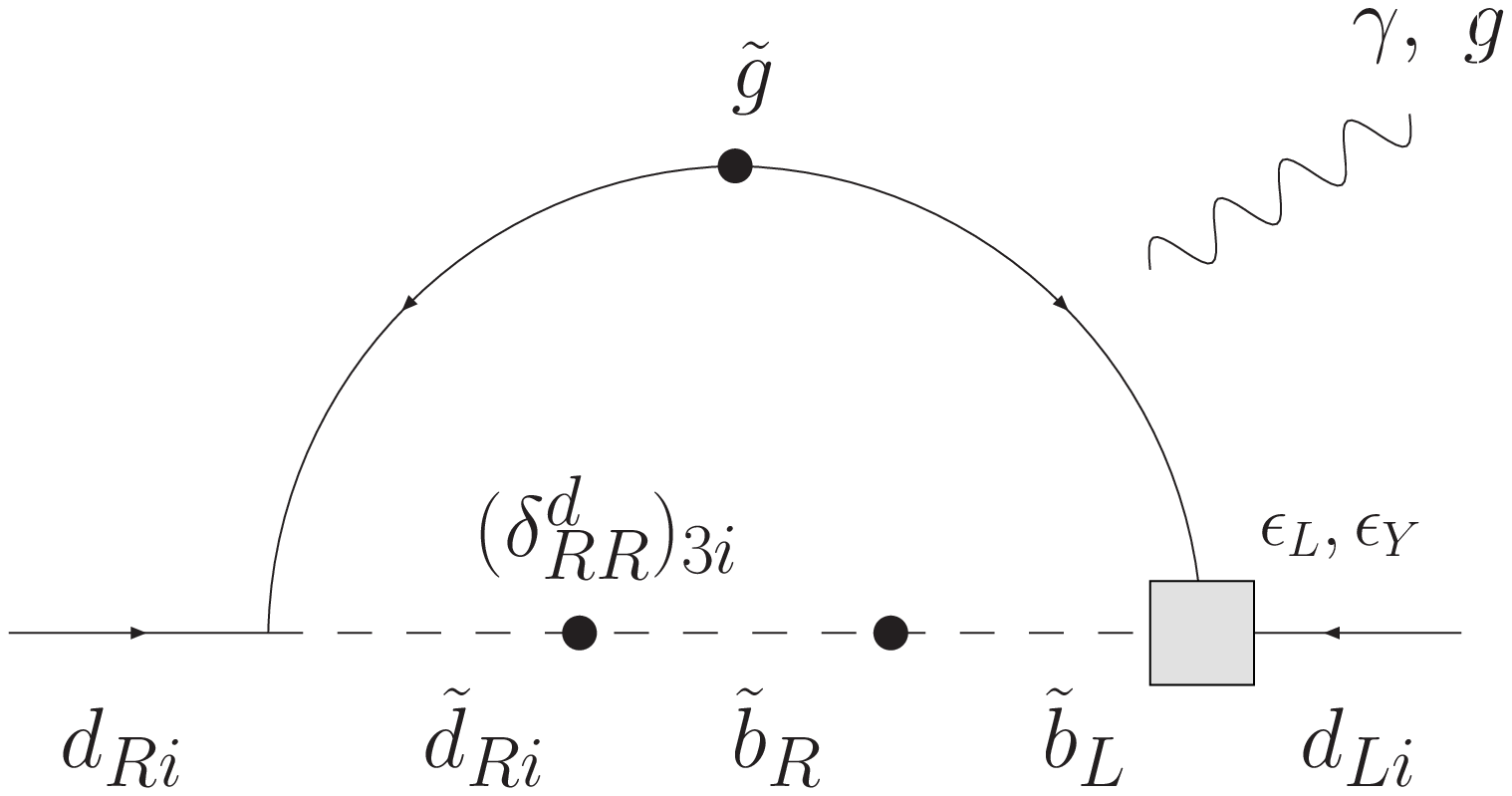} &
\includegraphics[scale=0.4]{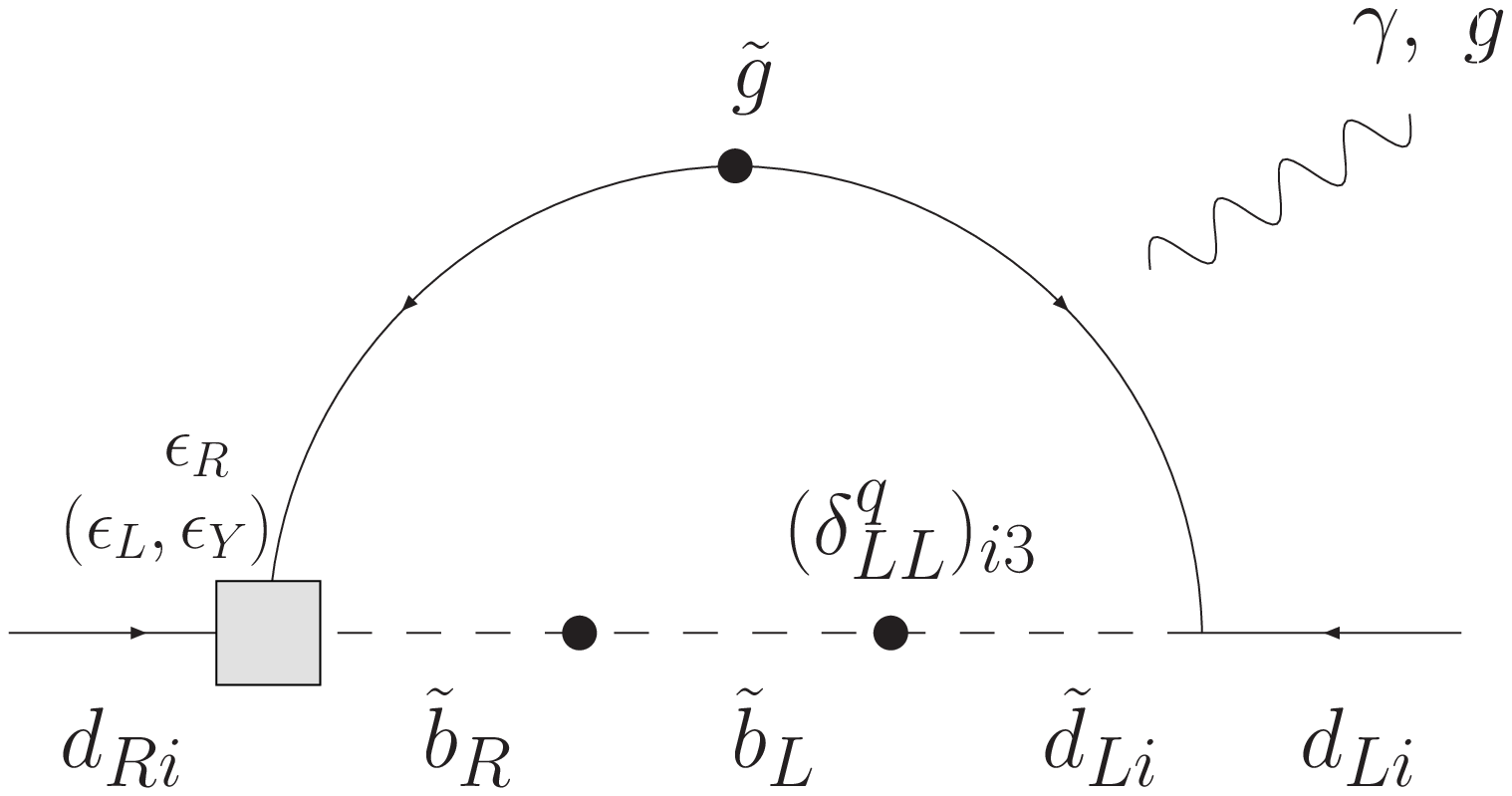} 
\end{tabular}
\vskip 0.5 cm
\begin{tabular}{ccc}
\includegraphics[scale=0.32]{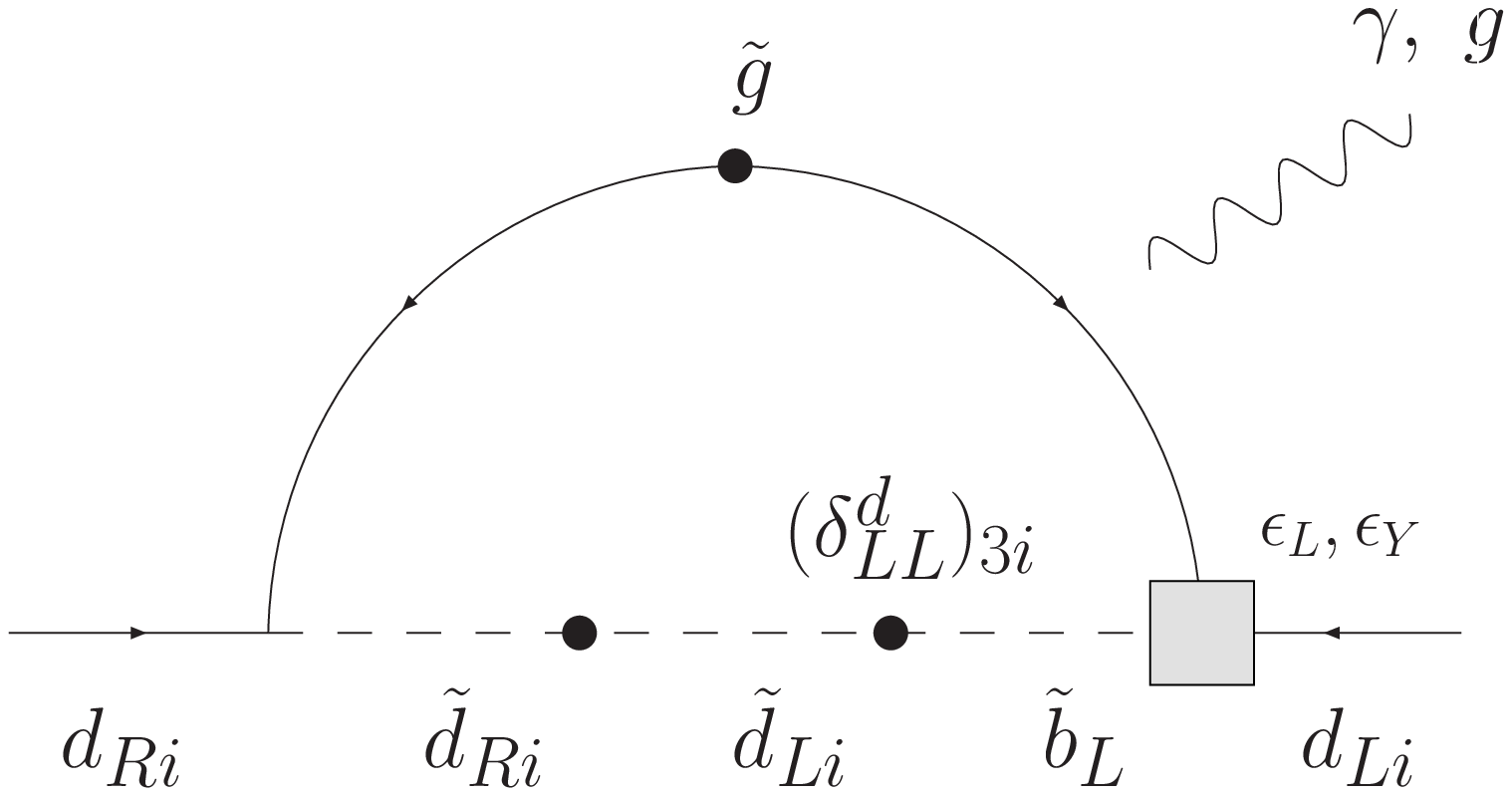} &
\includegraphics[scale=0.32]{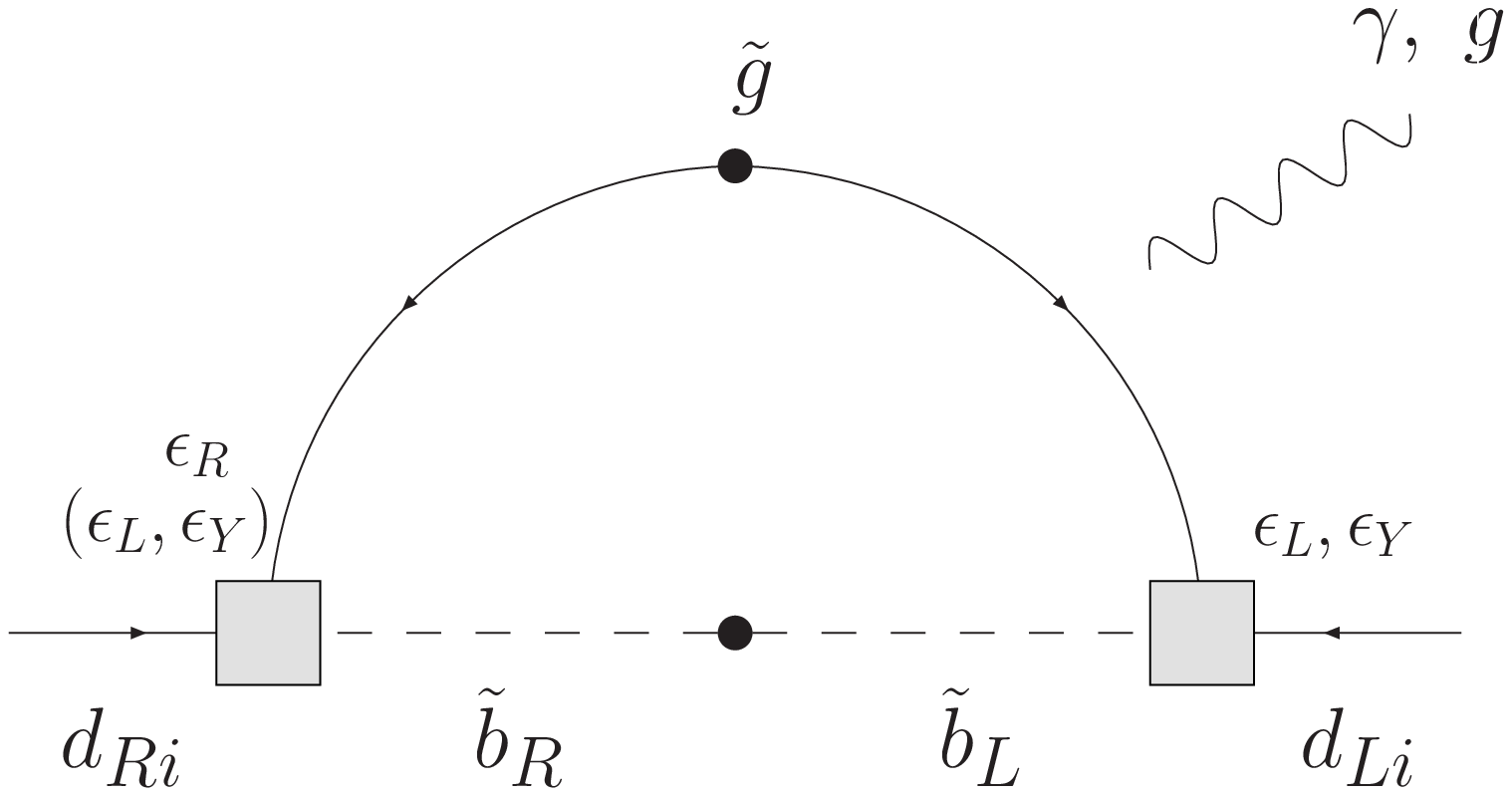} &
\includegraphics[scale=0.32]{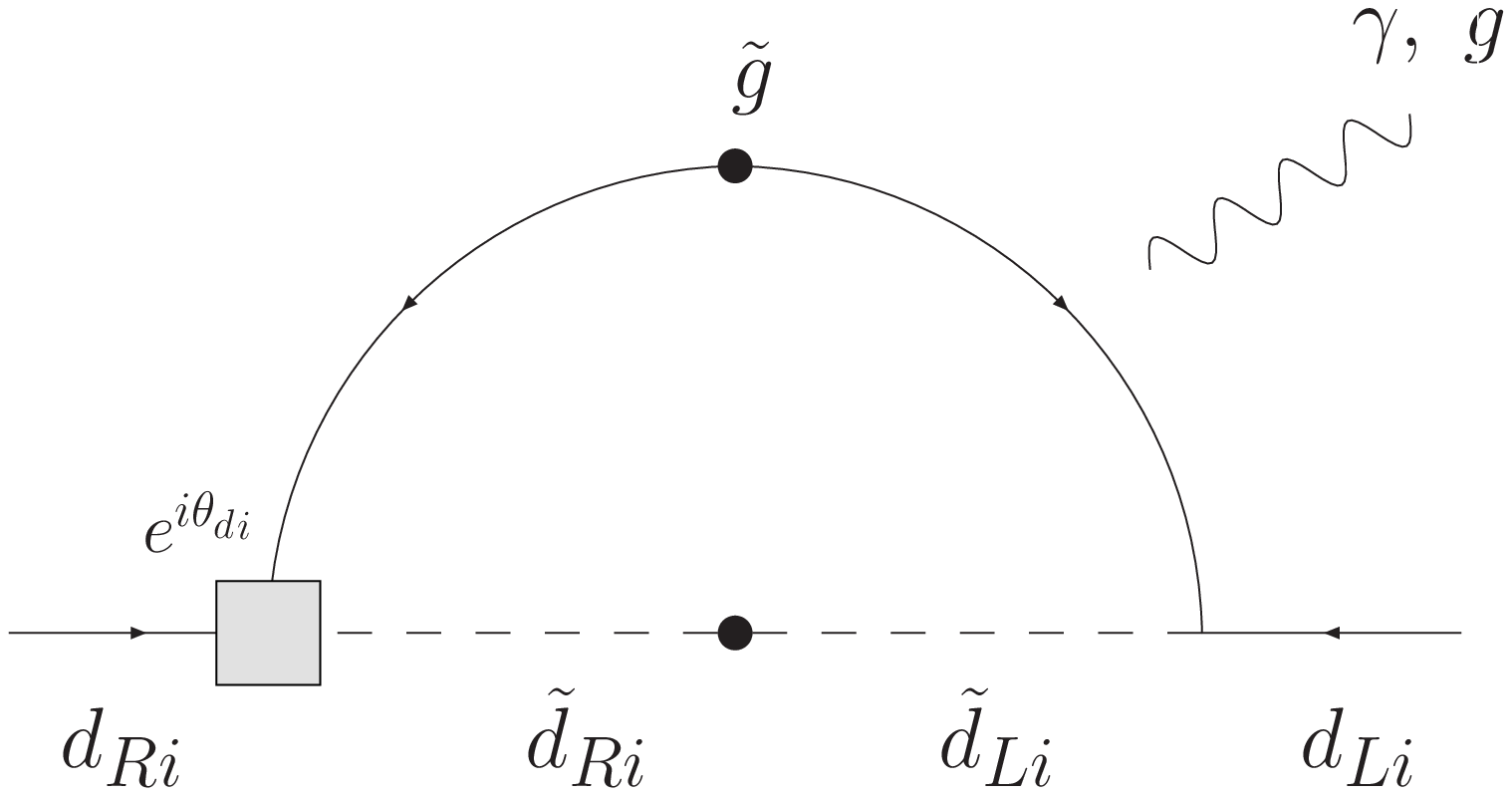} 
\end{tabular}
\caption{\label{Fig:gluino} Gluino-mediated contributions for down-type quark (C)EDMs
beyond the leading order. In each diagrams gray boxes indicate the effective couplings
generated by non-holomorphic Yukawa corrections.}
\end{center}
\end{figure}

The relevant gluino-mediated contributions entering the evaluation of the down-type quark
(C)EDMs at the BLO are reported in Fig.~\ref{Fig:gluino}. As we can see, the leading BLO
effects are basically obtained from the LO ones by means of the replacement of the tree-level
couplings with the effective couplings derived in the Appendices.

At the BLO, contributions proportional to all the JIs, $J^{(d_i)}_{LR}$, $J^{(d_i)}_{LL}$ 
and $J^{(d_i)}_{RR}$ naturally arise, in contrast to the LO case where only $J^{(d_i)}_{LR}$
and $J^{(d_i)}_{LL}$ contribute to the (C)EDMs. As a result, the total BLO gluino-mediated
contributions to the down-type quark (C)EDMs are given by
\begin{eqnarray}
 \left\{ \frac{d_{d_i}}{e},~d^c_{d_i} \right\}_{\tilde g}
 \!\!\!&=&\!\!\!
 -
 \frac{\alpha_s}{4\pi}
 \frac{m_b}{m^2}
 \frac{M_3 \mu}{m^2}
 \tgb~ 
 \Biggl(
  {\cal E}^{\tilde g}_1 ~
  {\rm Im}\left[(\delta^d_{LL})_{i3}(\delta^d_{RR})_{3i} \right]
 \nonumber\\
 && \qquad
  +~
  {\cal E}^{\tilde g}_2 ~
  {\rm Im}\left[ V^*_{3i}(\delta^d_{RR})_{3i} \right]
  + 
  {\cal E}^{\tilde g}_3 ~
  \frac{m_{d_i}}{m_b}
  {\rm Im}\left[ V^*_{3i}(\delta^d_{LL})_{3i} \right]
 \Biggr)\,,
\end{eqnarray}
where
\begin{eqnarray}
 {\cal E}^{\tilde g}_1 
 \!\!\!&=&\!\!\!
  \frac{f_{\tilde g}^{(3)}(x)}{1+\eps_3\tgb}
  +
  \frac{1}{3(1+\eps_3\tgb)}
  \left(
   \frac{\epsR\tgb}{1+\eps_3\tgb}
   +
   \frac{\epsL\tgb}{1+\epsb_3\tgb}
  \right)
  f_{\tilde g}^{(2)}(x)
 \nonumber\\
 &&
  +~
  \left(
   \frac{(1+r_i)\epsL\epsR \ttgb}{9(1+\eps_3\tgb)^2(1+\epsb_3\tgb)}
   -
   \frac{r_i \epsLR\tgb}{6(1+\eps_3\tgb)^2}
  \right)
  f_{\tilde g}^{(1)}(x)\,,
 \\
 {\cal E}^{\tilde g}_2
 \!\!\!&=&\!\!\!
  -
  \frac{\epsY \tgb}{(1+\eps_3\tgb)(1+\epsb_3\tgb)}
  f_{\tilde g}^{(2)}(x)
  -
  \frac{(1+r_i)\epsR\epsY \ttgb}{3(1+\eps_3\tgb)^2 (1+\epsb_3\tgb)}
  f_{\tilde g}^{(1)}(x)\,,
 \\
 {\cal E}^{\tilde g}_3
 \!\!\!&=&\!\!\!
  \left(
   \frac{r_i^2 \epsLY\tgb}{3(1+\eps_3\tgb)^2}
   -\frac{\bar{r}_i \epsLY\epsY\ttgb}
   {3(1+\epsb_3\tgb)^2(1+\eps_3\tgb)}
  \right)
  f_{\tilde g}^{(1)}(x)
 \nonumber\\
 &&
  +~
  \frac{\bar{r}_{i}\epsY\tgb}{(1+\epsb_3\tgb)(1+\eps_3\tgb)}
  f_{\tilde g}^{(2)}(x)\,,
\end{eqnarray}
and
\begin{eqnarray}
 f_{\tilde g}^{(i)}(x)
 = 
 \left\{ 
  - \frac{4}{9} f_0^{(i)}(x)\,,~
  - \frac{1}{6} f_0^{(i)}(x)
  + \frac{3}{2} f_1^{(i)}(x)
 \right\}
 \quad (i=1,2,3)\,,
\end{eqnarray}
where $x=M_3^2/m^2$ and the loop functions satisfy $f_{\tilde g}^{(1,2,3)}(1)=$ $\{2/27,5/18\}$, $\{-2/45,~-7/60\}$, $\{4/135,~11/180\}$ whereas their full expressions are listed in Appendix~\ref{App:loop_func}.
We also use the short hand notations $\epsL$, $\epsR$, $\eps_{LR}$, $\epsLY$, $z_Y$ and $z_L$
instead of $\epsL_{3i}$, $\epsR_{3i}$, $\epsLR_{ii}$, $\epsLY_{i}$, $\RE[z_{Y_i}]$ and 
$\RE[z_{L_i}]$, respectively.
The above loop factors are defined in Appendix~\ref{App:loop_factor} and for equal SUSY masses
and small MI parameters, they are given as $\epsLR=\epsL=\epsR=\epsb_3=\eps_i={\rm sign}(\mu M_3)
\times \alpha_s/3\pi$, $\epsY=\epsLY=-{\rm sign}(\mu A_t)\times y_t^2/32\pi^2$,
We also define $\eps_3=\epsb_3+\epsY$ and the $r_i$ and $\bar{r}_{i}$ parameters such that
\begin{eqnarray}
  r_i        = \frac{ 1+\eps_3\tgb }{|1+\eps_i\tgb+\eps^{(2)}_i\ttgb|}\,,\qquad\qquad
 \bar{r}_{i} = \frac{|1+\eps_i\tgb|}{|1+\eps_i\tgb+\eps^{(2)}_i\ttgb|}\,.
\end{eqnarray}
Terms proportional to ${\cal E}^{\tilde g}_1$, ${\cal E}^{\tilde g}_2$ and ${\cal E}^{\tilde g}_3$
contribute to the Jarlskog invariants $J^{(d_i)}_{LR}$, $J^{(d_i)}_{RR}$ and $J^{(d_i)}_{LL}$, respectively.

\subsection{Chargino Contribution to Down-type Quark (C)EDMs}
As discussed in previous sections, chargino contributions to the down-type quark
(C)EDMs arise already at the LO through a contribution proportional to $J^{(d_i)}_{LL}$,
hence suppressed by the light quark masses. At the BLO, new contributions proportional
to both $J^{(d_i)}_{RR}$ and $J^{(d_i)}_{LR}$ (thus proportional to the heaviest down-quark
mass $m_b$) arise.

\begin{figure}[t]
\begin{center}
\begin{tabular}{ccc}
\includegraphics[scale=0.32]{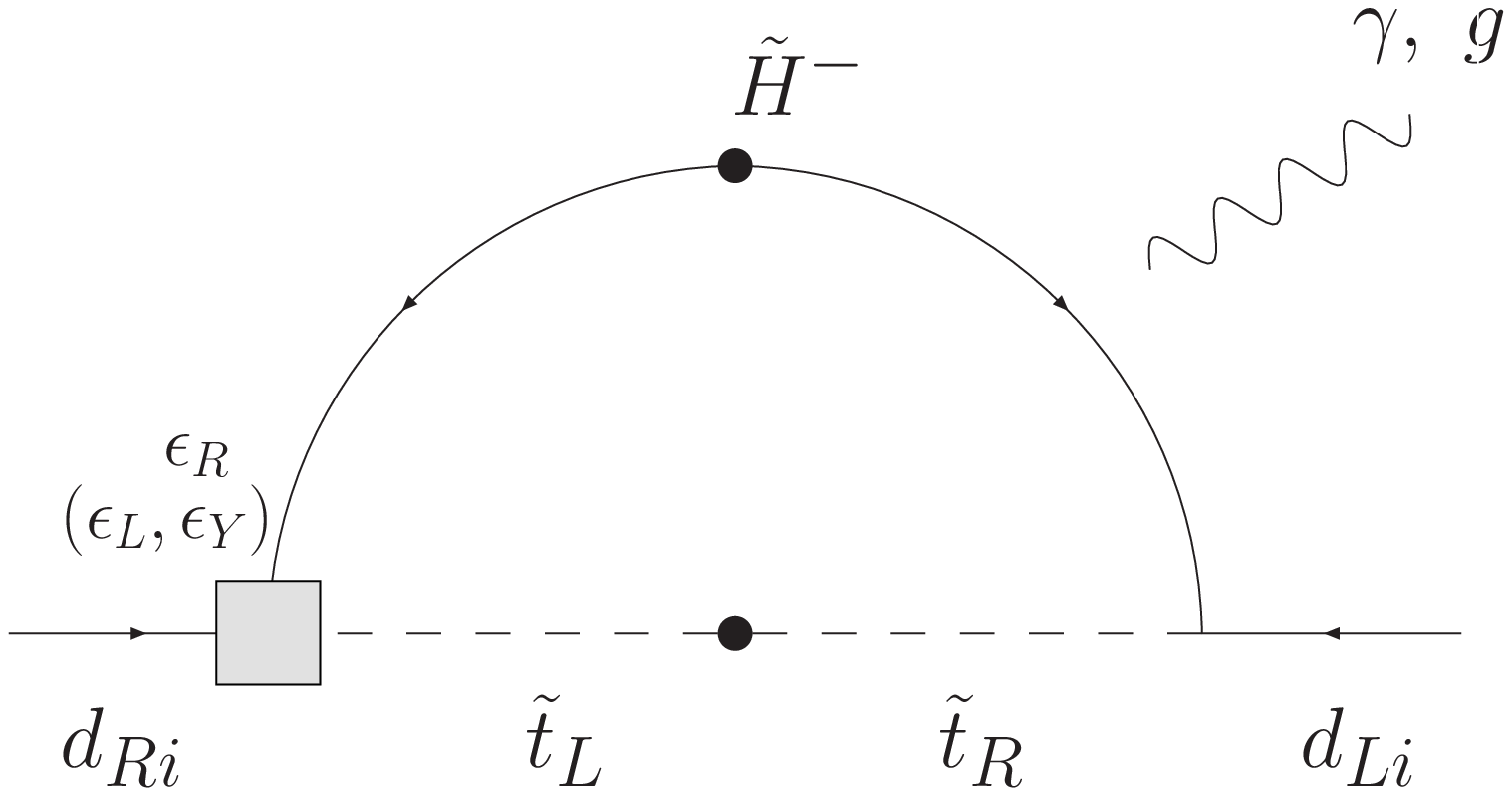} &
\includegraphics[scale=0.32]{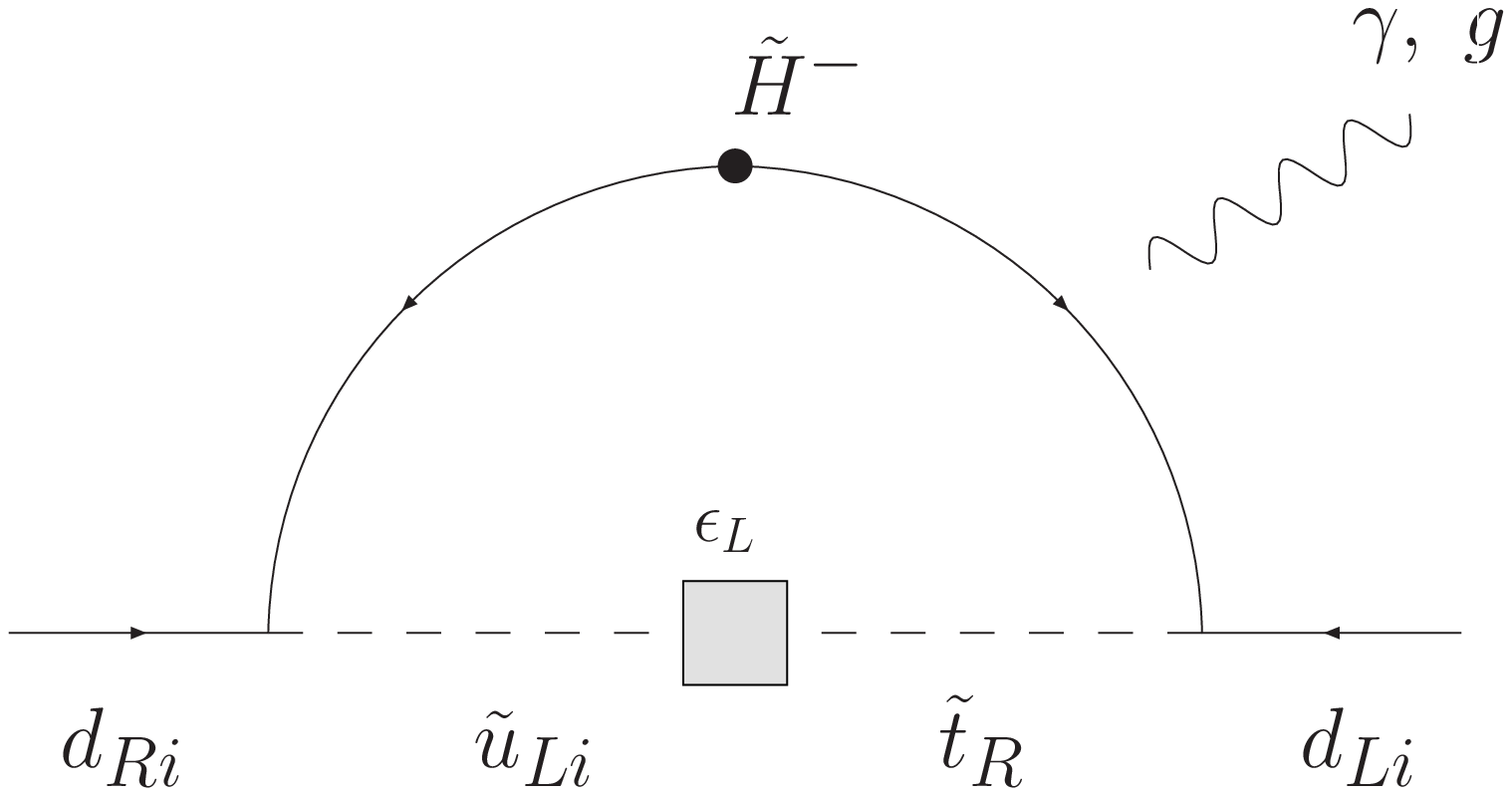} &
\includegraphics[scale=0.32]{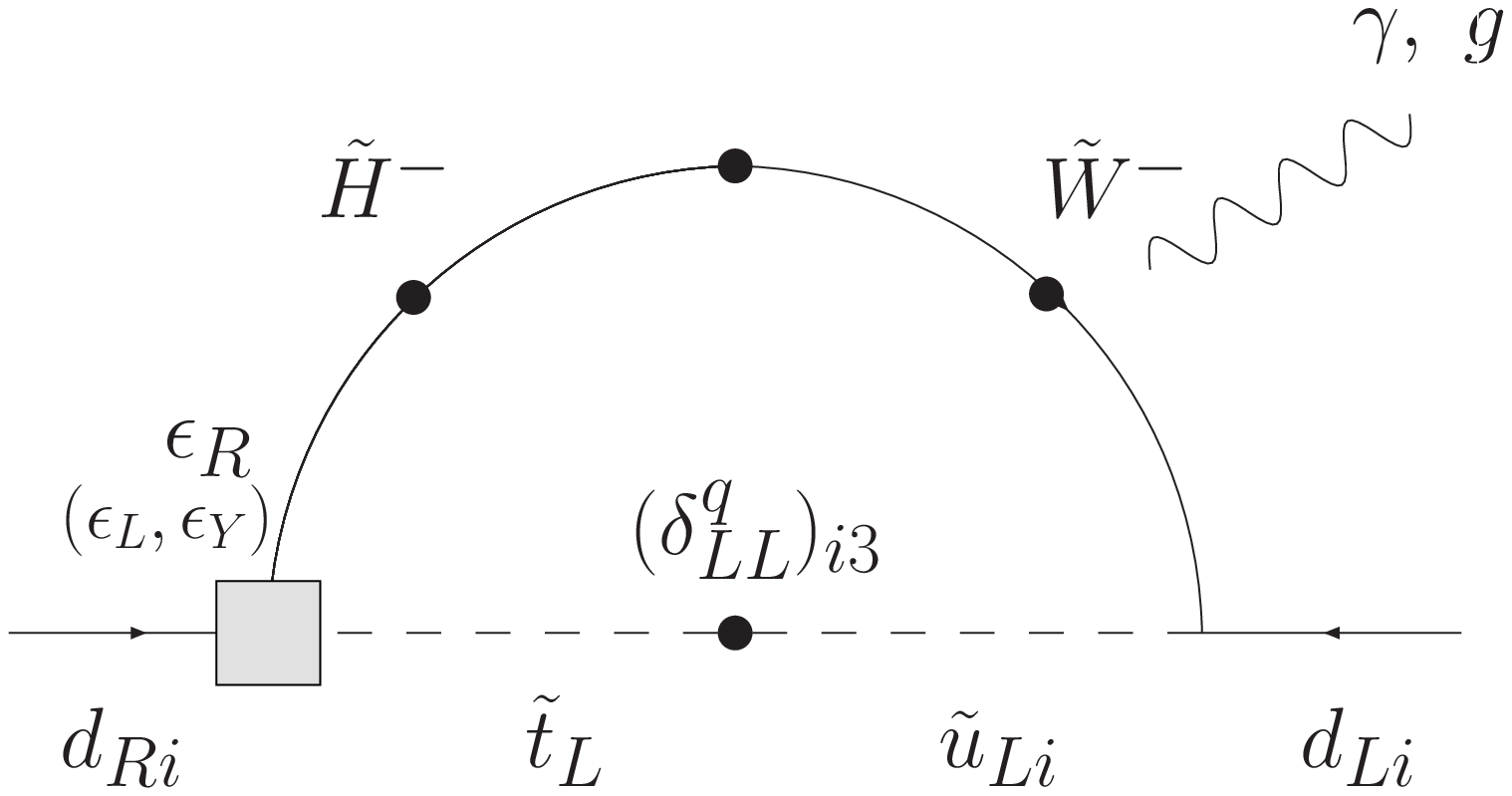} 
\end{tabular}
\vskip 0.5 cm
\begin{tabular}{ccc}
\includegraphics[scale=0.32]{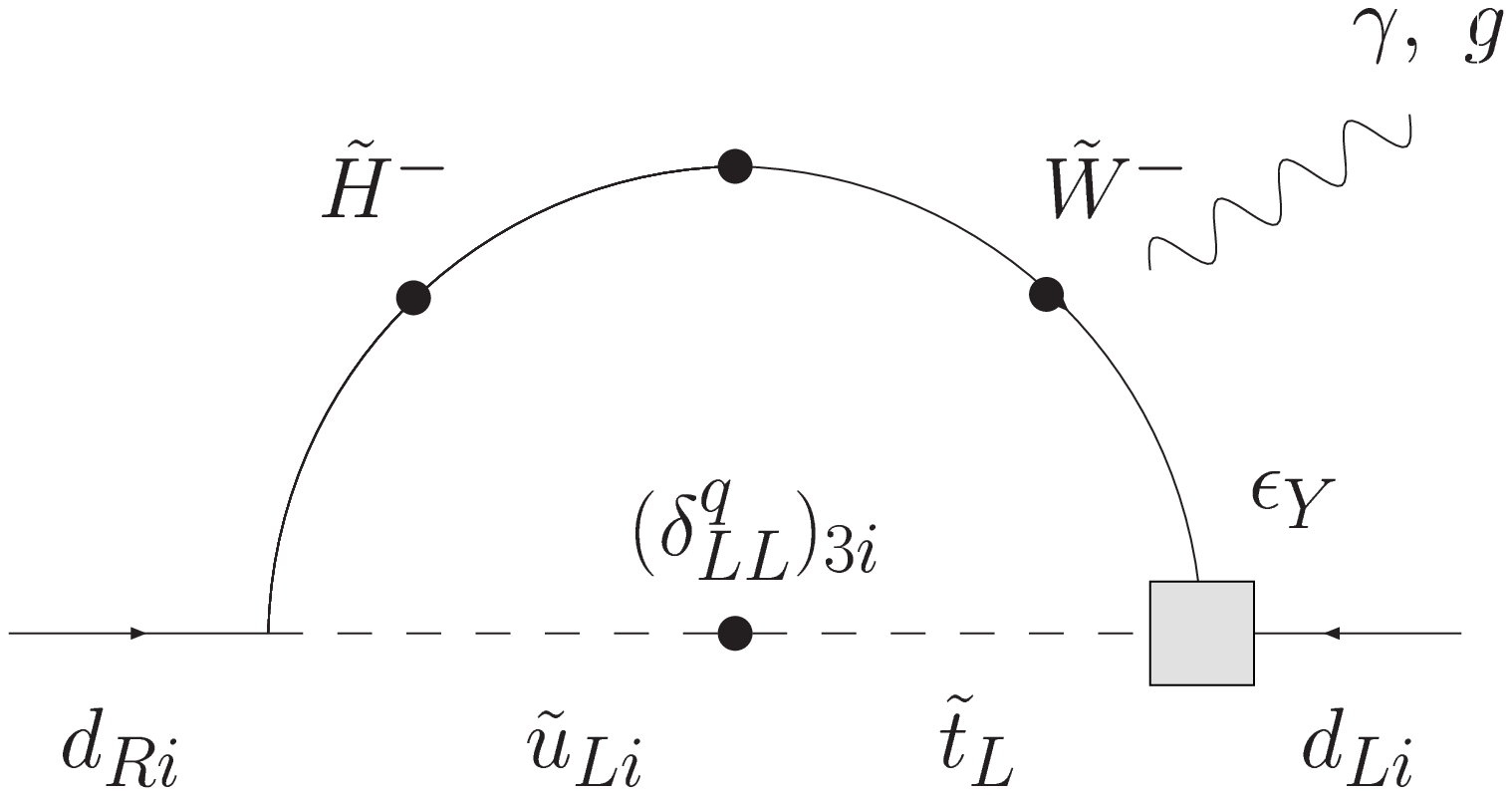} &
\includegraphics[scale=0.32]{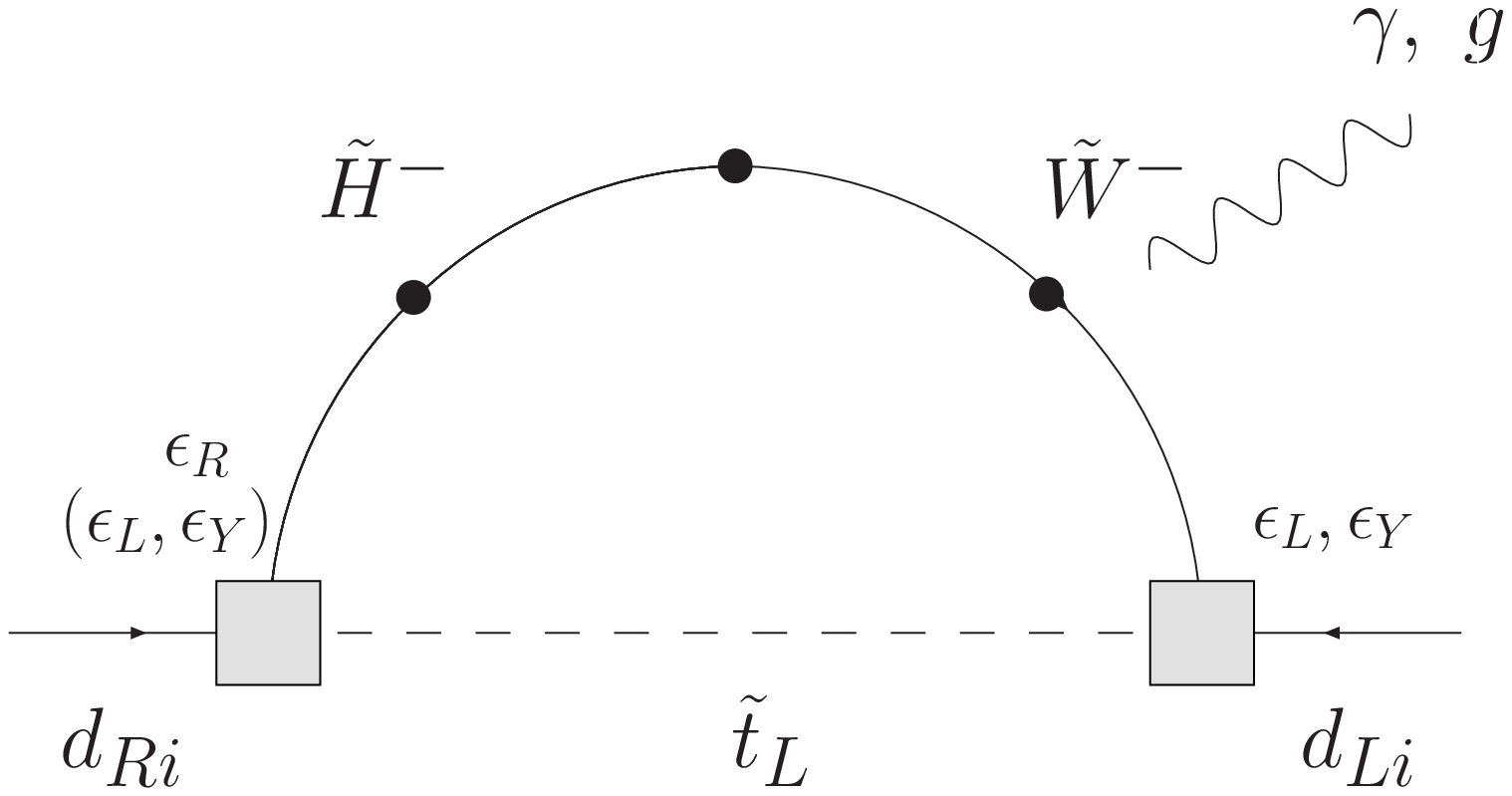} &
\includegraphics[scale=0.32]{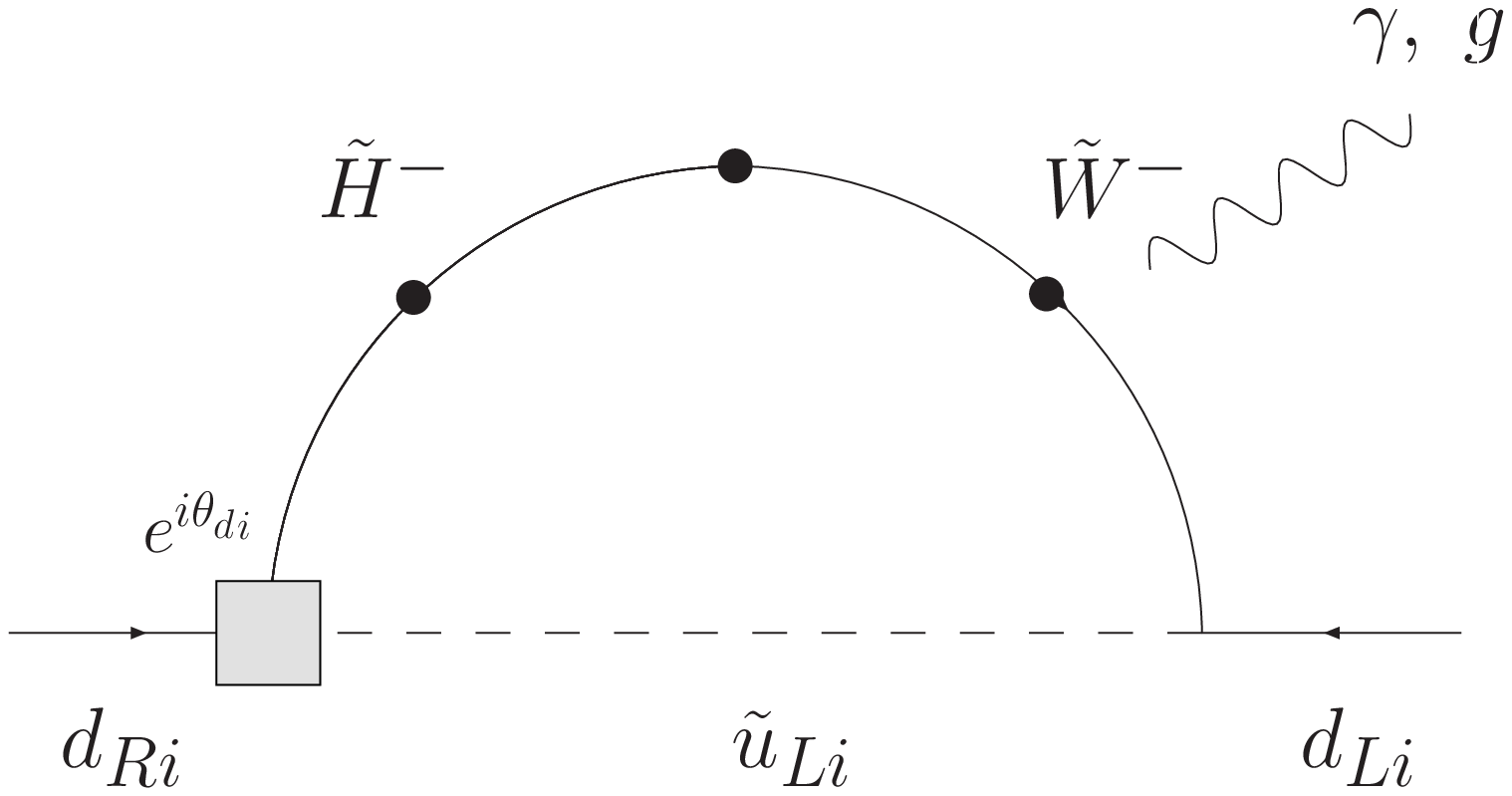}
\end{tabular}
\caption{\label{Fig:chargino} Chargino-mediated contributions for down-type quark (C)EDMs
beyond the leading order. The effective couplings with $\tan\beta$-enhanced corrections
are depicted by gray boxes.}
\end{center}
\end{figure}

Combining the BLO contributions from the upper diagrams of Fig.~\ref{Fig:chargino} with the
LO contribution of Eq.~(\ref{Eq:LO_Higgsino}), the total charged Higgsino contribution reads
\begin{eqnarray}
 \left\{ \frac{d_{d_i}}{e},~d^c_{d_i} \right\}_{{\tilde H}^\pm}
 \!\!\!\!\!&=&\!\!\!
 \frac{\alpha_2}{16\pi}
 \frac{m_b}{m^2}
 \frac{m_t^2}{m_W^2}
 \frac{A_t \mu}{m^2}
 \tgb
 \Biggl(
  {\cal E}^{{\tilde H}^\pm}_2
  {\rm Im}\left[ V^*_{3i}(\delta^d_{RR})_{3i} \right]
 \nonumber\\
 &&\qquad
  +~
  {\cal E}^{{\tilde H}^\pm}_3
  \frac{m_{d_i}}{m_b}
  {\rm Im}\left[ V^*_{3i}(\delta^d_{LL})_{3i} \right]
 \Biggr)\,,
\label{Eq:FEDM_Higgsino}
\end{eqnarray}
where
\begin{eqnarray}
 {\cal E}^{{\tilde H}^\pm}_2 
 \!\!\! &=& \!\!\!
 \frac{\epsR\tgb}{3(1+\eps_3\tgb)^2}
 f_{{\tilde H}^\pm}^{(1)}(y)\,,
 \\
 {\cal E}^{{\tilde H}^\pm}_3
 \!\!\! &=& \!\!\!
 \frac{r_i}{(1+\eps_3\tgb)}
  f_{{\tilde H}^\pm}^{(2)}(y)
  +
  \frac{\bar{r}_i(\epsLY+\epsL)\tgb}{3(1+\epsb_3\tgb)(1+\eps_3\tgb)}
  f_{{\tilde H}^\pm}^{(1)}(y)\,,
\end{eqnarray}
and
\begin{eqnarray}
 f_{{\tilde H}^\pm}^{(i)}(y) = \left\{\frac{2}{3}f_0^{(i)}(y)-f_1^{(i)}(y),~
 f_0^{(i)}(y)\right\}\quad (i=1,2)\,,
\end{eqnarray}
where $y=\mu^2/m^2$ and $ f_{{\tilde H}^\pm}^{(1,2)}(1)=\{-5/18,-1/6\},\,\{2/15,~1/10\}$.
The charged Higgsino contributions play a very important role in the calculation of
the EDMs resulting often dominant over the LO gluino-mediated contribution.
Additionally, there exist also contributions from charged Higgsino-wino mixing diagrams
as shown in the lower diagrams of Fig.~\ref{Fig:chargino}. The final expression of
all these contributions reads
\begin{eqnarray}
 \left\{ \frac{d_{d_i}}{e},~d^c_{d_i} \right\}_{{\tilde H}^\pm{\tilde W}^\pm}
 \!\!\!\!\!\!\!\!&=&\!\!\!\!\!
 \frac{\alpha_2}{8\pi}
 \frac{m_b}{m^2}
 \frac{\mu M_2}{m^2}
 \tgb
 \Biggl(
  {\cal E}^{{\tilde H}^\pm{\tilde W}^\pm}_1~
  {\rm Im}\left[ (\delta^d_{LL})_{i3}(\delta^d_{RR})_{3i} \right]
 \nonumber\\
 && 
  +~
  {\cal E}^{{\tilde H}^\pm{\tilde W}^\pm}_2
  {\rm Im}\left[ V^*_{3i}(\delta^d_{RR})_{3i} \right]
  +
  {\cal E}^{{\tilde H}^\pm{\tilde W}^\pm}_3
  \frac{m_{d_i}}{m_b}
  {\rm Im}\left[ V^*_{3i}(\delta^d_{LL})_{3i} \right]
 \Biggr)\,,
\end{eqnarray}
where
\begin{eqnarray}
 {\cal E}^{{\tilde H}^\pm{\tilde W}^\pm}_1 
 \!\!\!\!\!\!\! &=& \!\!\!
  \frac{\epsR\tgb}{3(1+\eps_3\tgb)^2}
  g_{\tilde{H}^\pm\tilde{W}^\pm}^{(1)}(x,y)
 \nonumber\\
 && \quad
  +~
  \left(
   \frac{(1+r_i)\epsR\epsL\ttgb}{9(1+\eps_3\tgb)^2(1+\epsb_3\tgb)}
   -
   \frac{r_i \epsLR\tgb}{6(1+\eps_3\tgb)^2}
  \right)
  g_{{\tilde H}^\pm{\tilde W}^\pm}^{(0)}(x,y)\,,
 \\
 {\cal E}^{{\tilde H}^\pm{\tilde W}^\pm}_2 
 \!\!\!\!\!\!\! &=& \!\!\!
 -
 \frac{(1+r_i)\epsR\epsY\ttgb}{3(1+\eps_3\tgb)^2(1+\epsb_3\tgb)}
 g_{{\tilde H}^\pm{\tilde W}^\pm}^{(0)}(x,y)\,,
 \\
 {\cal E}^{{\tilde H}^\pm{\tilde W}^\pm}_3
 \!\!\!\!\!\!\! &=& \!\!\!
  \left(
   \frac{r_i^2 \epsLY \tgb}{3(1+\eps_3\tgb)^2}
   -\frac{\bar{r}_i\epsLY\epsY \ttgb}
   {3(1+\epsb_3\tgb)^2(1+\eps_3\tgb)}
  \right)
 g_{{\tilde H}^\pm{\tilde W}^\pm}^{(0)}(x,y)
 \nonumber\\
 && \quad
 +~
 \frac{\bar{r}_i\epsY\tgb}{(1+\epsb_3\tgb)(1+\eps_3\tgb)}
 g_{{\tilde H}^\pm{\tilde W}^\pm}^{(1)}(x,y)\,,
\end{eqnarray}
and
\begin{eqnarray}
 g_{{\tilde H}^\pm{\tilde W}^\pm}^{(i)}(x,y)
 = 
 \left\{ 
  \frac{2}{3} g_0^{(i)}(x,y)
  - g_1^{(i)}(x,y)\,,~
  g_0^{(i)}(x,y)
 \right\}
 \quad (i=0,1)\,,
\end{eqnarray}
where $x=M_2^2/m^2$, $y=\mu^2/m^2$ and $g_{{\tilde H}^\pm{\tilde W}^\pm}^{(0,1)}(1,1)=$ $\{-11/18,~-1/6\}$, $\{13/45,~2/15\}$.
The explicit expressions of the loop functions are presented in Appendix~\ref{App:loop_func}.

\subsection{Neutralino Contribution to Down-type Quark (C)EDMs}

\begin{figure}[thbp]
\begin{center}
\begin{tabular}{ccc}
\includegraphics[scale=0.32]{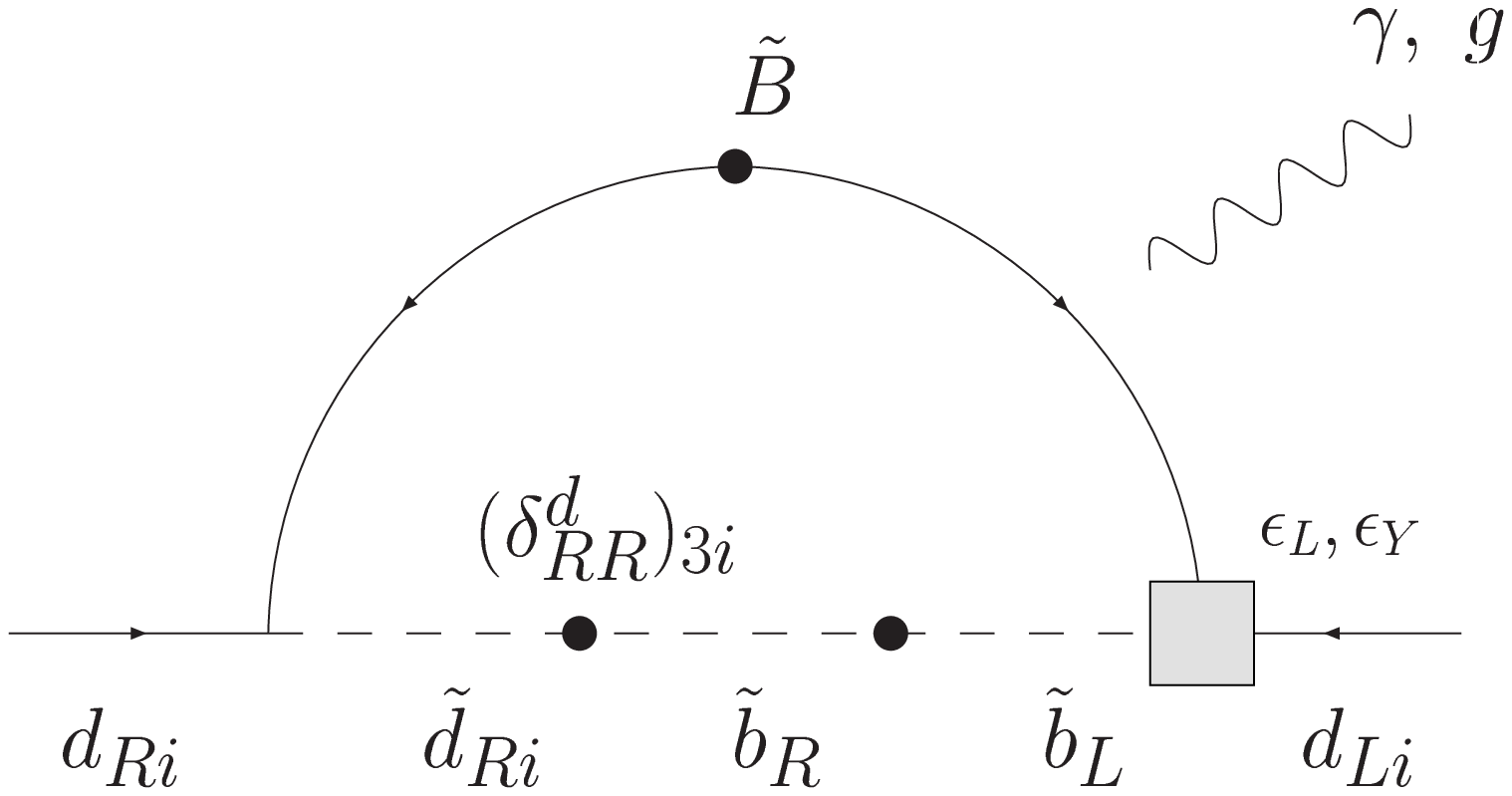} &
\includegraphics[scale=0.32]{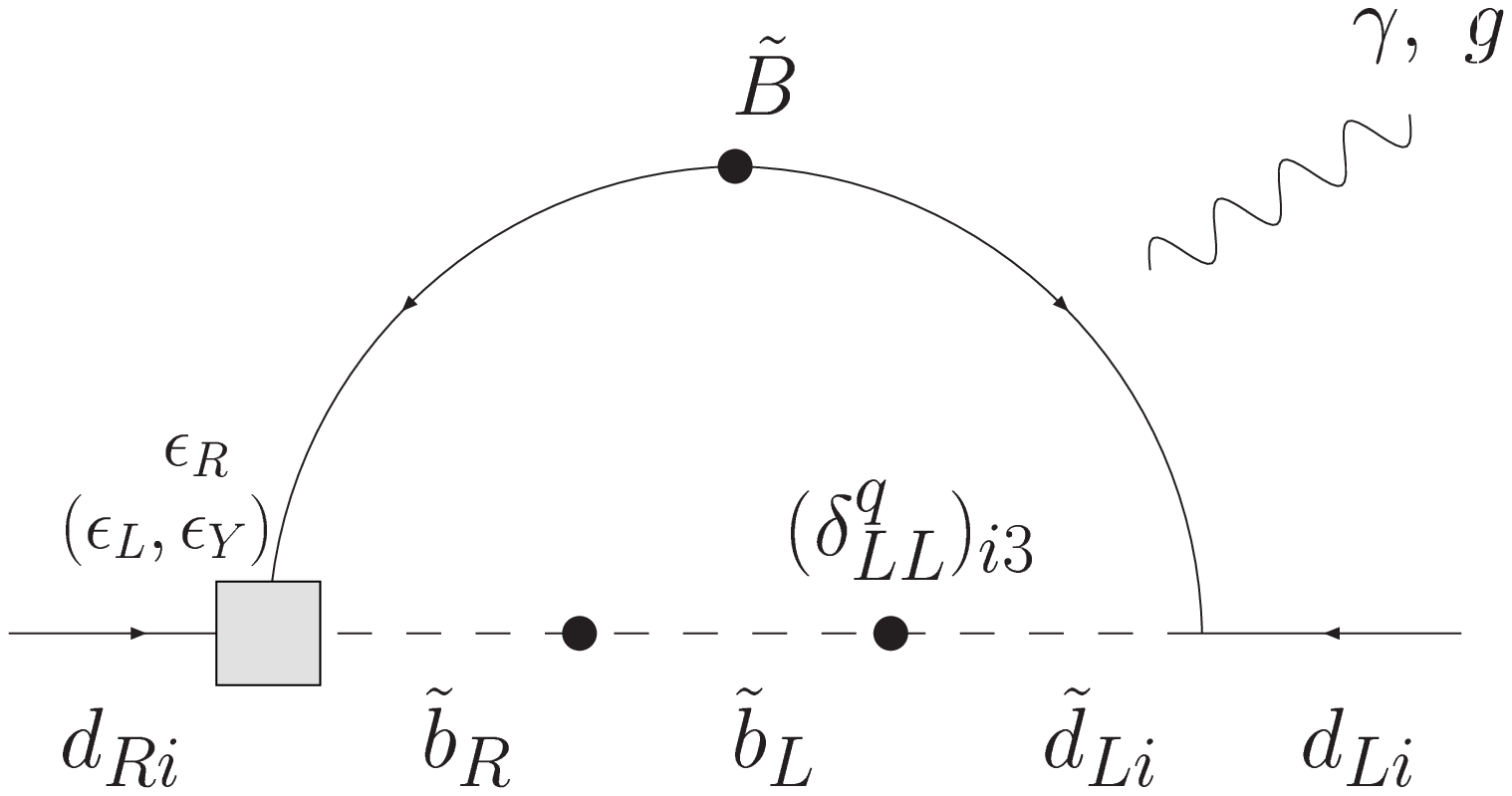} &
\includegraphics[scale=0.32]{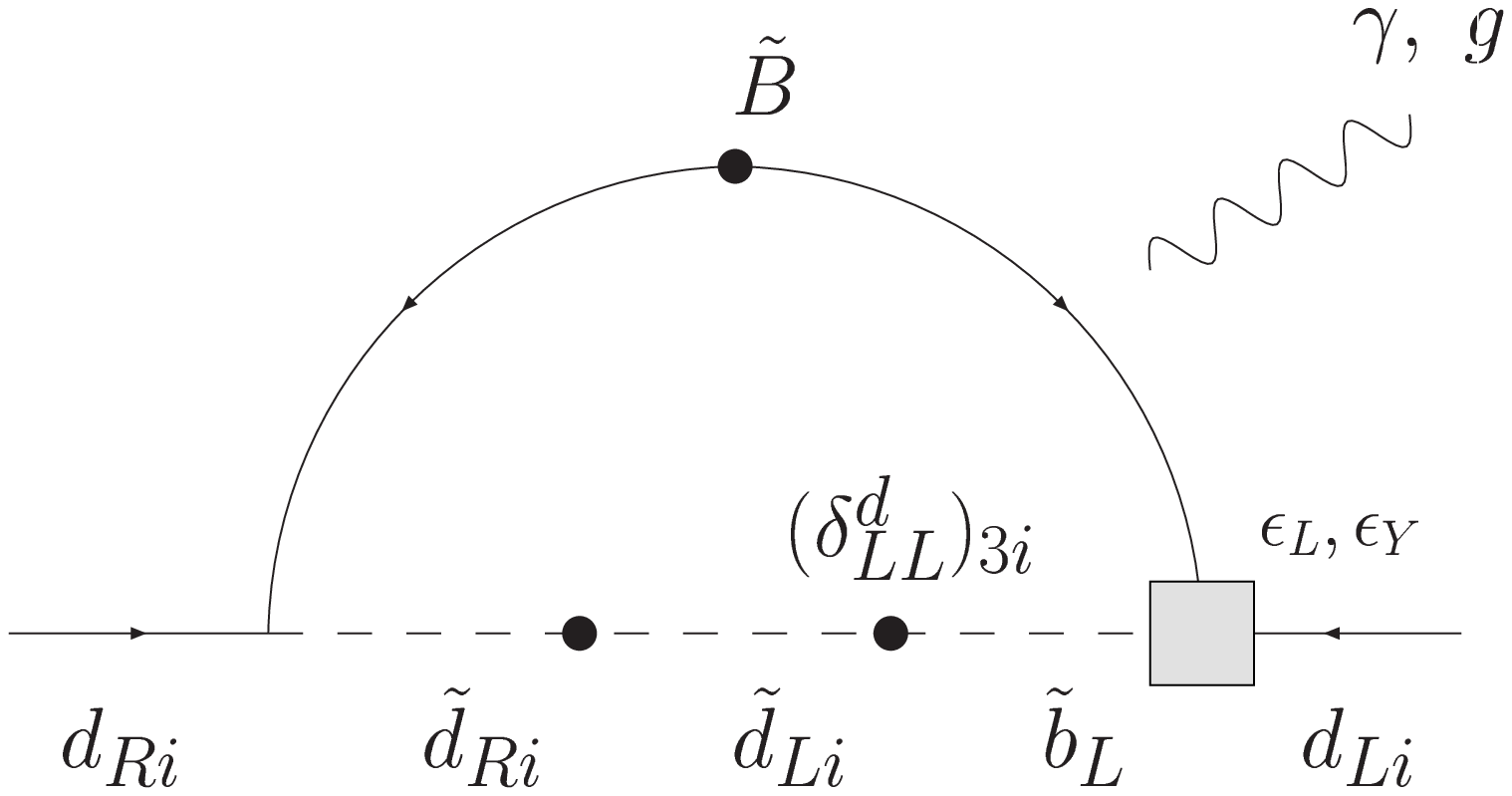} 
\end{tabular}
\vskip 0.5 cm
\begin{tabular}{ccc}
\includegraphics[scale=0.32]{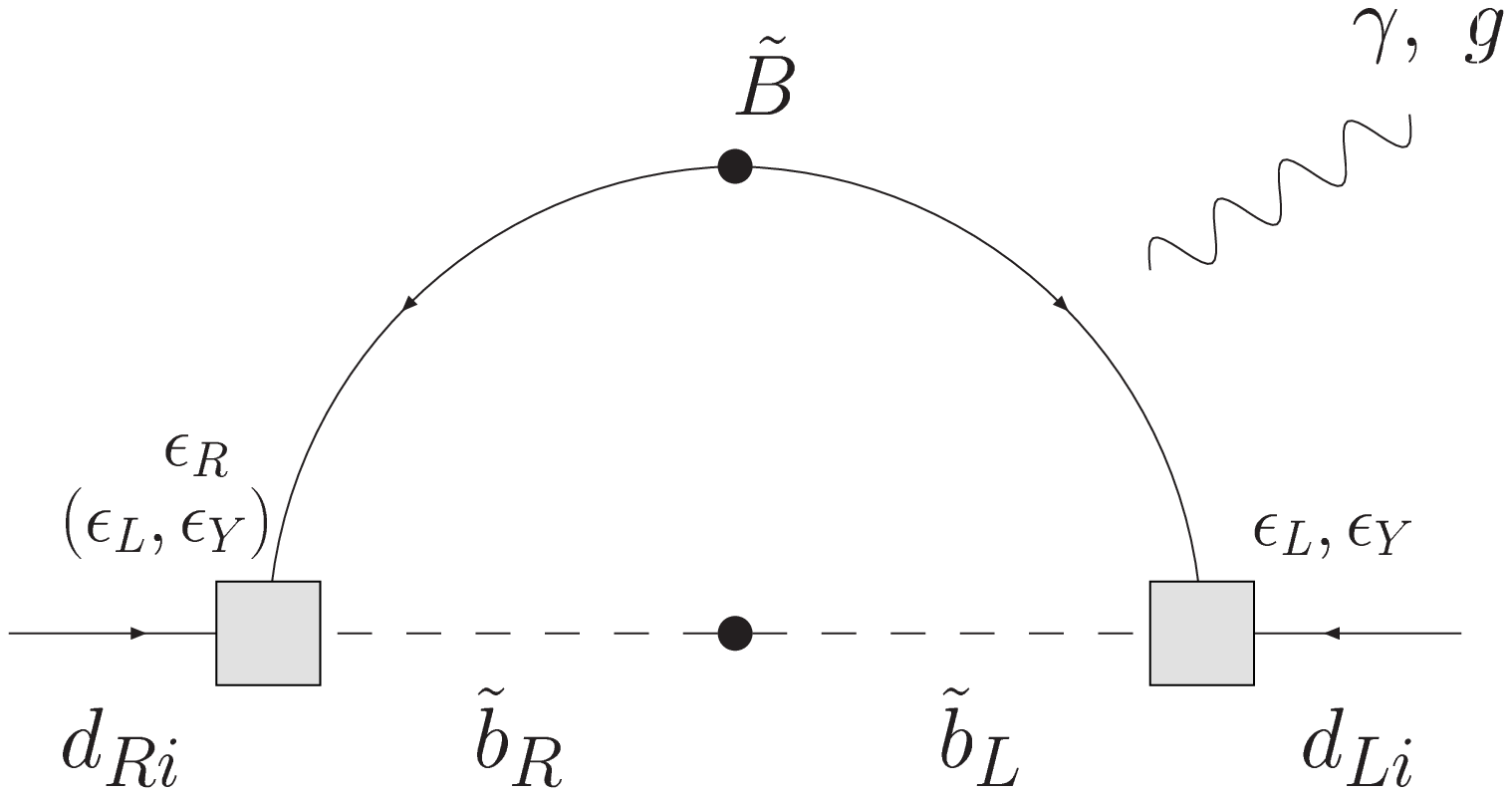} &
\includegraphics[scale=0.32]{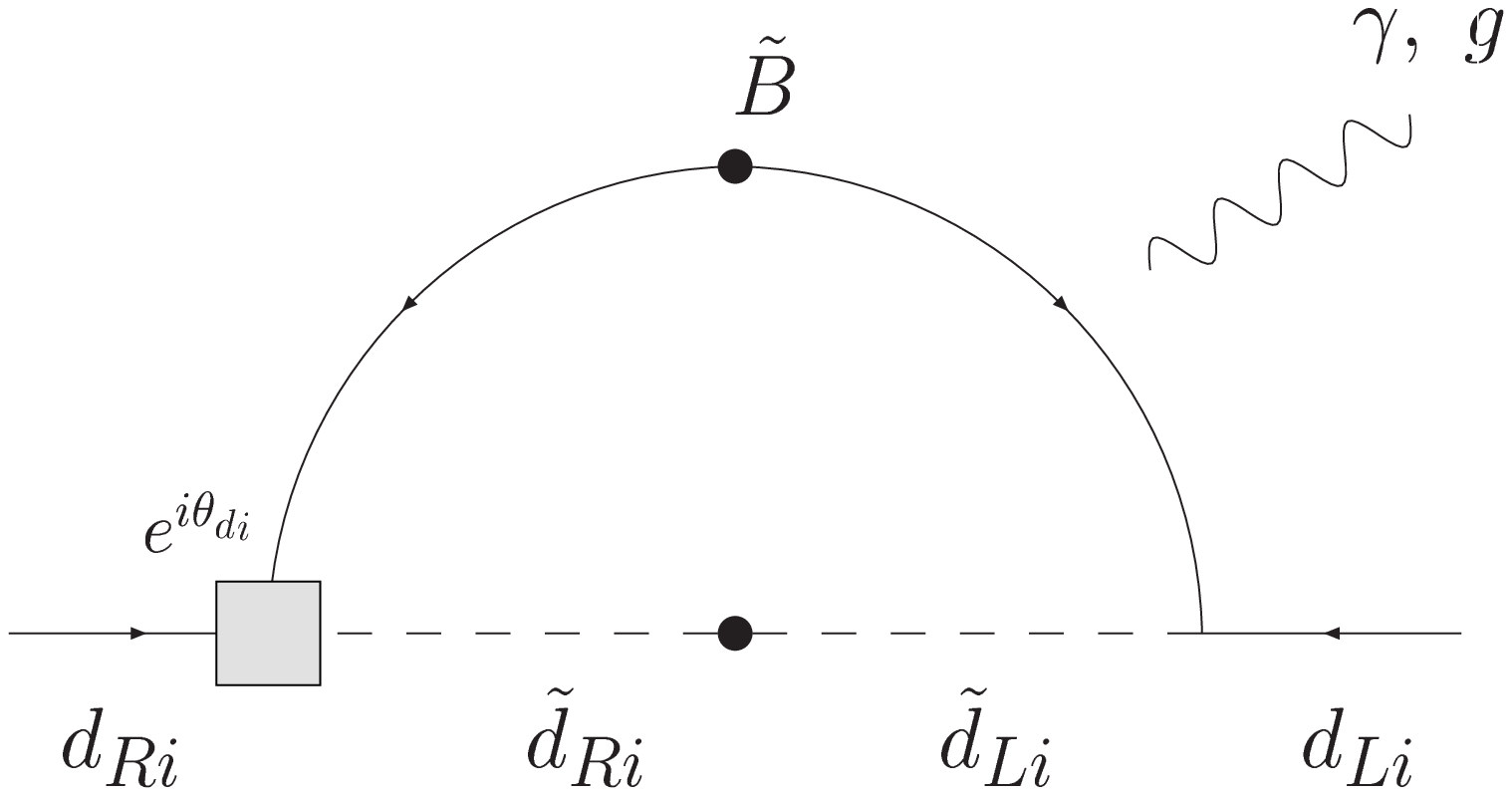} &
\includegraphics[scale=0.32]{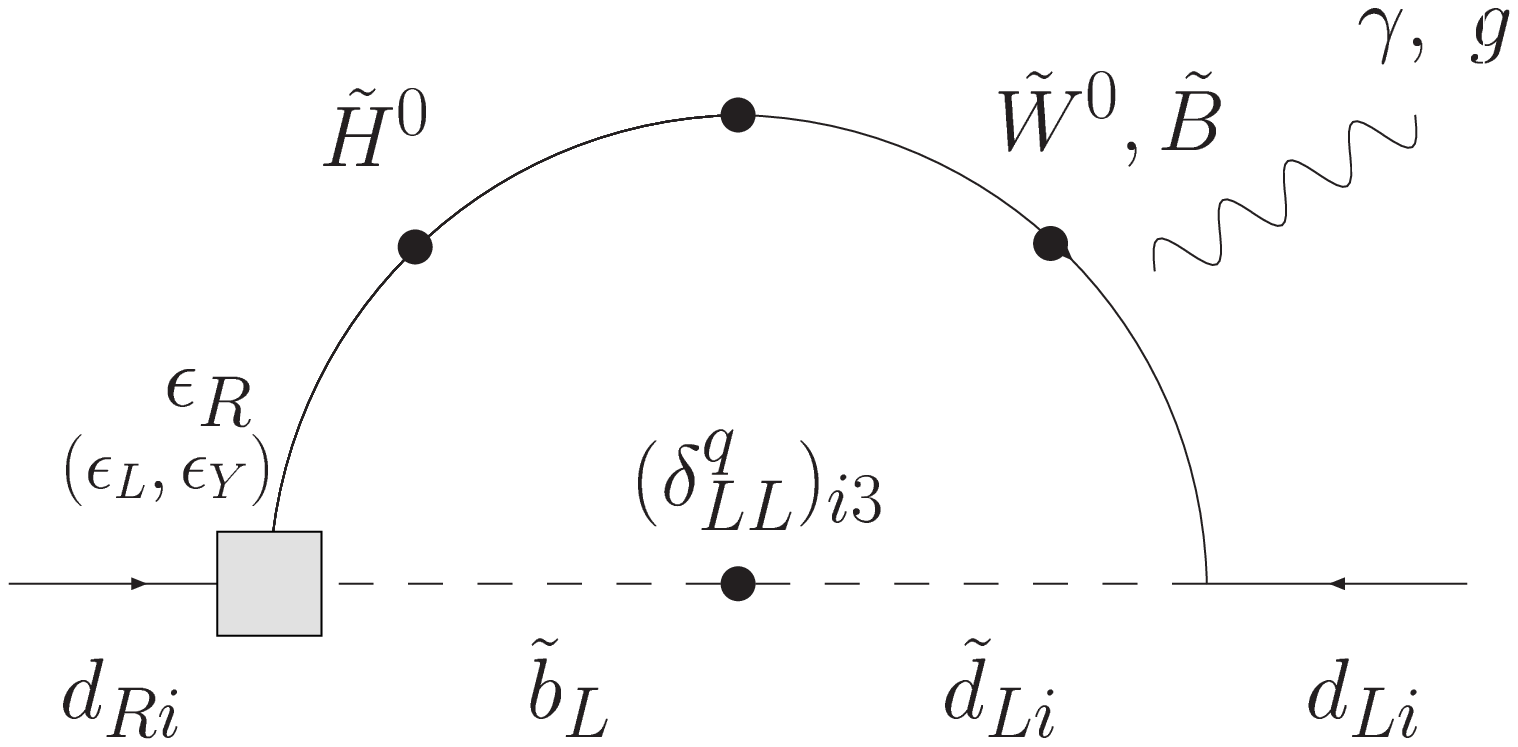} 
\end{tabular}
\vskip 0.5 cm
\begin{tabular}{ccc}
\includegraphics[scale=0.32]{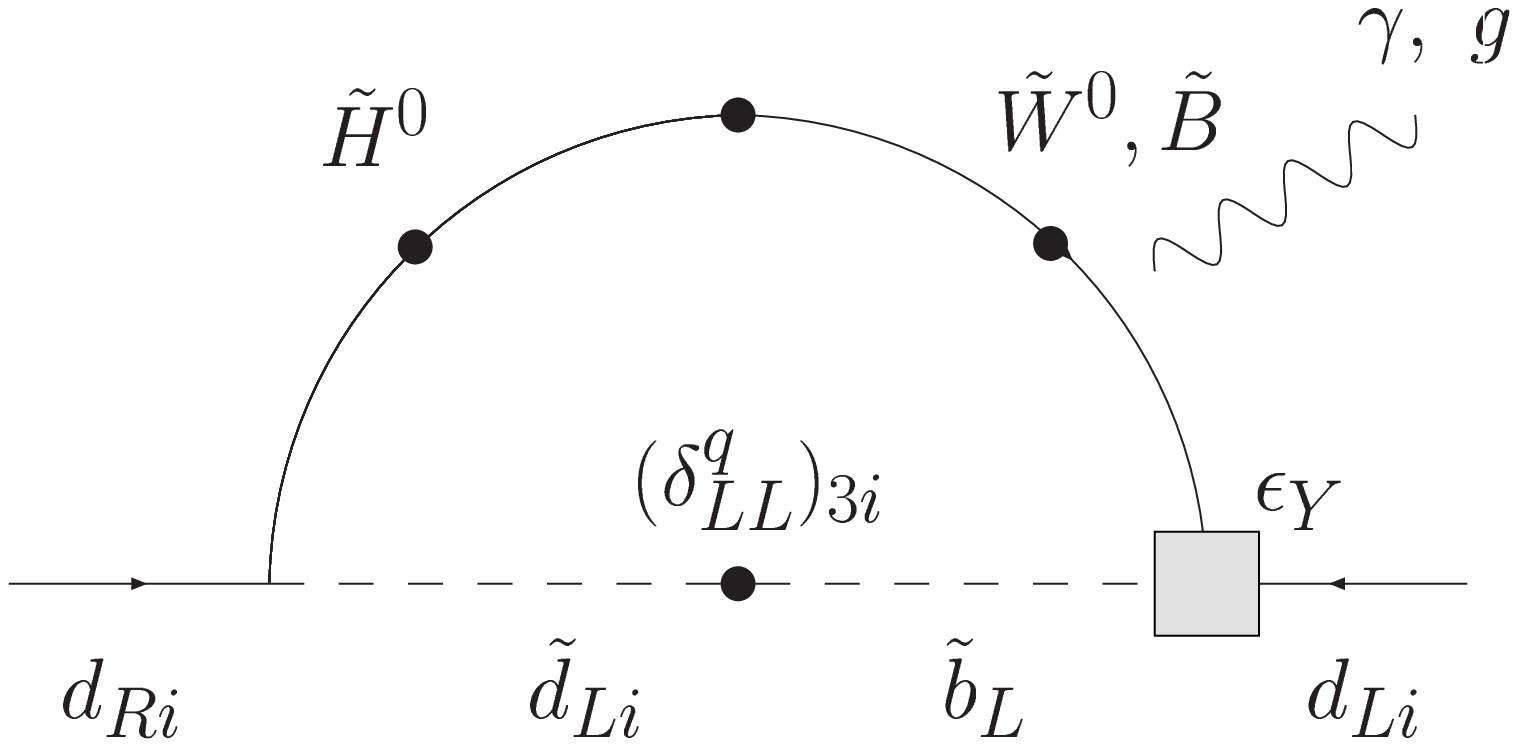} &
\includegraphics[scale=0.32]{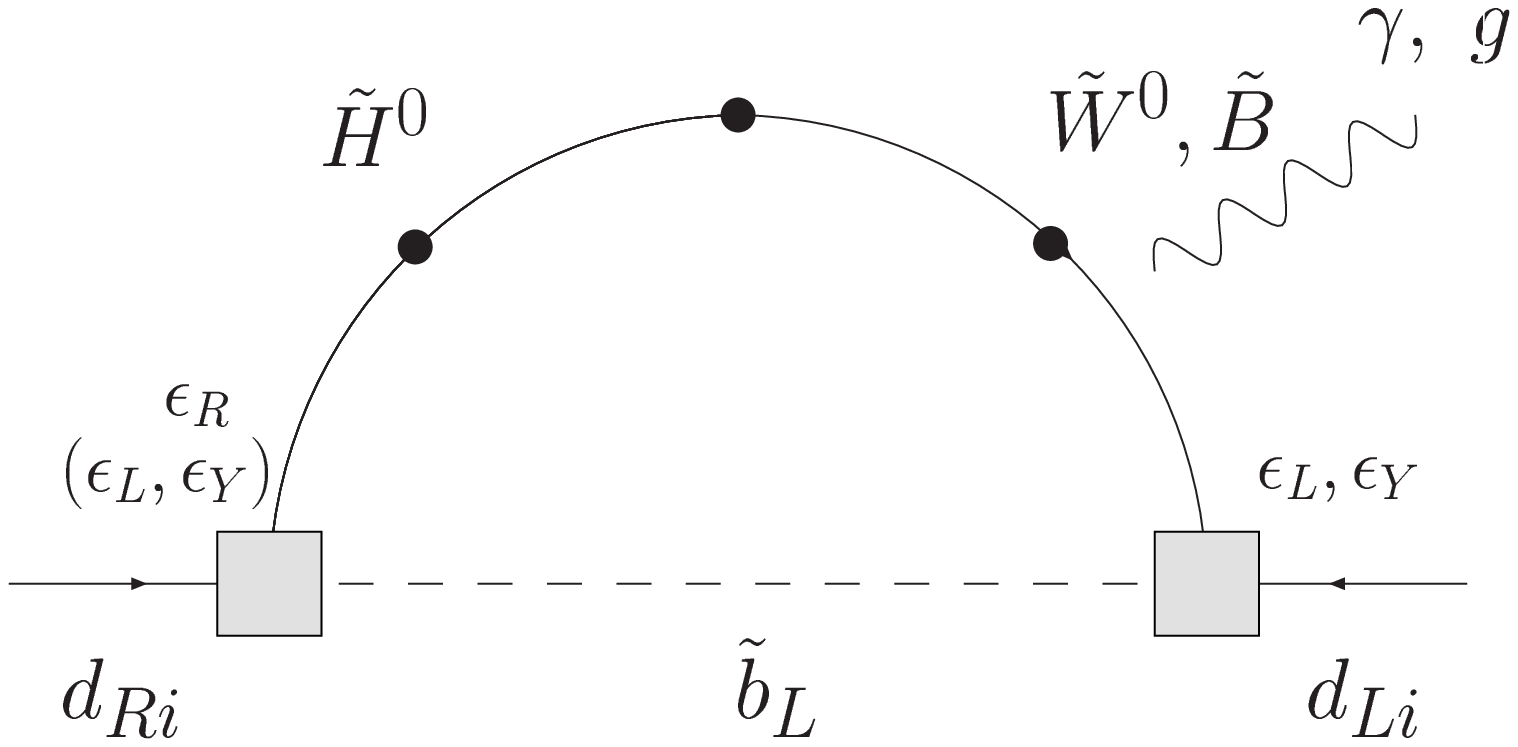} &
\includegraphics[scale=0.32]{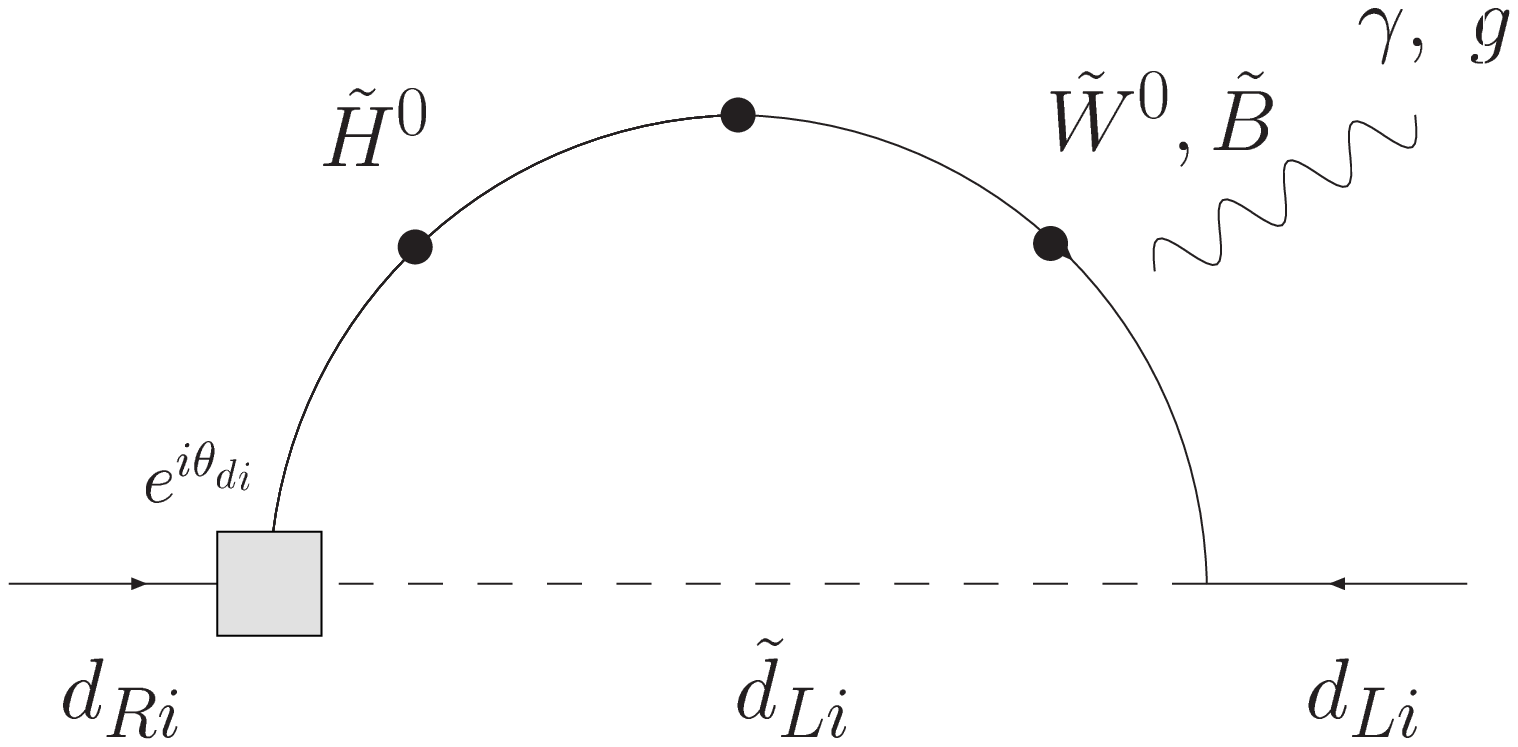} 
\end{tabular}
\vskip 0.5 cm
\begin{tabular}{ccc}
\includegraphics[scale=0.32]{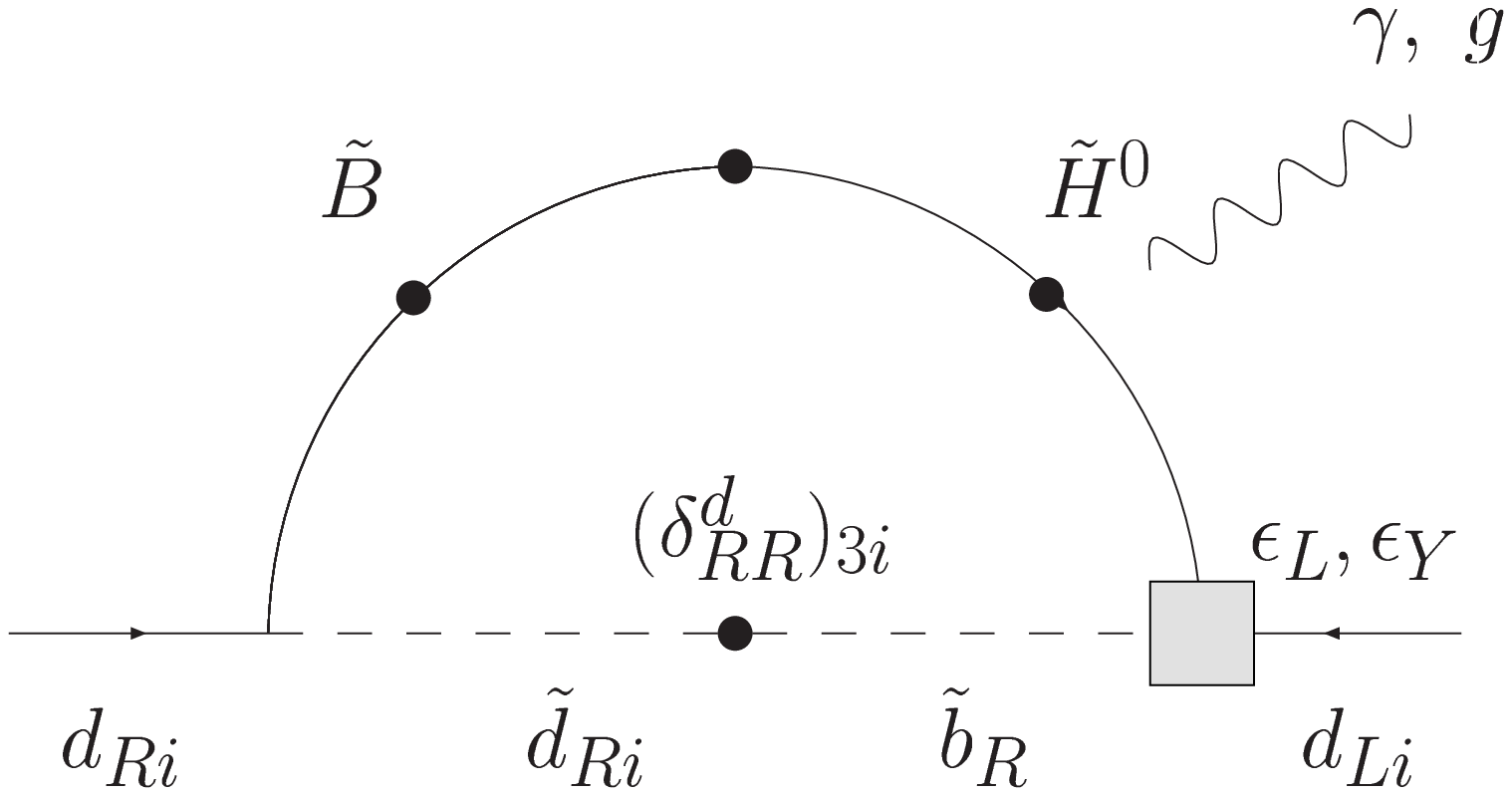} &
\includegraphics[scale=0.32]{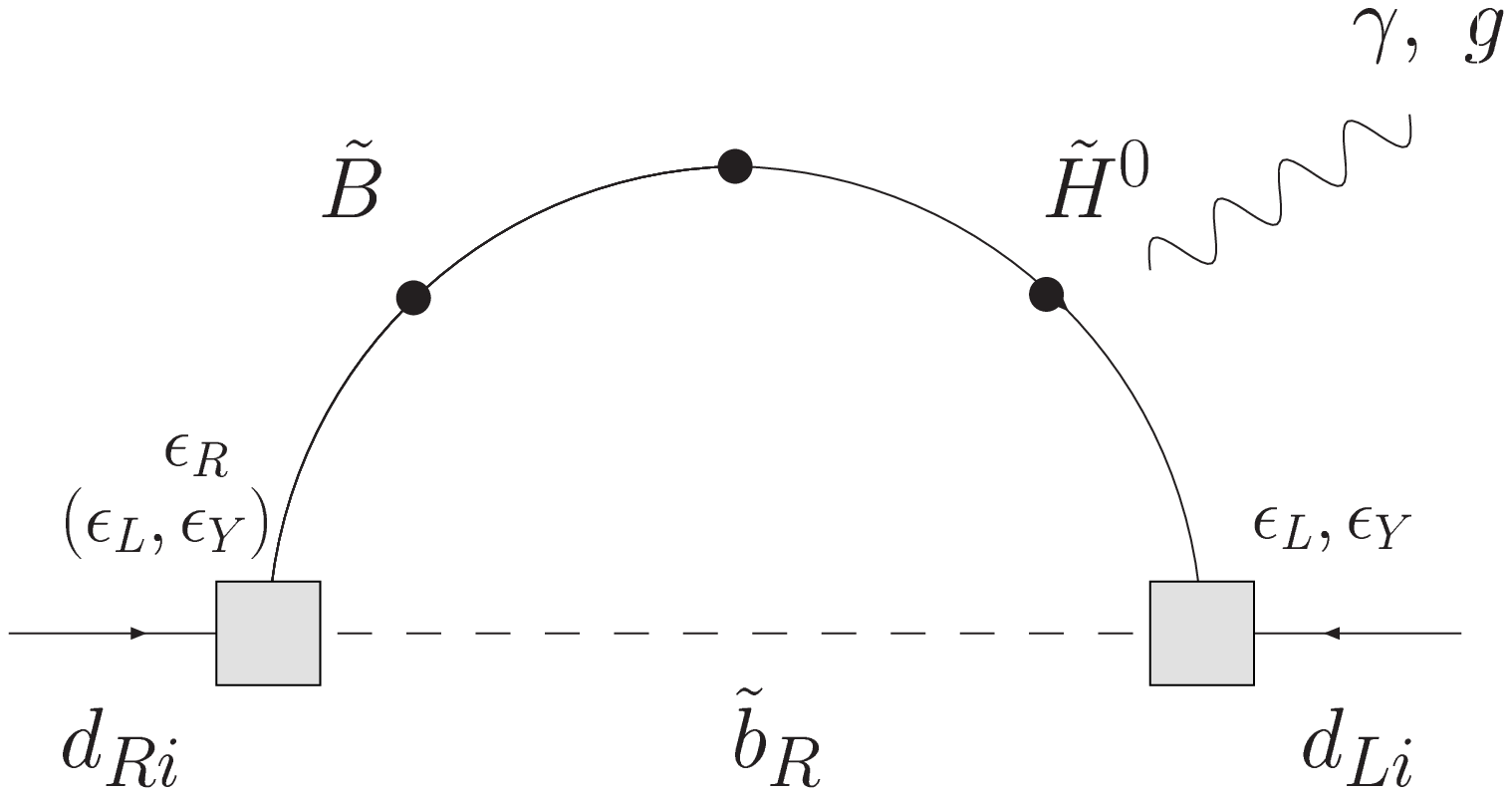} &
\includegraphics[scale=0.32]{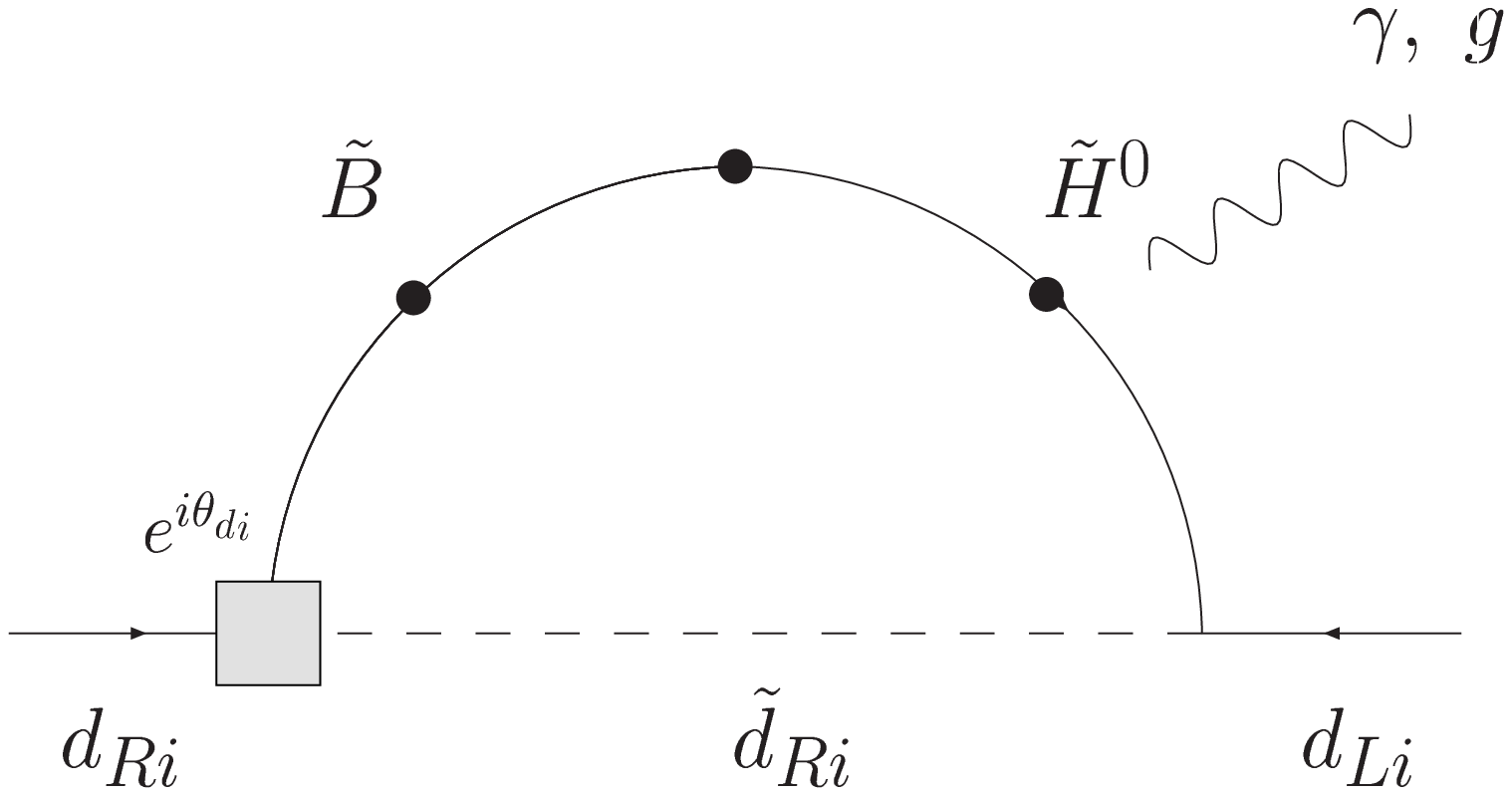}
\end{tabular}
\caption{\label{Fig:neutralino} Neutralino-mediated contributions for down-type quark (C)EDMs
beyond the leading order. Gray boxes indicate the effective couplings with possible loop factors.}
\end{center}
\end{figure}

Neutralino-mediated contributions are divided into three classes: pure bino-mediated diagrams,
neutral wino-Higgsino mixing diagrams and bino-Higgsino diagrams. The relevant diagrams for
the neutralino contributions are reported in Fig.~\ref{Fig:neutralino}. The pure bino-mediated contributions are given by
\begin{eqnarray}
 \left\{ \frac{d_{d_i}}{e},~d^c_{d_i} \right\}_{\tilde B}
 \!\!\!&=&\!\!\!
 \frac{\alpha_Y Y_L Y_R}{4\pi}
 \frac{m_b}{m^2}
 \frac{\mu M_1}{m^2}
 \tgb~ 
 \Biggl(
  {\cal E}^{\tilde B}_1 ~
  {\rm Im}\left[ (\delta^d_{LL})_{i3}(\delta^d_{RR})_{3i} \right]
 \nonumber\\
 && \qquad
  +~
  {\cal E}^{\tilde B}_2 ~
  {\rm Im}\left[ V^*_{3i}(\delta^d_{RR})_{3i} \right]
  + 
  {\cal E}^{\tilde B}_3 ~
  \frac{m_{d_i}}{m_b}
  {\rm Im}\left[ V^*_{3i}(\delta^d_{LL})_{3i} \right]
 \Biggr)\,,
\end{eqnarray}
where
\begin{eqnarray}
 {\cal E}^{\tilde B}_1 
 \!\!\!&=&\!\!\!
 \frac{f_{\tilde B}^{(3)}(x)}{1+\eps_3\tgb}
 +
 \frac{1}{3(1+\eps_3\tgb)}
 \left(
  \frac{\epsR\tgb}{1+\eps_3\tgb}
  +
  \frac{\epsL\tgb}{1+\epsb_3\tgb}
 \right)
 f_{\tilde B}^{(2)}(x)
 \nonumber\\
 &&
  +~
  \left(
   \frac{(1+r_i)\epsL\epsR \ttgb}{9(1+\eps_3\tgb)^2(1+\epsb_3\tgb)}
   -
   \frac{r_i \epsLR\tgb}{6(1+\eps_3\tgb)^2}
  \right)
  f_{\tilde B}^{(1)}(x)\,,
 \\
 {\cal E}^{\tilde B}_2 
 \!\!\!&=&\!\!\!
 -
 \frac{\epsY \tgb}{(1+\eps_3\tgb)(1+\epsb_3\tgb)}
 f_{\tilde B}^{(2)}(x)
 -
 \frac{(1+r_i)\epsR\epsY \ttgb}{3(1+\eps_3\tgb)^2 (1+\epsb_3\tgb)}
 f_{\tilde B}^{(1)}(x)\,,
 \\
 {\cal E}^{\tilde B}_3
 \!\!\!&=&\!\!\!
  \left(
   \frac{r_i^2 \epsLY \tgb}{3(1+\eps_3\tgb)^2}
   -\frac{\bar{r}_i\epsLY\epsY\ttgb}
   {3(1+\epsb_3\tgb)^2(1+\eps_3\tgb)}
  \right)
 f_{\tilde B}^{(1)}(x)
 \nonumber\\
 &&
 +~
 \frac{\bar{r}_i\epsY\tgb}{(1+\epsb_3\tgb)(1+\eps_3\tgb)}
 f_{\tilde B}^{(2)}(x)\,,
\end{eqnarray}
and
\begin{eqnarray}
 f_{\tilde B}^{(i)}(x)
 = 
 \left\{ 
  - \frac{1}{3} f_0^{(i)}(x),~
  f_0^{(i)}(x)
 \right\}
 \quad (i=1,2,3)\,,
\end{eqnarray}
where $x=M_1^2/m^2$ and $f_{\tilde B}^{(1,2,3)}(1)=$ $\{1/18,-1/6\}$, $\{-1/30,1/10\}$, $\{1/45,-1/15\}$, respectively. Here, the hypercharges $Y_L$ and $Y_R$ are defined as
$Y_L=1/6$ and $Y_R=1/3$.

Next, the contributions of neutral wino-Higgsino mixing diagrams read
\begin{eqnarray}
 \left\{ \frac{d_{d_i}}{e},~d^c_{d_i} \right\}_{{\tilde H}^0{\tilde W}^0}
 \!\!\!\!\!\!\!\! &=& \!\!\!
 \frac{\alpha_2}{16\pi}
 \frac{m_b}{m^2}
 \frac{\mu M_2}{m^2}
 \tgb
 \Biggl(
  {\cal E}^{{\tilde H}^0{\tilde W}^0}_1~
  {\rm Im}\left[ (\delta^d_{LL})_{i3}(\delta^d_{RR})_{3i} \right]
 \nonumber\\
 && \qquad
 +~
  {\cal E}^{{\tilde H}^0{\tilde W}^0}_2
  {\rm Im}\left[ V^*_{3i}(\delta^d_{RR})_{3i} \right]
  +
  {\cal E}^{{\tilde H}^0{\tilde W}^0}_3
  \frac{m_{d_i}}{m_b}
  {\rm Im}\left[ V^*_{3i}(\delta^d_{LL})_{3i} \right]
 \Biggr)\,,
\end{eqnarray}
where
\begin{eqnarray}
 {\cal E}^{{\tilde H}^0{\tilde W}^0}_1 
 \!\!\!\!\!\!\! &=& \!\!\!
  \frac{\epsR\tgb}{3(1+\eps_3\tgb)^2}
  g_{\tilde{H}^0\tilde{W}^0}^{(1)}(x,y)
 \nonumber\\
 && \quad
  +~
  \left(
   \frac{(1+r_i)\epsR\epsL\ttgb}{9(1+\eps_3\tgb)^2(1+\epsb_3\tgb)}
   -
   \frac{r_i \epsLR\tgb}{6(1+\eps_3\tgb)^2}
  \right)
  g_{{\tilde H}^0{\tilde W}^0}^{(0)}(x,y)\,,
\end{eqnarray}
\begin{eqnarray}
 {\cal E}^{{\tilde H}^0{\tilde W}^0}_2 
 \!\!\!\!\!\!\! &=& \!\!\!
  -
  \frac{(1+r_i)\epsR\epsY\ttgb}{3(1+\eps_3\tgb)^2(1+\epsb_3\tgb)}
  g_{{\tilde H}^0{\tilde W}^0}^{(0)}(x,y)\,,
 \\
 {\cal E}^{{\tilde H}^0{\tilde W}^0}_3 
 \!\!\!\!\!\!\! &=& \!\!\!
  \left(
   \frac{r_i^2 \epsLY \tgb}{3(1+\eps_3\tgb)^2}
   -\frac{\bar{r}_i \epsLY\epsY \ttgb}
   {3(1+\epsb_3\tgb)^2(1+\eps_3\tgb)}
  \right)
 g_{{\tilde H}^0{\tilde W}^0}^{(0)}(x,y)
  \nonumber\\
  && \
 +~
  \frac{\bar{r}_i \epsY\ttgb}{(1+\epsb_3\tgb)(1+\eps_3\tgb)}
  g_{{\tilde H}^0{\tilde W}^0}^{(1)}(x,y)\,,
\end{eqnarray}
and
\begin{eqnarray}
 g_{{\tilde H}^0{\tilde W}^0}^{(i)}(x,y)
 = 
 \left\{ 
  -\frac{1}{3} g_0^{(i)}(x,y)\,,~
  g_0^{(i)}(x,y)
 \right\}
 \quad (i=0,1)\,,
\end{eqnarray}
where $x=M_2^2/m^2,~y=\mu^2/m^2$ and $g_{{\tilde H}^0{\tilde W}^0}^{(0,1)}(1,1)=\{1/18,-1/6\}$, $\{-2/45,2/15\}$.

Finally, the contributions from bino-Higgsino mixing diagrams read
\begin{eqnarray}
 \left\{ \frac{d_{d_i}}{e},~d^c_{d_i} \right\}_{{\tilde H}{\tilde B}}
 \!\!\!\!\!\!\!\! &=& \!\!\!
 \frac{\alpha_Y}{8\pi}
 \frac{m_b}{m^2}
 \frac{\mu M_1}{m^2}
 \tgb
 \Biggl(
  {\cal E}^{{\tilde H}{\tilde B}}_1~
  {\rm Im}\left[ (\delta^d_{LL})_{i3}(\delta^d_{RR})_{3i} \right]
 \nonumber\\
 && \qquad
 +~
  {\cal E}^{{\tilde H}{\tilde B}}_2
  {\rm Im}\left[ V^*_{3i}(\delta^d_{RR})_{3i} \right]
  +
  {\cal E}^{{\tilde H}{\tilde B}}_3
  \frac{m_{d_i}}{m_b}
  {\rm Im}\left[ V^*_{3i}(\delta^d_{LL})_{3i} \right]
 \Biggr)\,,
\end{eqnarray}
where
\begin{eqnarray}
 {\cal E}^{{\tilde H}{\tilde B}}_1 
 \!\!\!\!\!\!\! &=& \!\!\!
  \left(
   \frac{\epsR\tgb}{3(1+\eps_3\tgb)^2} Y_L
   +
   \frac{\epsL\tgb}{3(1+\eps_3\tgb)(1+\epsb_3\tgb)} Y_R
  \right)
  g_{\tilde{H}^0\tilde{B}^0}^{(1)}(x,y)
 \nonumber\\
 && \qquad
  +~
  \left(
   \frac{(1+r_i)\epsR\epsL\ttgb}{9(1+\eps_3\tgb)^2(1+\epsb_3\tgb)}
   -
   \frac{r_i \epsLR\tgb}{6(1+\eps_3\tgb)^2}
  \right) (Y_L + Y_R)
  g_{{\tilde H}{\tilde B}}^{(0)}(x,y)\,,
 \\
 {\cal E}^{{\tilde H}{\tilde B}}_2 
 \!\!\!\!\!\!\! &=& \!\!\!
  -
  \frac{Y_R \epsY\tgb}{(1+\eps_3\tgb)(1+\epsb_3\tgb)} 
  g_{{\tilde H}{\tilde B}}^{(1)}(x,y)
  -
  \frac{(Y_L+Y_R)(1+r_i)\epsR\epsY\ttgb}{3(1+\eps_3\tgb)^2(1+\epsb_3\tgb)} 
  g_{{\tilde H}{\tilde B}}^{(0)}(x,y)\,,
 \\
 {\cal E}^{{\tilde H}{\tilde B}}_3 
 \!\!\!\!\!\!\! &=& \!\!\!
  \left(
   \frac{r_i^2 \epsLY \tgb}{3(1+\eps_3\tgb)^2}
   -\frac{\bar{r}_i (Y_L+Y_R)\epsLY\epsY \ttgb}
   {3(1+\epsb_3\tgb)^2(1+\eps_3\tgb)}
  \right)
  g_{{\tilde H}{\tilde B}}^{(0)}(x,y)
  \nonumber\\
  && \
  +~
  \frac{\bar{r}_i Y_L\epsY\tgb}{(1+\epsb_3\tgb)(1+\eps_3\tgb)}
  g_{{\tilde H}{\tilde B}}^{(1)}(x,y)\,,
\end{eqnarray}
and
\begin{eqnarray}
 g_{{\tilde H}{\tilde B}}^{(i)}(x,y)
 = 
 \left\{ 
  -\frac{1}{3} g_0^{(i)}(x,y),~
  g_0^{(i)}(x,y)
 \right\}
 \quad (i=0,1)\,,
\end{eqnarray}
where $x=M_1^2/m^2, y=\mu^2/m^2$ and $g_{{\tilde H}{\tilde B}}^{(0,1)}(1,1)=\{1/18,-1/6\}$, $\{-2/45,2/15\}$, respectively.

\subsection{Higgs Contribution to Down-type Quark (C)EDMs}
\begin{figure}[t]
\begin{center}
\begin{tabular}{cc}
\includegraphics[scale=0.4]{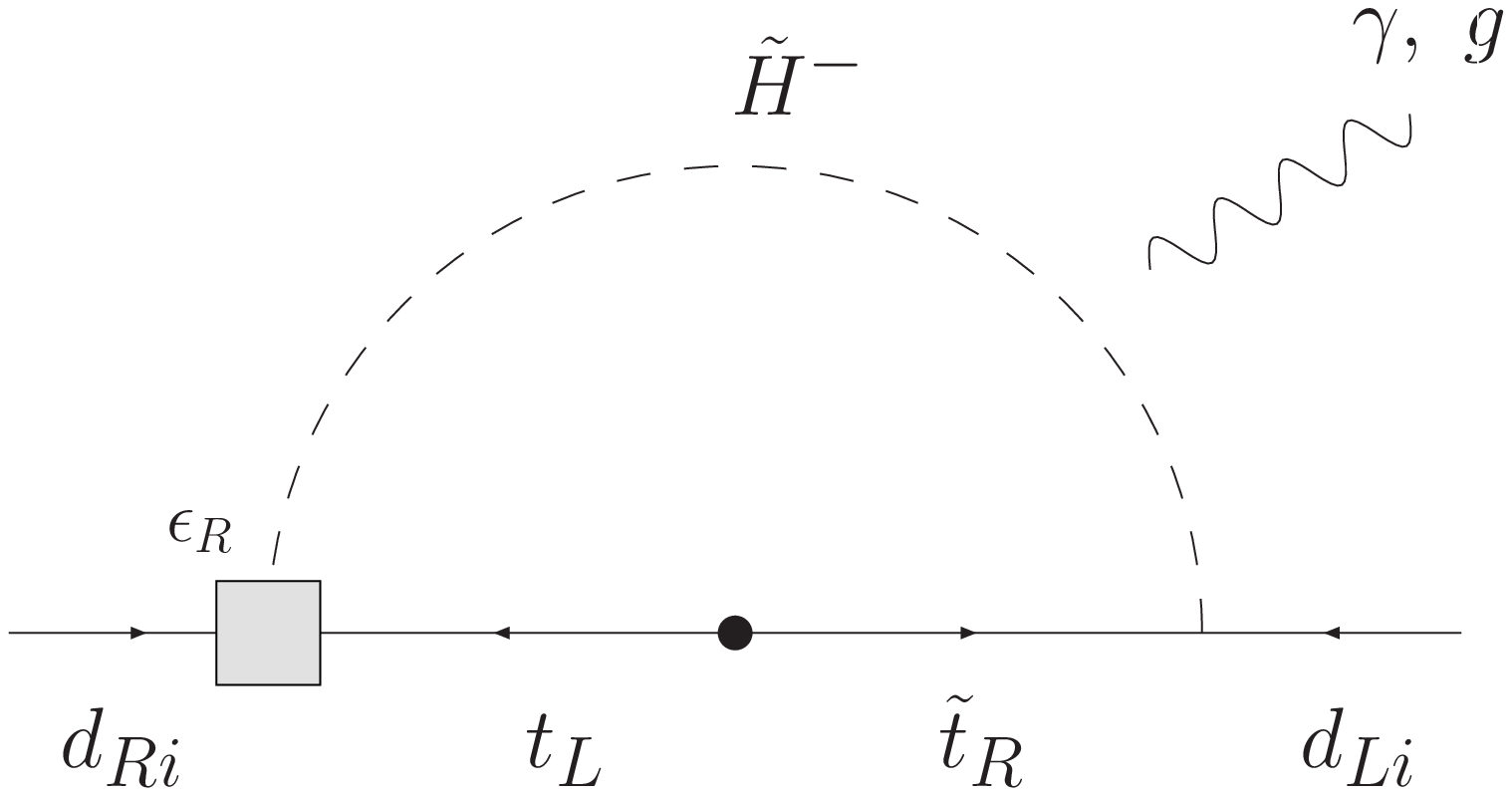} &
\includegraphics[scale=0.4]{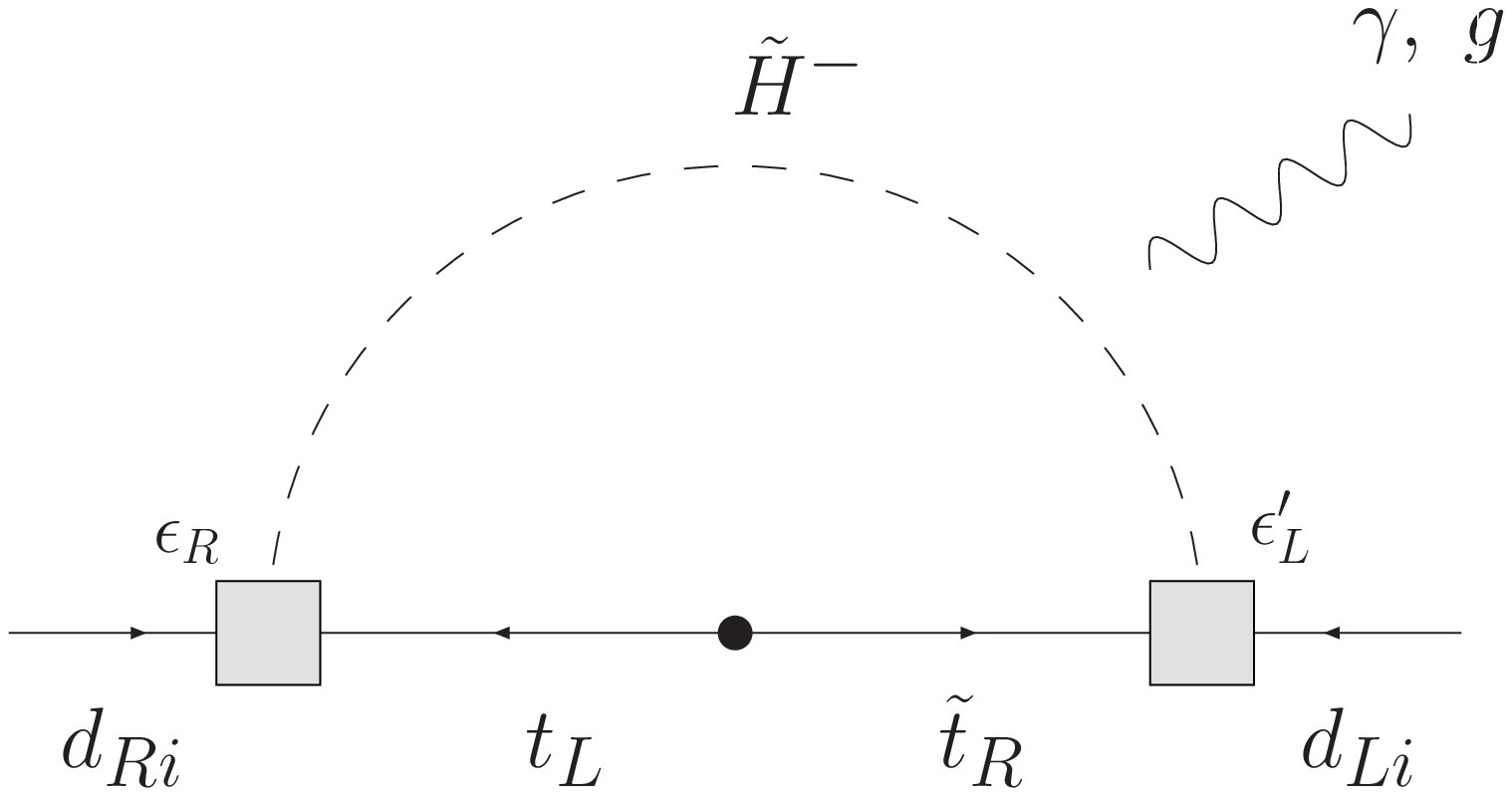}
\end{tabular}
\caption{\label{Fig:CHiggs} Left (Right): Charged Higgs mediated contributions to
down-type quark (C)EDMs at the leading order for the case of $J^{(d_i)}_{RR}$ ($J^{(d_i)}_{LR}$).
The effective couplings with $\tan\beta$-enhanced corrections are depicted by gray boxes.}
\end{center}
\end{figure}
The importance of the charged-Higgs mediated contributions to the hadronic (C)EDMs was already
discussed in Ref.~\cite{Hisano:2006mj} where, on the other hand, only the leading contributions of
order $\mathcal{O}(\epsilon\tgb)$ were presented. The full expression for the charged Higgs-mediated contributions, beyond the leading-order expansion in $\mathcal{O}(\epsilon\tgb)$, reads
\begin{eqnarray}
 \left\{ \frac{d_{d_i}}{e},~d^c_{d_i} \right\}_{H^\pm}
 \!\!\!&=&\!\!\!
 -
 \frac{\alpha_2}{16\pi}
 \frac{m_b}{M_{H^\pm}^2}
 \frac{m_t^2}{m_W^2}
 \Biggl(
  {\cal E}^{H^\pm}_1~
  {\rm Im}\left[ (\delta^d_{LL})_{i3}(\delta^d_{RR})_{3i} \right]
  \nonumber\\
  &&
  \qquad
  +~
  {\cal E}^{H^\pm}_2~
  {\rm Im}\left[ V^*_{3i}(\delta^d_{RR})_{3i} \right]
  +
  \frac{m_{d_i}}{m_b}
  {\cal E}^{H^\pm}_3~
  {\rm Im}\left[ V^*_{3i}(\delta^d_{LL})_{3i} \right]
 \Biggr)\,,
\label{Eq:edmH}
\end{eqnarray}
where
\begin{eqnarray}
 {\cal E}^{H^\pm}_1 
 \!\!\!&=&\!\!\!
  \frac{\epsR\tgb}{9(1+\eps_3\tgb)^2}
  \left(
   \epsLp\tgb
   -
   \frac{\epsL\epsYp\ttgb}{1+\epsb_3\tgb}
  \right)
  f_{H^\pm}^{(0)}(z)\,,
 \\
 {\cal E}^{H^\pm}_2 
 \!\!\!&=&\!\!\!
  \frac{\epsR\tgb}{3(1+\eps_3\tgb)^2}
  \left(
   1 - \epsbp_3\tgb
   + 
   \frac{\epsY\epsYp\ttgb}{1+\epsb_3\tgb}
  \right)
  f_{H^\pm}^{(0)}(z)\,,
 \\
 {\cal E}^{H^\pm}_3 
 \!\!\!&=&\!\!\!
  \Bigg[
   \left(
    \frac{\bar{r}_i \epsL\tgb}{3(1+\epsb_3\tgb)(1+\eps_3\tgb)}
	+
    \frac{r_i \epsLY\tgb}{3(1+\eps_3\tgb)^2}
   \right)
   \left(
    1 - \epsbp_3\tgb
    +
    \frac{\epsY\epsYp\ttgb}{1+\epsb_3\tgb}
   \right)
 \nonumber\\
 && 
   \qquad
   - 
   \frac{ r_i + r_i\epsb_3\tgb + \bar{r}_i \epsY\tgb}{3(1+\epsb_3\tgb)(1+\eps_3\tgb)}
   \left(
    \epsLp\tgb
    -
    \frac{\epsL\epsYp\ttgb}{1+\epsb_3\tgb}
   \right)
  \Bigg]
  f_{H^\pm}^{(0)}(z)
\label{Eq:CH_LL}\,,
\end{eqnarray}
and
\begin{eqnarray}
 f_{H^\pm}^{(0)}(z)
 =
 \left\{
  - f_0^{(0)}(z)
  + \frac{2}{3} f_1^{(0)}(z),~
  f_1^{(0)}(z)
 \right\}\,,
\end{eqnarray}
where $z=m_t^2/M_{H^+}^2$ and $f_{H^\pm}^{(0)}(1) = \{-7/9,-2/3\}$. The loop factors with
primes come from the vertex correction for charged Higgs with left-handed down quarks, and
these are given as $\epsLp=\epsbp_3={\rm sign}(\mu M_3)\times \alpha_s/3\pi$,
$\epsYp=-{\rm sign}(\mu A_b)\times y_b^2/32\pi^2$ for equal SUSY masses.

We note that, since the loop function $f_1^{(0)}(z)$ contains a large logarithmic and it behaves
as $f_1^{(0)}(z)\simeq 3+\log z$ for $z\ll 1$, both EDMs and CEDMs have a slow decoupling with
the charged Higgs masses going as $d^{(c)}_{d_i}\sim z\log\,z$.  Even for moderate $\tan\beta$
values and charged Higgs masses $M_{H^\pm} <2 \,{\rm TeV}$, the down quark (C)EDMs remain within
the reach of sensitivities of future planned experiments, {\it i.e.} $(d_d/e)_{H^\pm}$ and
$(d^c_d)_{H^\pm}\sim {\cal O}(10^{-(26-27)}) \,{\rm cm}$ for $|(\delta^d_{RR})_{13}|=(0.2)^3$.
Another important feature of the charged Higgs contribution is that it does not  decouple for
$\msusy\to\infty$, as long as the Higgs mass doesn't decouple.

Finally, additional contributions to the (C)EDMs come from the neutral-Higgs sector. However,
the $H$ and $A$ contributions have opposite signs and their net effect to the (C)EDMs is really
small even in the most favorable case where a mass splitting between $H$ and $A$ of order 10$\%$
was allowed.

Before concluding this section, we would like to note that the {\it flavored} (C)EDMs are highly correlated with FCNC observables. As an interesting example, let us mention that, irrespective
to the particular choice for the SUSY spectrum, the $\tilde{\chi}^\pm$ and $H^\pm$ contributions
to the EDMs are related to the NP contributions entering $B\to X_s\gamma$ as
\beq
 (d_{d})_{\tilde{\chi}^\pm}+ (d_{d})_{H^\pm}
 \simeq
  -e\frac{\alpha_2}{4\pi}\frac{m_b}{m^2_W}
  \frac{\epsR\tgb}{1+\eps_3\tgb}\,{\rm Im}\left[ V^*_{3i}(\delta^d_{RR})_{3i} \right]\,C_7\,,
  \label{correlation}
\eeq
where $C_7 = C^{\tilde{\chi}^\pm}_{7}(M_W) + C^{H^\pm}_{7}(M_W)$ is defined as
${\cal B}(B\!\to\! X_s \gamma)\simeq 3.15 - 8\,C_7 - 1.9\,C_8$~\cite{Misiak:2006zs}.
A detailed exploration of the intriguing correlation and interplay among FCNC processes
and flavored (C)EDMs would deserve a dedicated study that goes beyond the scope of the
present work.

\subsection{Comparison of Various Contributions for Down-type Quark (C)EDMs}

As we have discussed in the previous sections, the down-type quark (C)EDMs receive 
the first contributions already at the LO level through the exchange of gluinos and 
charginos. Moreover, at the BLO, many additional contributions, proportional to the
various JIs, are generated. The purpose of this section is to provide a detailed
numerical comparison of the various effects.
\begin{figure}[tp]
\begin{center}
\begin{tabular}{cc}
\includegraphics[scale=0.55]{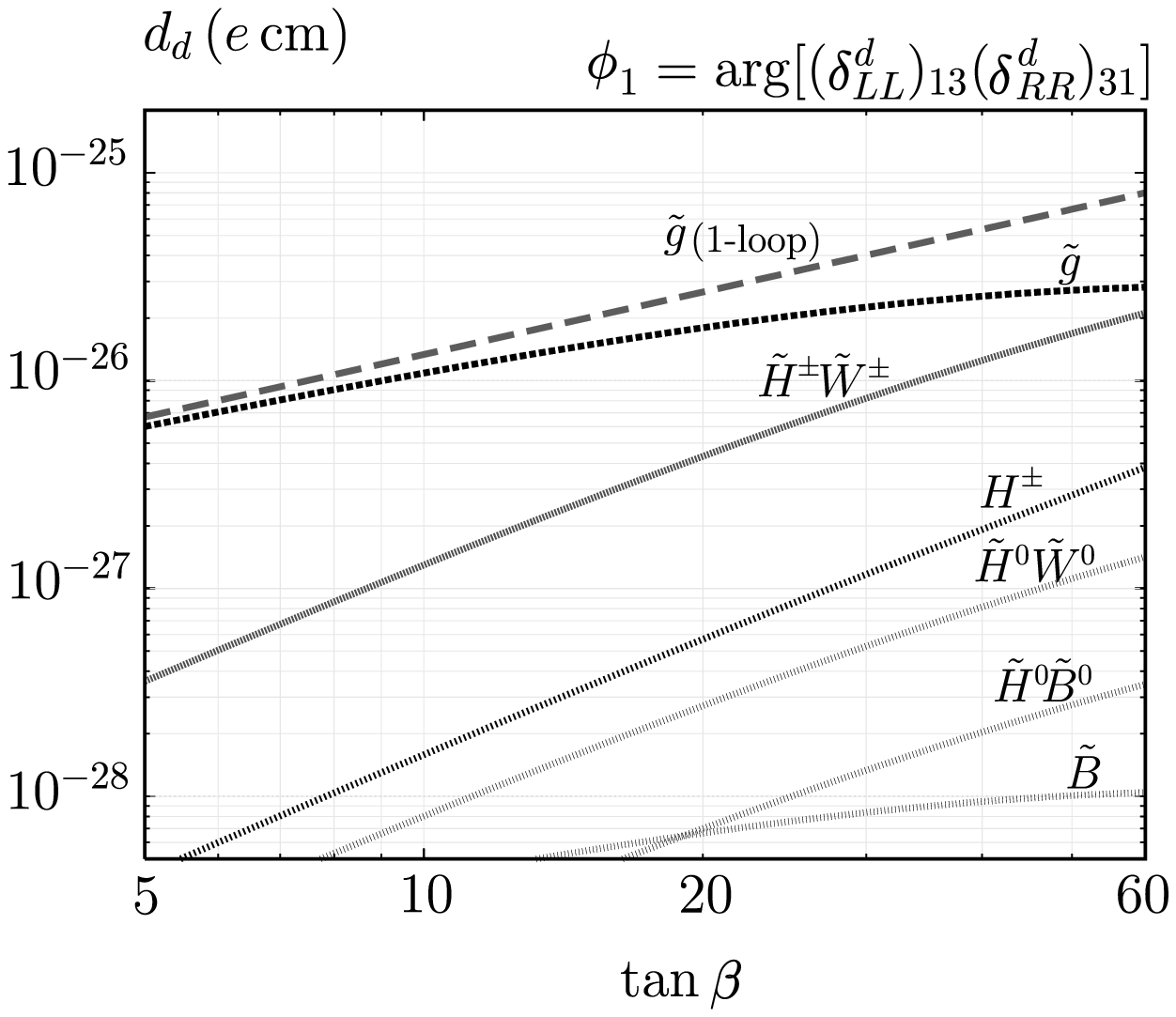} &
\includegraphics[scale=0.55]{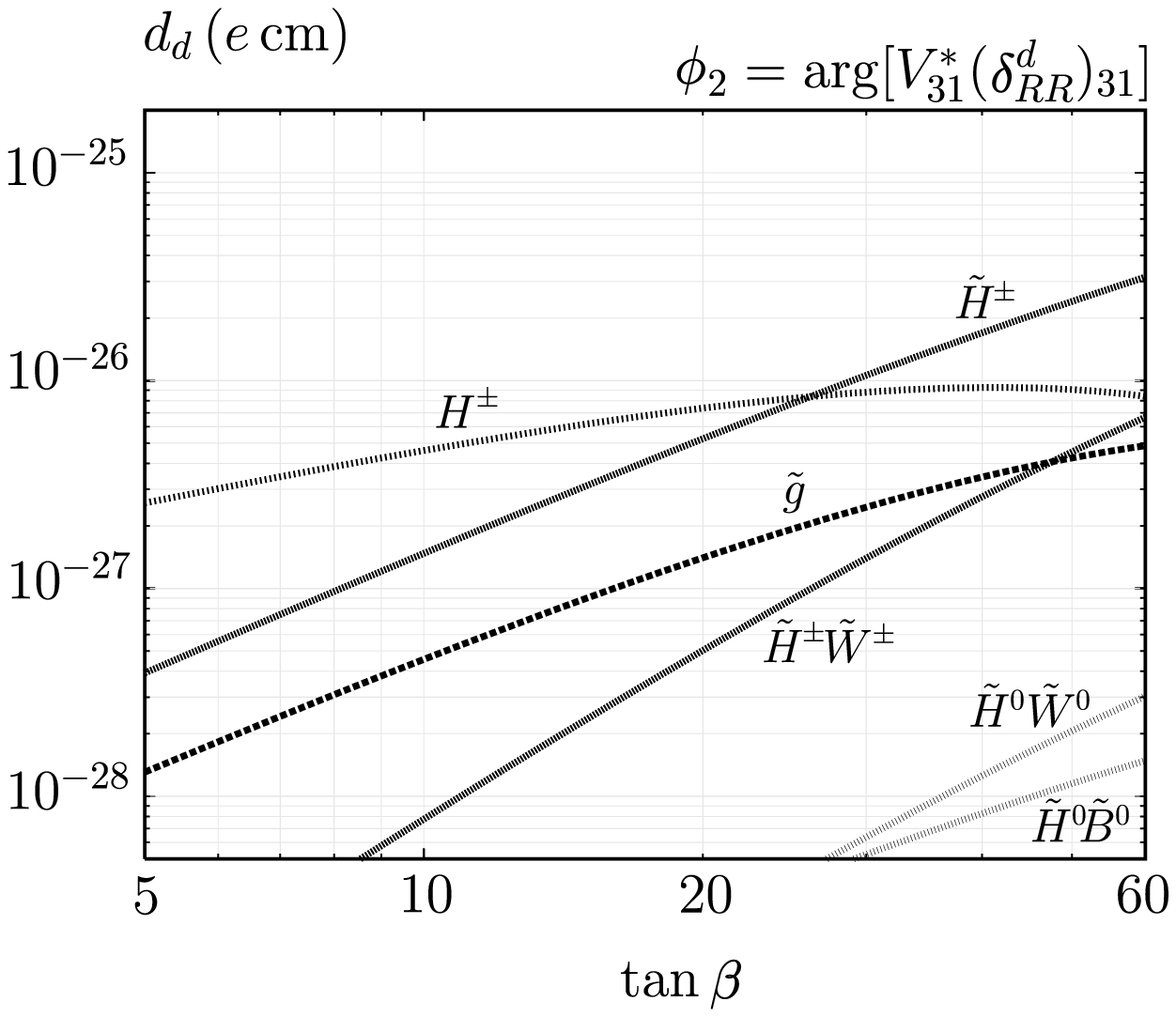} \\
(a) & (b) 
\end{tabular}
\vskip 0.5 cm
\begin{tabular}{cc}
\includegraphics[scale=0.55]{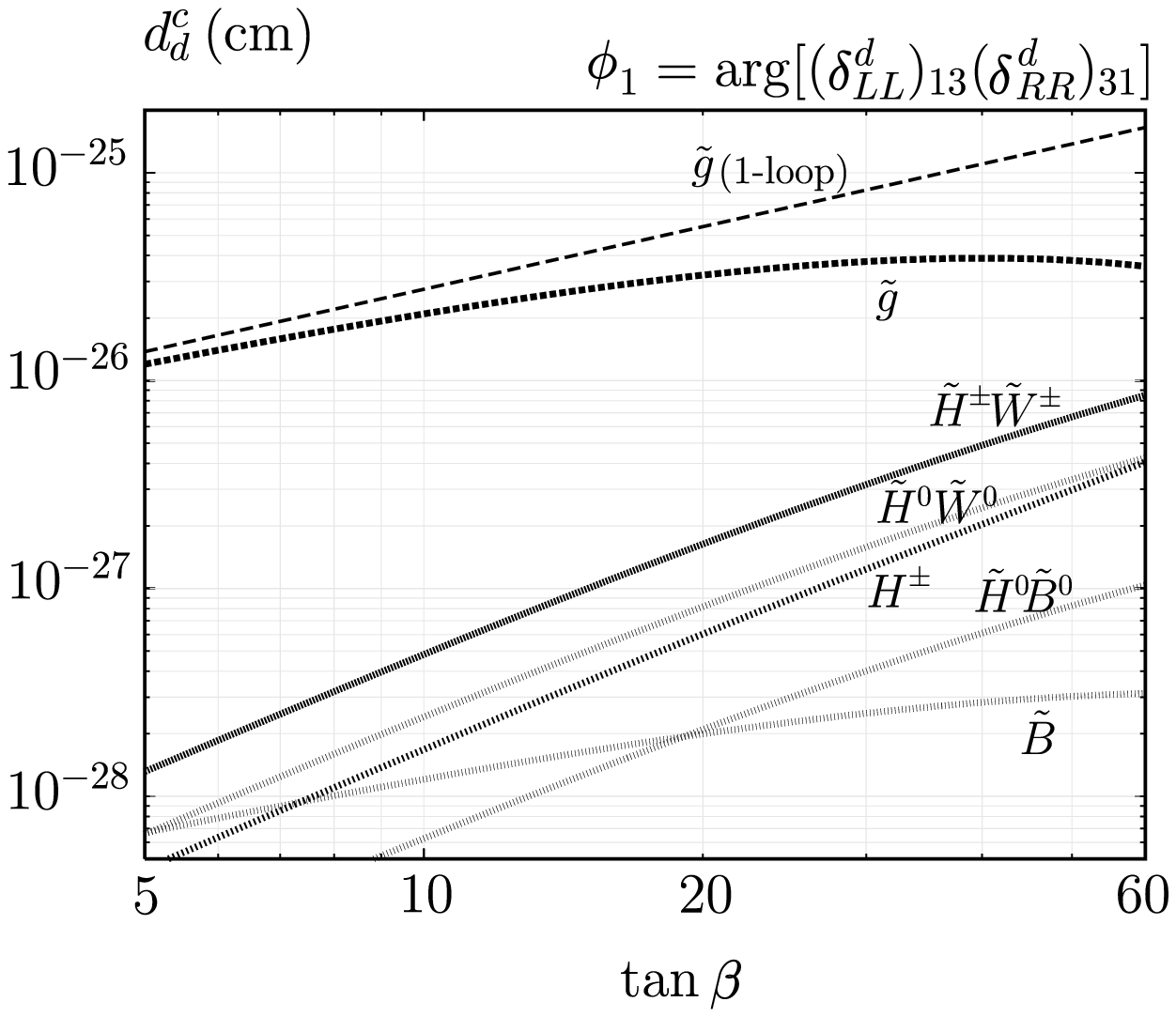} &
\includegraphics[scale=0.55]{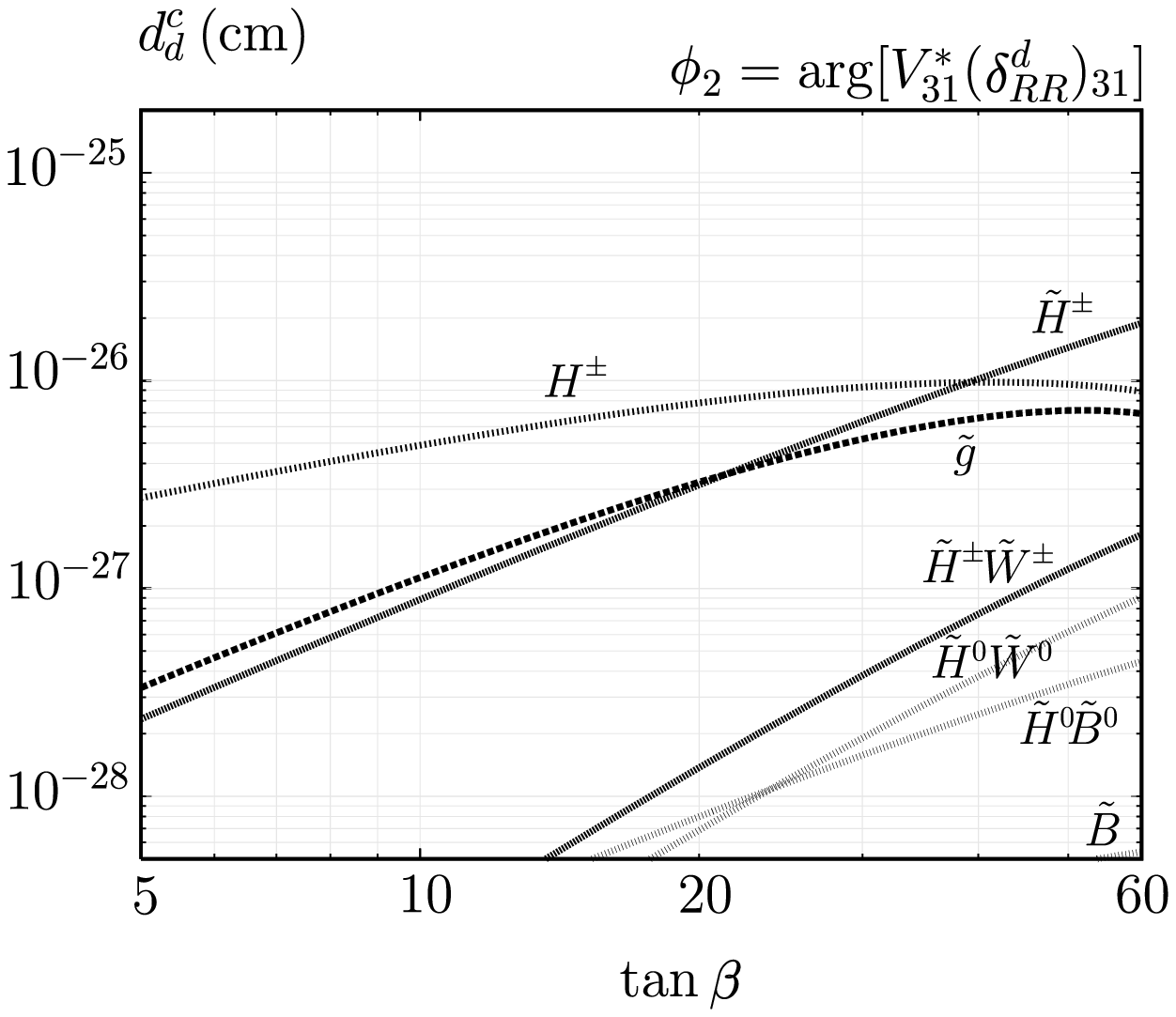} \\
(c) & (d)
\end{tabular}
\caption{Various contributions to the down quark EDM in (a), (b) and CEDM in (c), (d) as 
functions of $\tan\beta$. Here, the masses of SUSY particles are taken to be $1{\rm TeV}$ 
while $M_{H^\pm}=500\,{\rm GeV}$.  Contributions from the phase
$\phi_1={\rm arg}[(\delta^q_{LL})_{13}(\delta^d_{RR})_{31}]$ is presented in (a) and (c)
and those from the phase $\phi_2 = {\rm arg}[V^*_{31}(\delta^d_{RR})_{31}]$ in (b) and (d).
In all of these figures, maximum phases are assumed and only the absolute values of each
contribution are plotted.}
\label{Fig:FEDM_comparison}
\end{center}
\end{figure}
In the following, we discuss the contributions from $J^{(d_i)}_{LR}$ and $J^{(d_i)}_{RR}$
which are typically more important than the contributions from $J^{(d_i)}_{LL}$ thanks to the
enhancement by the heaviest Yukawa couplings.

In Fig.~\ref{Fig:FEDM_comparison} (a) [(c)] and (b) [(d)], we show various contributions 
to the down quark EDM [CEDM] as generated by the phases
$\phi_1={\rm Im}\left[(\delta^d_{LL})_{13}(\delta^d_{RR})_{31}\right]$ and
$\phi_2={\rm Im}\left[V^*_{31}(\delta^d_{RR})_{31}\right]$. We choose CKM-like mixing angles
$|(\delta^d_{LL})_{31}|=|(\delta^d_{RR})_{31}|=(0.2)^3$ and we assume maximum phases; moreover,
we consider a common soft SUSY mass $m_{\rm SUSY}=1\, {\rm TeV}$, $M_{H^\pm}=500\,{\rm GeV}$
and a positive $\mu$ term.

As shown in Fig.~\ref{Fig:FEDM_comparison} (a) and (c), the gluino effects are the dominant ones
as long as $\phi_1$ is the only non zero phase. We note that the LO gluino effects have a linear dependence with $\tan\beta$, however, the inclusion of higher order corrections acts to reduce
sizably the LO effects, specially in the large $\tan\beta$ regime.

The wino-Higgsino mixing contributions turn out to provide the most important effects after the
gluino ones. In particular, they almost grow as $\tan^2\beta$ and become comparable to the
gluino effects at around $\tan\beta\sim 60$. However, in realistic mass spectra, the charginos
tend to be lighter than the gluino, thus, the wino-Higgsino contribution becomes more important
compared to what Fig.~\ref{Fig:FEDM_comparison}~(a) and Fig.~\ref{Fig:FEDM_comparison}~(c) show.

Turning to the EDMs from $\phi_2$, which are generated only at the BLO, we find that the 
contributions from charged Higgs and Higgsino are dominant for any value of $\tan\beta$.
In particular, for low to moderate values of $\tan\beta$, the charged Higgs contributions
(growing almost linearly with $\tan\beta$) provide the dominant effects while from moderate
to large values of $\tan\beta$, the Higgsino contributions dominate thanks to their quadratic
dependence on $\tan\beta$.
Fig.~\ref{Fig:FEDM_comparison}~(b) and Fig.~\ref{Fig:FEDM_comparison}~(d) also show that the
linear dependence on $\tan\beta$ of the charged Higgs effects is partially lost at large
$\tan\beta$ due to higher order effects. Contributions from the gluino are typically subdominant
but still not completely negligible.

Next, we compare charged Higgs and Higgsino contributions from $\phi_2$ with the gluino contribution 
from $\phi_1$. Their relative size is controlled by the ratio $|V_{31}^*/(\delta^{d}_{LL})_{13}|$,
which we set to be 1 in Fig.~\ref{Fig:FEDM_comparison}. In concrete models, as for instance the
CMSSM or SUSY $SU(5)$ models with right handed neutrinos, $(\delta^d_{LL})$ is generated by radiative correction through the CKM matrix and the top Yukawa coupling, and it turns out typically that $(\delta^d_{LL})_{13}\simeq-cV^*_{31}$ with $c\sim {\cal O}(0.1)$. As a result, the BLO charged
Higgs and Higgsino contributions can be dominant over the LO gluino contributions.

All the above considerations for the EDMs apply also to the CEDMs with the only difference being
that the chargino contributions for the CEDMs are less important than the EDM ones because of the
absence of some constructive interferences among amplitudes.

\subsection{CP-violating four-Fermi interactions}

It is well-known that CP-violating four-Fermi interactions can generate atomic and hadronic
EDMs through the effective (CP-violating) couplings of the neutral Higgs bosons with fermions~\cite{Barr:1991yx,Lebedev:2002ne,Demir:2003js}.
These contributions to the EDMs are proportional to $\tan^3\beta$ and become important only
for large $\tan\beta$ values and for Higgs masses much lighter than the soft masses.
In Refs.~\cite{Barr:1991yx,Lebedev:2002ne,Demir:2003js}, flavor-conserving but CP-violating
phases were assumed.

As we will see, CP-violating four-Fermi interactions can be also generated by flavor effects.
In fact, as discussed in the Appendices, flavor-mixing effects contribute to both
flavor-violating and flavor-conserving CP-violating couplings of the Higgses with the fermions
(see Eq.~(\ref{Eq:NHiggs_coupling_MI})).

The relevant CP-odd four-Fermi interactions generated by flavor effects have the following form
\begin{eqnarray}
 {\cal L}_{eff} = \sum_{f_i,f_j=e,d,s,b} C_{f_i f_j}
 \bar{\psi}_{f_i}\psi_{f_i}\bar{\psi}_{f_j} i \gamma_5 \psi_{f_j}\,,
\end{eqnarray}
with the coefficients $C_{f_i f_j}$ defined as

\begin{eqnarray}
 C_{f_i f_j}
 &\simeq&
  -\frac{g_2^2 m_{f_i} m_{f_j}}{4 m_W^2}
  \frac{\tgb^2}{m_A^2}~
  \IM 
  \left[
   \left({C_{f_L}^H}\right)_{ii}^* \left({C_{f_L}^H}\right)_{jj}
  \right]
 \nonumber\\
 &\equiv&
  -\frac{g_2^2 m_{f_i} m_{f_j}}{4 m_W^2}
  \frac{\tgb^2}{m_A^2}~
  \frac{A_{f_i}-A_{f_j}}
  {|1+\eps_i\tgb+\eps^{(2)}_i\ttgb||1+\eps_j\tgb+\eps^{(2)}_j\ttgb|}\,.
\end{eqnarray}
$\left({C_{f_L}^H}\right)_{ii}$, which is given in Appendix~\ref{Sec:effective_vertex},
is the effective coupling of the neutral Higgs $H$ with the fermion $f$ and $ A_{f_i}$ 
is such that
\begin{eqnarray}
 A_{f_i}
 &=&
  \frac{\mb_b}{\mb_{d_i}}
  \left[ 
   \frac{r_i\epsLR\tgb}{6(1+\eps_3\tgb)}
   -
   \frac{(1+r_i)\epsL\epsR\ttgb}{9(1+\epsb_3\tgb)(1+\eps_3\tgb)}
  \right]
  {\rm Im}\left[ (\delta^d_{LL})_{i3}(\delta^d_{RR})_{3i} \right]
  \nonumber\\
  &&
  +
  \frac{\mb_b}{\mb_{d_i}}
   \frac{(1+r_i)\epsR\epsY\ttgb}{3(1+\epsb_3\tgb)(1+\eps_3\tgb)}
  {\rm Im} \left[ V^*_{3i}(\delta^d_{RR})_{3i} \right]\,,
\label{4fermion}
\end{eqnarray}
for $f_i=d,s$ and $A_b\simeq 0$. If $f_i=e$, the following expression arises
\begin{eqnarray}
 A_e
 &=&
  \frac{\mb_\tau}{\mb_{e}}
  \left[ 
   \frac{r^e_i\eps^e_{LR}\tgb}{6(1+\eps^e_3\tgb)}
   -
   \frac{(1+r^e_i)\eps^e_L\eps^e_R\ttgb}{9(1+\eps^e_3\tgb)(1+\eps^e_3\tgb)}
  \right]
  {\rm Im}\left[ (\delta^e_{LL})_{i3}(\delta^e_{RR})_{3i} \right]\,,
\end{eqnarray}
where the loop factors ${\epsilon^e_L}$, ${\epsilon^e_R}$ and ${\epsilon^e_{LR}}$ are
defined in Appendix~\ref{App:loop_factor} and for equal SUSY masses it turns out that 
$\eps^e_3=\eps^e_L=-{\rm sign}(\mu M_2)\times 3\alpha_2/16\pi$,
$\epsilon^e_{LR}={\rm sign}(\mu M_1)\times \alpha_Y/8\pi$ and $\eps^e_R=0$; moreover,
$r^e_i=(1+\epsilon^e_3\tgb)/|1+\eps^e_i\tgb+\eps^{e(2)}_i\ttgb|$.

The most prominent effect of the above CP-odd four-Fermi interaction compared to the
flavor conserving case of Refs.~\cite{Barr:1991yx,Lebedev:2002ne,Demir:2003js} is that,
thanks to the flavor effects, we can always pick up the heaviest Yukawa couplings in
the Higgs couplings with fermions.
Let us note that in Eq.~(\ref{4fermion}) we have neglected the contributions proportional
to $J^{d_i}_{LL}$ because they are not enhanced by the heaviest Yukawa coupling, hence,
they can never reach an experimentally interesting value.

\begin{figure}[tp]
\begin{center}
\begin{tabular}{c}
\includegraphics[scale=0.6]{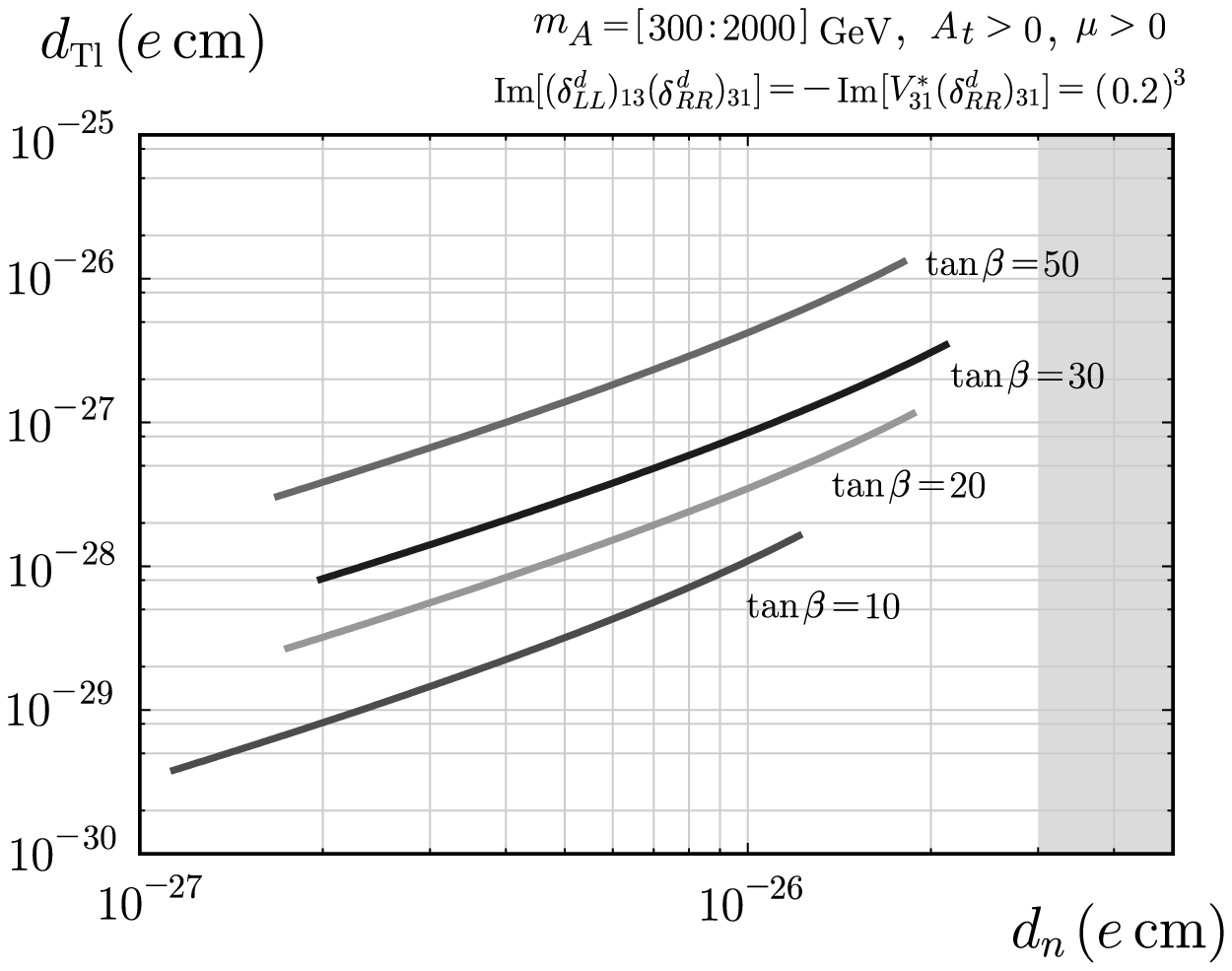}
\end{tabular}
\caption{$d_{\rm{Tl}}$ vs $d_n$ in the 2HDM limit of the MSSM. Here, we assume that
the EDMs are generated by the flavor dependent CP phases in the squark mass matrices.
The shaded region is excluded by the experimental bound on $d_n$}
\label{Fig:4edm}
\end{center}
\end{figure}

In order to understand the impact of the {\it flavored} four-Fermi interactions on the
atomic and hadronic EDMs, it is convenient to consider the two Higgs doublet model (2HDM)
limit of the MSSM, where the soft sector is assumed to be much heavier than the Higgs sector.
In this limit, $d_e$ is negligible and $d_{\rm{Tl}}$ can be only generated by the CP-odd 
four-Fermi interactions (see Eq.~(\ref{Eq:dTl})) while the main contribution to $d_n$ arises 
from the charged-Higgs mediated diagrams.
In Fig.~\ref{Fig:4edm}, we show the predictions for $d_{\rm{Tl}}$ and $d_n$ in the 2HDM limit
of the MSSM setting $A_e=0$ and assuming a common heavy SUSY mass for the soft sector.
Here, we assume that the EDMs are generated by the flavor dependent CP phases in the squark 
mass matrices.
The shaded region in Fig.~\ref{Fig:4edm} is excluded by the current experimental bound on $d_n$.
Considering the current experimental bounds on $d_{\rm{Tl}}<9\times 10^{-25}e$~cm~\cite{Regan:2002ta}
and $d_n<3\times 10^{-26} e$~cm~\cite{expedm}, we conclude that $d_n$ represents the
most powerful probe of the 2HDM limit scenario, when CP violation is generated by flavor effects
in squark mass matrices.

\subsection{Charged lepton EDMs}

As already discussed in Section~\ref{Sec:Jarlskog}, the dominant contributions to the lepton
EDMs arise at one-loop level through the invariant $J^{(e_i)}_{LR}$, that is generated by the
exchange of binos and sleptons.

On general ground, BLO corrections to the charged lepton EDMs are not so important as for the
down quarks case, as the gluino and Higgsino-mediated contributions are absent in the leptonic
sector.

In the following, we present the complete formulae for the charged lepton EDMs at the BLO, which
include the exchange of bino/sleptons, bino-Higgsino/sleptons and wino-Higgsino/sneutrinos
\begin{eqnarray}
\label{Eq:lEDM}
 \frac{d_{e_i}}{e}
 \!\!\!&=&\!\!\!
 \frac{\alpha_Y}{8\pi}
 \frac{m_\tau}{m^2}
 \frac{\mu M_1}{m^2}
 \tgb~ 
  \left(
   {\cal E}^{\tilde B}_1
    + {\cal E}^{{\tilde H}{\tilde B}}_1 
    + \frac{\alpha_2 M_2}{\alpha_Y M_1}
    {\cal E}^{{\tilde H}{\tilde W}}_1~
  \right)
  {\rm Im}\left[ (\delta^e_{LL})_{i3}(\delta^e_{RR})_{3i} \right]\,,
\end{eqnarray}
where
\begin{eqnarray} \label{Eq:lEDM_VC}
 {\cal E}^{\tilde B}_1 
 \!\!\!&=&\!\!\!
  \frac{f_0^{(3)}(x_1)}{1+\eps^e_3\tgb}
  +
  \frac{(\epsilon^e_{R}+\epsilon^e_{L})\tgb}{3(1+\eps^e_3\tgb)^2}
  f_0^{(2)}(x_1)
  +
  \left[
   \frac{(1+r_i)\epsilon^e_{L}\epsilon^e_{R} \ttgb}{9(1+\eps^{e}_3\tgb)^3}
   -
   \frac{r_i \epsilon^e_{LR}\tgb}{6(1+\eps^{e}_3\tgb)^2}
  \right]
  f_0^{(1)}(x_1),
 \nonumber \\
 {\cal E}^{{\tilde H}{\tilde B}}_1 
 \!\!\!\!\!\!\! &=& \!\!\!
  \frac{(\epsilon^e_{R}-2\epsilon^e_{L})\tgb}{6(1+\eps^{e}_3\tgb)^2} 
  g_0^{(1)}(x_1,y)
  +
  \left[
   \frac{r_i \epsilon^e_{LR}\tgb}{12(1+\eps^{e}_3\tgb)^2}
   -
   \frac{(1+r_i)\epsilon^e_{R}\epsilon^e_{L}\ttgb}{18(1+\eps^{e}_3\tgb)^3}
  \right] 
  g_0^{(0)}(x_1,y),
 \nonumber \\
 {\cal E}^{{\tilde H}{\tilde W}}_1 
 \!\!\!\!\!\!\! &=& \!\!\!
  \frac{\epsilon^e_{R}\tgb}{3(1+\eps^{e}_3\tgb)^2}
  g_{{\tilde H}{\tilde W}}^{(1)}(x_2,y)
  +
  \left[
   \frac{(1+r_i)\epsilon^e_{R}\epsilon^e_{L}\ttgb}{9(1+\eps^{e}_3\tgb)^3}
   -
   \frac{r_i \epsilon^e_{LR}\tgb}{6(1+\eps^{e}_3\tgb)^2}
  \right]
  g_{{\tilde H}{\tilde W}}^{(0)}(x_2,y)\,,
\end{eqnarray}
and the loop functions for wino-Higgsino mixing diagrams are given as
\begin{eqnarray}
 g^{(i)}_{{\tilde H}{\tilde W}} (x_2,y)
 = -g_1^{(i)}(x_2,y) -\frac{1}{2}g_0^{(i)}(x_2,y)
 \quad (i=0,1)\,,
\end{eqnarray}
where $x_2=M_2^2/m^2$, $x_1=M_1^2/m^2$ and $y=\mu^2/m^2$, with $m$ being a common
slepton/sneutrino mass.

\begin{figure}[tp]
\begin{center}
\begin{tabular}{c}
\includegraphics[scale=0.6]{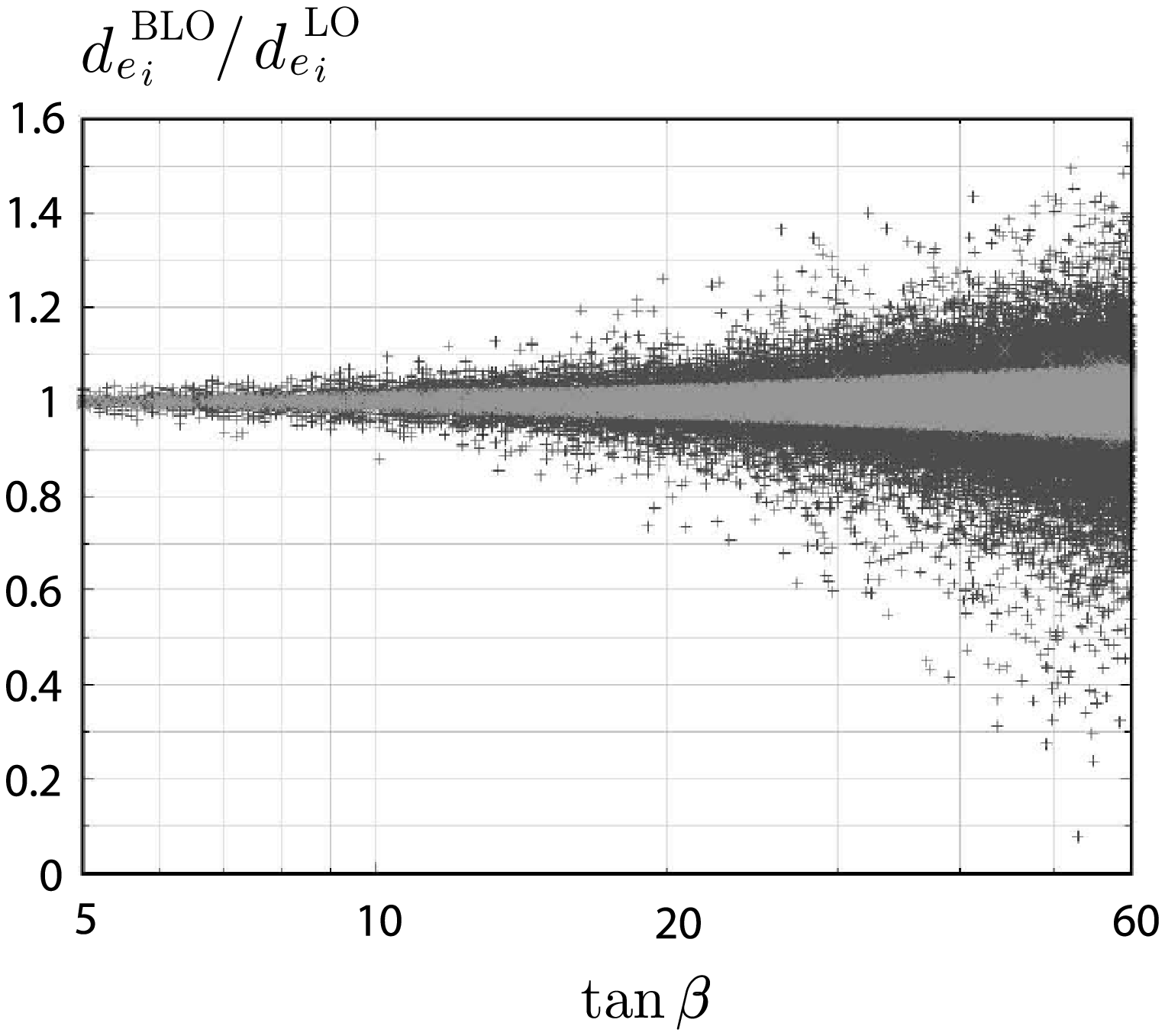} 
\end{tabular}
\caption{
$d^{\rm{BLO}}_{e_i}/d^{\rm{LO}}_{e_i}$ vs ${\rm tan}\beta$ taking the $\mu$ term, the slepton
and the gaugino masses randomly in the range of $[200:1000]$~\rm{GeV}. Light gray points satisfy
the GUT relation for the gaugino masses, $M_2=2M_1$, while dark gray points do not.}
\label{Fig:ledm}
\end{center}
\end{figure}

In Fig.~\ref{Fig:ledm}, we plot the leptonic EDMs evaluated at the BLO normalized to the
LO estimate as a function of $\tan\beta$, scanning over the relevant SUSY masses in the
range $[200:1000]$~\rm{GeV}. The inner region of Fig.~\ref{Fig:ledm} (light gray) corresponds
to the points satisfying the GUT relation between the gaugino masses, $M_2=2M_1$, while
the points in the larger region (dark gray) do not satisfy such a GUT relation.

As shown by Fig.~\ref{Fig:ledm}, BLO corrections to the leptonic EDMs can provide only
modest contributions, at the level of ${\cal O}(10)\%$ for large values of $\tan\beta$,
if the GUT relation between the gaugino masses is assumed. In contrast, if we do not
impose such a GUT relation, large BLO effects can arise in the large $\tan\beta$ regime
at the level of $0.5\lesssim d^{\rm{BLO}}_{e_i}/d^{\rm{LO}}_{e_i}\lesssim 1.5$.
Such a behavior of $d^{\rm{BLO}}_{e_i}/d^{\rm{LO}}_{e_i}$ with the gaugino masses can be
understood noting the different dependence of $d^{\rm{BLO}}_{e_i}$ and $d^{\rm{LO}}_{e_i}$
on the gaugino masses; in particular, when $M_2 \ll M_1$, it turns out that the LO
contributions have a faster decoupling property with $M_1$ compared to the BLO contributions.
%


\section{Flavored EDMs in SUSY GUT models} \label{Sec:GUT}

In this section, we discuss the predictions for the (C)EDMs in some specific SUSY
models as SUSY $SU(5)$ with right-handed neutrinos. As we will show, in this class
of models, the inclusion of the BLO contributions presented in this work turn out 
to be crucial to get the correct predictions for the (C)EDMs.

Since SUSY GUT models present rich flavor structures, they typically predict non-negligible
{\it flavored} (C)EDMs~\cite{gutedm} as well as non-negligible rates for various flavor-violating phenomena~\cite{gut_FCNC}.
Moreover, since within SUSY GUTs leptons and quarks sit into same multiplets, the flavor
violation in the squark and slepton sectors may be correlated~\cite{gut_FCNC}, 
if the SUSY-breaking effects are mediated at such a high energy scale at least.

A detailed exploration of the intriguing correlation and interplay among FCNC processes
in the leptonic and hadronic sectors with the flavored (C)EDMs would be very desirable,
but this analysis would require a dedicated study that goes beyond the scope of the present
work. On the contrary, the major aim of this section is to stress the importance of BLO effects
for the {\it flavored} EDMs in well motivated SUSY scenarios as SUSY $SU(5)$ with right-handed 
neutrinos.

Let us first review the flavor structures in the squark and slepton mass matrices in the SUSY
$SU(5)$ GUT with right-handed neutrinos. The Yukawa interactions for quarks and leptons and the
Majorana mass terms for the right-handed neutrinos are given by the following superpotential,
\begin{eqnarray}
W&=& 
\frac14 {y_u}_{ij}\Psi_i\Psi_j H
+\sqrt{2} {y_d}_{ij}\Psi_i\Phi_j\overline{H}
+{y_{\nu}}_{ij} \overline{N}_i\Phi_j {H}
+M_{ij}\overline{N}_i\overline{N}_j,
\label{Eq:W_GUT}
\end{eqnarray}
where $\Psi$ and $\Phi$ stand for {\bf 10}- and {$\bf\bar{5}$}-dimensional multiplets,
respectively, and $\overline{N}$ refers to the right-handed neutrinos. $H$ ($\overline{H}$)
is a {\bf 5}-({$\bf \bar{5}$}-) dimensional Higgs multiplet. After removing the unphysical
degrees of freedom, the Yukawa couplings and the right-handed neutrino mass matrix of
Eq.~(\ref{Eq:W_GUT}) can be written as
\begin{eqnarray}
{y_u}_{ij} &=&
 \left( V^T \hat{y}_u {\rm e}^{i \hat{\phi}_u } V \right)_{ij},
\nonumber\\
{y_d}_{ij} &=&
 y_{d_i} \delta_{ij},
\nonumber\\
{y_\nu}_{ij} &=&
 {\rm e}^{-i \hat{\phi}_{\bar \nu}}
 W^\dagger {\rm e}^{-i \hat{\phi}_\nu}
 \hat{y}_{\nu} U^\dagger,
\nonumber\\
M_{ij} &=&
M_{\nu_i} \delta_{ij}.
\label{Eq:GUT_yukawa}
\end{eqnarray}
In this paper, hat and bar symbols refer to diagonal matrices and to {\it bare} quantities,
respectively. In Eq.~(\ref{Eq:GUT_yukawa}), $\phi_{u}$, $\phi_{d}$, $\phi_{\nu}$ and
$\phi_{\bar\nu}$ are the physical CP-violating phases satisfying $\sum_i \phi_{f_i}=0$
$(f=u, d, \nu $ and ${\bar \nu})$. The unitary matrix $V$ is the CKM matrix in the
extension of the SM to the SUSY $SU(5)$ GUT, and each unitary matrices $U, V$ and $W$
have only a phase. In the limit where $W=1$, which we assume, $U$ corresponds to the
MNS matrix. In this case, the light neutrino mass eigenvalues are given as
\begin{eqnarray}
m_{\nu_{i}}
&=&
\frac{y_{\nu_i}^2}{M_{\nu_i}} \langle H_f \rangle^2\,,
\label{Eq:seesaw}
\end{eqnarray}
where $H_f$ is a doublet Higgs in $H$. The colored Higgs multiplets $H_c$ and $\overline{H}_c$
are introduced in $H$ and $\overline{H}$ respectively as $SU(5)$ partners of the Higgs doublets
of the MSSM. They have new flavor-violating interactions in Eq.~(\ref{Eq:W_GUT}).
If the SUSY-breaking terms in the MSSM are generated by dynamics above the colored Higgs masses,
such as in the gravity mediation, the squark mass terms may get sizable corrections by the colored
Higgs interactions. In the mSUGRA scenario, low-energy flavor-violating SUSY-breaking terms are radiatively induced, and they are qualitatively given as
\begin{eqnarray}
(m_{{\tilde u}_L}^2)_{ij} 
 &=& 
  - \frac{2}{(4\pi)^2} (V\hat{y}_d^2 V^\dagger)_{ij}(3m_0^2+A_0^2)
    \lsp 2\ln \frac{\mpl}{\mgut} + \ln \frac{\mgut}{\msusy} \rsp,
 \nonumber\\
 (m_{{\tilde u}_R}^2)_{ij} 
 &=& - \frac{4}{(4\pi)^2} 
    (e^{-i\hat{\phi}_u}V^* \hat{y}_d^2 V^T e^{i\hat{\phi}_u})_{ij}
    (3m_0^2+A_0^2)
    \ln \frac{\mpl}{\mgut},
 \nonumber\\
 (m_{{\tilde d}_L}^2)_{ij} 
 &=& - \frac{2}{(4\pi)^2} (V^\dagger \hat{y}_u^2 V)_{ij}(3m_0^2+A_0^2)
    \lsp 3\ln \frac{\mpl}{\mgut} + \ln \frac{\mgut}{\msusy} \rsp,
 \nonumber\\
 (m_{{\tilde d}_R}^2)_{ij} 
 &=& - \frac{2}{(4\pi)^2} 
    (e^{i\hat{\phi}_{d}}U^*\hat{y}_{\nu}^2 U^T e^{-i\hat{\phi}_{d}})_{ij}
    (3m_0^2+A_0^2)
    \ln \frac{\mpl}{\mgut},
 \nonumber\\
 (m_{{\tilde l}_L})_{ij} 
 &=& - \frac{2}{(4\pi)^2} 
    ( U \hat{y}_\nu e^{i\hat{\phi}_\nu} W )_{ik}
    ( W^\dagger e^{-i\hat{\phi}_\nu}\hat{y}_\nu U^\dagger )_{kj}
    (3m_0^2+A_0^2)
    \ln \frac{\mpl}{M_{\nu_k}},
 \nonumber\\ 
 (m_{{\tilde e}_R}^2)_{ij} 
 &=& - \frac{6}{(4\pi)^2} (V^T \hat{y}_u^2 V^*)_{ij}
    (3m_0^2+A_0^2)
    \ln \frac{\mpl}{\mgut},
\label{Eq:SU5RN_FV}
\end{eqnarray}
with $i\ne j$; we have also assumed the colored Higgs mass to be the
GUT scale $M_{\rm GUT}$ and $m_0$ and $A_0$ are the universal scalar
mass and trilinear couplings, respectively
As shown by Eqs.~(\ref{Eq:SU5RN_FV}), the off-diagonal components of the right-handed
squark and left-handed slepton mass matrices are induced by the neutrino Yukawa couplings,
and they depend on the various CP-violating phases of the model. The existence of flavor
violation in the right-handed down-type squarks is essential to generate the contributions
to the down-type quark (C)EDMs proportional to the Jarlskog invariants $J^{(d_i)}_{LR}$ and $J^{(d_i)}_{RR}$, which are enhanced by the largest down quark mass $m_b$. In this case, as stressed in the previous section, not only the gluino but also chargino and charged Higgs are expected to contribute significantly to the (C)EDMs.

Although in the previous section we have reported the analytical expressions for the down-type
quark (C)EDMs in the MI approximation, our numerical analysis is performed using the mass
eigenstates. Moreover, we also include the leading QCD corrections and we derive the (C)EDMs
at the hadron scale ($\sim 1\,{\rm GeV}$)~\cite{Degrassi:2005zd}.

\begin{figure}
\begin{center}
\begin{tabular}{cc}
\includegraphics[scale=0.55]{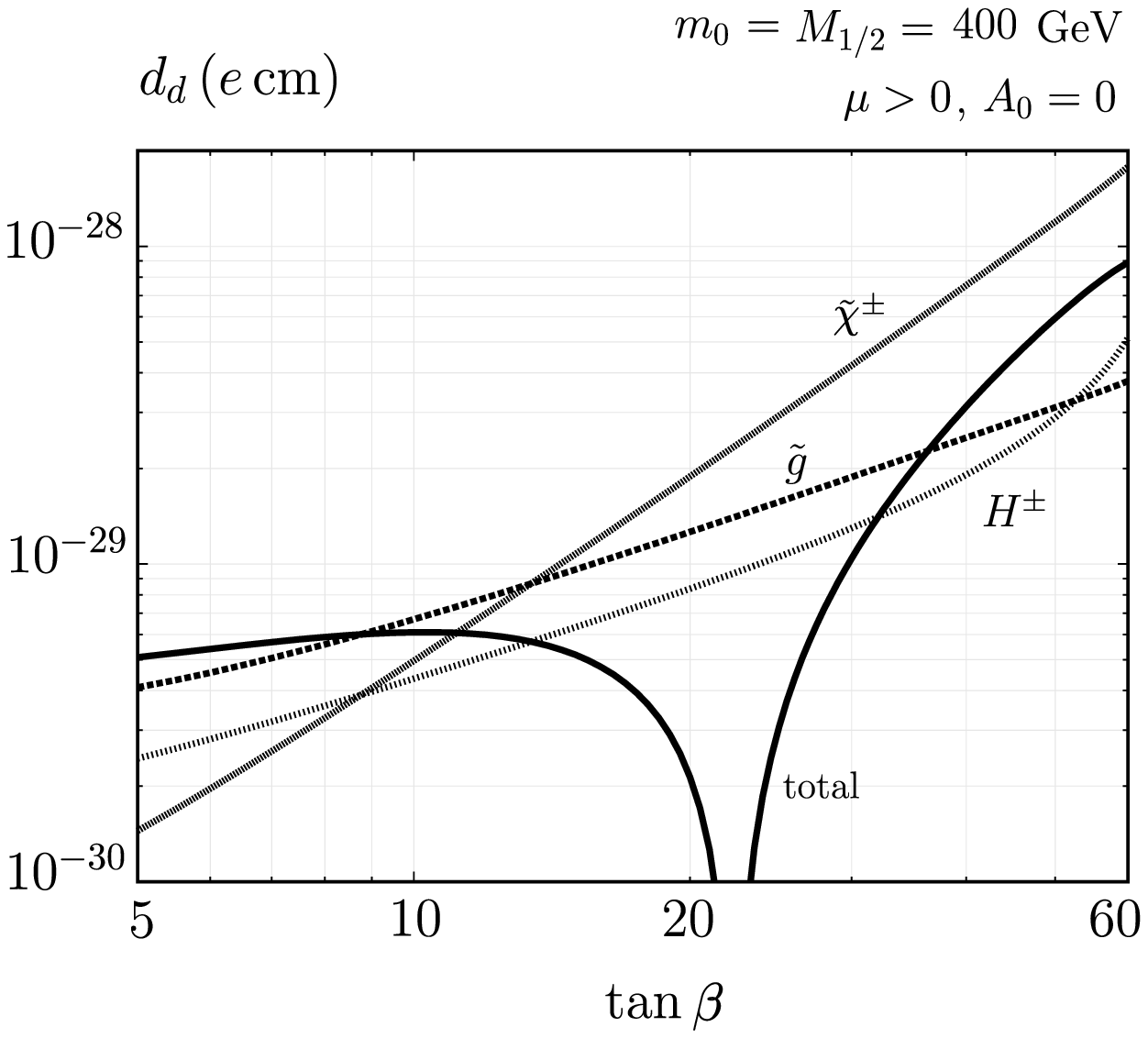} &
\includegraphics[scale=0.55]{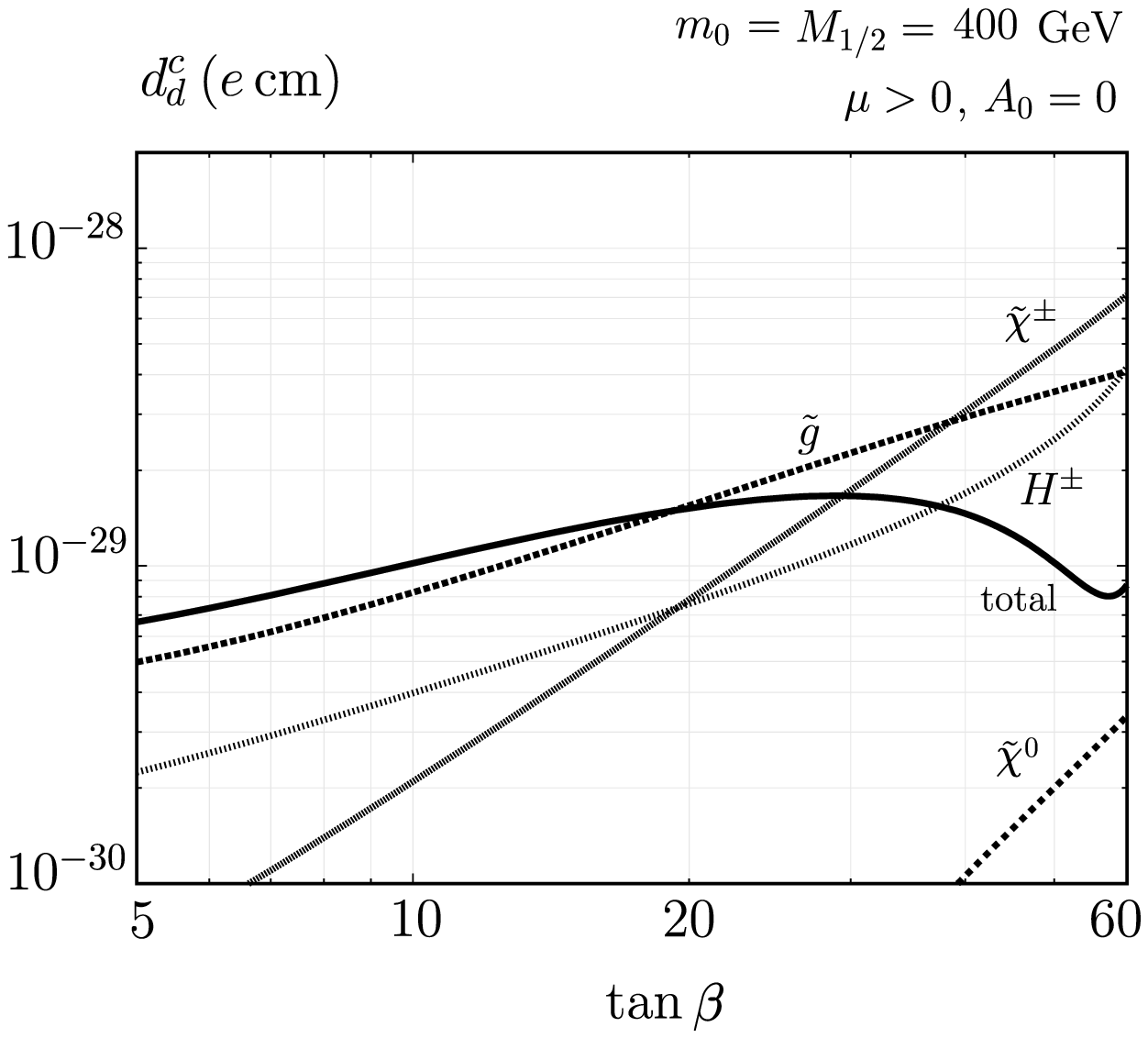} 
\end{tabular}
\caption{\label{Fig:dep_tanbeta} The down quark EDM (left) and CEDM (right) in the
$SU(5)$ GUT model with right-handed neutrinos as functions of $\tan\beta$.
Various contributions to the down quark EDM and CEDM are also shown. Here, maximum
phases are assumed and only the absolute values of each contribution are plotted.
The input parameters are given as $m_0=M_{1/2}=400\,{\rm GeV}$, $A_0=0$ and $\mu>0$.
The predicted values are approximately proportional to the mass of right-handed tau
neutrino and the (1,3) component of the MNS matrix, which are taken to be
$M_{\nu_3}=10^{14}\,{\rm GeV}$ and $U_{e3}=0.01$, respectively. The region for
$\tan\beta>35$ is already excluded by the current experimental limit on the LFV process, $Br(\mu\rightarrow e \gamma)<1.2\times 10^{-11}$.}
\end{center}
\end{figure}

In Fig.~{\ref{Fig:dep_tanbeta}}, we show the various contributions to
the down quark (C)EDM, setting $m_0=M_{1/2}=400\, {\rm GeV}$, $A_0=0$
and $\mu>0$. For the neutrino sectors, we assume a hierarchical mass
structure for light neutrino masses and take $m_{\nu_3}=0.05{\rm eV}$,
$M_{\nu_3}=10^{14}\,{\rm GeV}$ and $U_{e3}=0.01$.

For the down quark EDM in Fig.~\ref{Fig:dep_tanbeta} (left), the chargino and
charged Higgs contributions are typically comparable to the gluino ones despite
the fact that they are generated only at the BLO. This happens because {\it i)}
the charged Higgs, chargino and stop masses (entering the BLO contributions)
are generally lighter than the gluino and the other squark masses (entering the
LO contributions) in most of the MSSM parameter space; {\it ii)}
$|(\delta^{d}_{LL})_{13}| < |V_{31}|$, when $\delta^{d}_{LL}$ is radiatively-induced 
via the CKM matrix (as it happens in our model); {\it iii)} the mass functions for
the chargino (charged Higgs) contributions tend to become larger than those for the
gluino contributions due to positive interference of diagrams (due to the existence
of large logarithm).

Moreover, when $\tan\beta$ is large, the BLO effects become more
significant and they dominate over the LO ones. In the chargino case,
this is explained by the explicit $\tan\beta$-dependence of
$(d_{d})_{{\tilde\chi^{\pm}}}\sim\tgb^2$, to be compared with
$(d_{d})_{\tilde g}\sim\tgb$.  In the charged Higgs case, in spite of
the same explicit $\tan\beta$-dependence of $(d_{d})_{\tilde g}$ and
$(d_{d})_{H}$, $(d_{d})_{H}>(d_{d})_{\tilde g}$ for increasing
$\tan\beta$, since $m_H$ is highly reduced by large RG effects driven
by $y^2_b\sim y^2_t\sim1$.

Notice that, the corner of the MSSM parameter space where the BLO
$H^\pm$ effects are particularly enhanced compared to the LO ones,
corresponds to the so-called {\em A-funnel} region (where $m_A\simeq
2\,m_{\rm LSP}$), satisfying the WMAP constraints.  In the mSUGRA
scenario, for $\mu>0$, the chargino contributions have an opposite
sign compared to the gluino and charged Higgs contributions, leading
to a large cancellation between them.

From Fig.~\ref{Fig:dep_tanbeta} we note that the chargino contribution
to the CEDM is much lesser than the corresponding contribution to the
EDM; this is because the EDM has positive interferences among diagrams
that are absent in the CEDM case.

\begin{figure}[tp]
\begin{center}
\begin{tabular}{cc}
\includegraphics[scale=1.2]{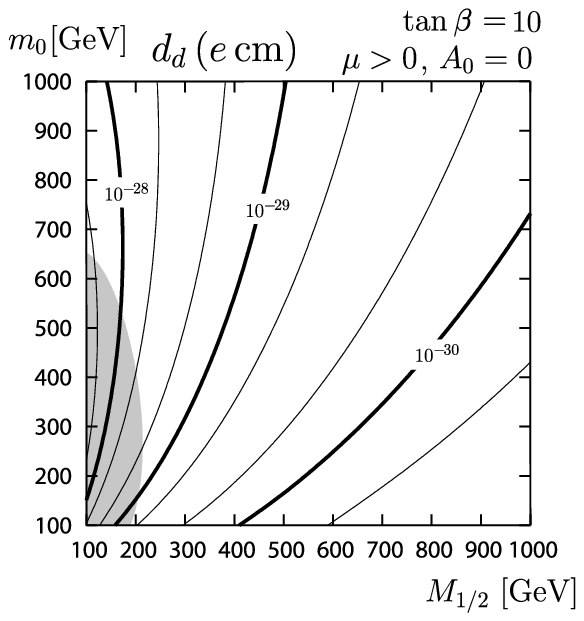} &
\includegraphics[scale=1.2]{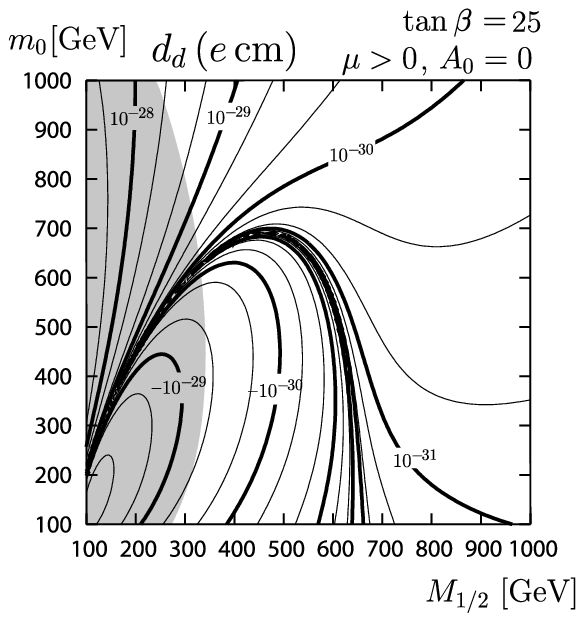} \\
(a) & (b) 
\end{tabular}
\vskip 0.5 cm
\begin{tabular}{cc}
\includegraphics[scale=1.2]{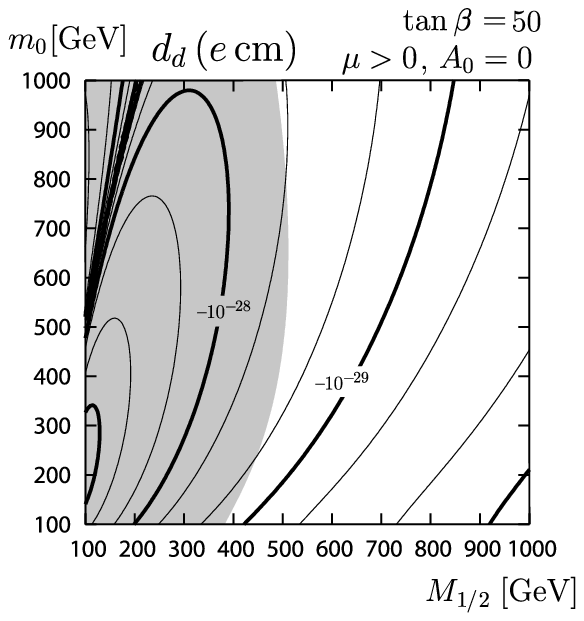} &
\includegraphics[scale=1.2]{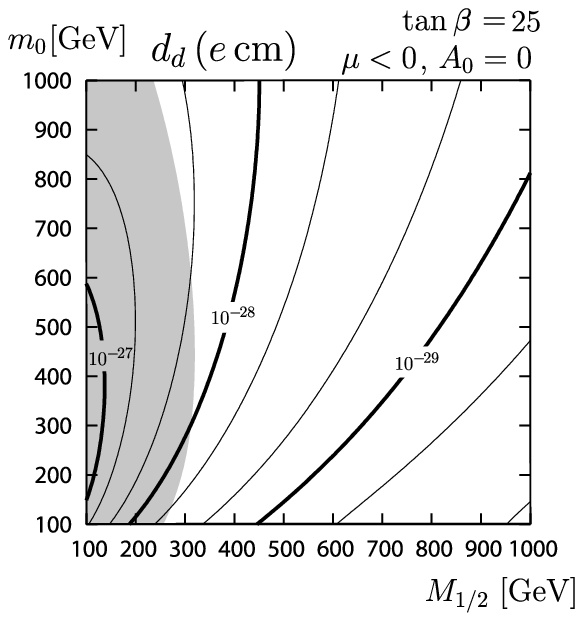} \\
(c) & (d)
\end{tabular}
\caption{\label{Fig:dep_m0_mhalf} Predictions of the down quark EDM in SUSY $SU(5)$ GUT
  model with right-handed neutrinos in the ($m_0$, $M_{1/2}$) plane. The input parameters
  $m_0$, $M_{1/2}$ and $A_0$ are given at the Planck scale and $A_0$ is set to $A_0=0$.
  The sign of the $\mu$ term is taken always positive but in (d) and $\tan\beta$ is taken
  to be 10, 25, 50 and 25 for (a), (b), (c) and (d), respectively. The predicted values for
  the EDM are approximately proportional to the mass of right-handed tau neutrino and to the
  (1,3) component of the MNS matrix, which are taken to be $M_{N_3}=10^{14}\,{\rm GeV}$ and
  $U_{e3}=0.01$, respectively. The grey region is excluded by the current experimental bound
  $Br(\mu \rightarrow e\gamma)<1.2\times 10^{-11}$.  }
\end{center}
\end{figure}

The predictions for the down quark EDM are investigated in more detail
in Fig.~\ref{Fig:dep_m0_mhalf}.  In particular, Figs.~(a),~(b) and (c)
describe the value of the EDM in the ($M_{1/2},m_0$) plane for $\mu>0$
and $\tan\beta=10, 25$ and $50$, respectively. Fig.~(d) refers to the
case in which $\tan\beta=25$ and $\mu<0$.

For relatively small values of $\tan\beta$ (see Fig.~\ref{Fig:dep_m0_mhalf}(a)),
the chargino contribution is not very large compared to the gluino and charged
Higgs ones. On the other hand, while increasing $\tan\beta$, the chargino
contributions become more important and they dominate over the other contributions
in many regions of the parameter space, see Fig.~\ref{Fig:dep_m0_mhalf}(b).
Chargino and charged Higgs contributions are specially important in the region
where $M_{1/2}\gsim m_0$.  Although the off-diagonal components of the squark
mass matrices mainly depend on the values of $m_0$ and $A_0$, as can be seen in
Eq.~(\ref{Eq:SU5RN_FV}), their diagonal components are mainly determined by $M_{1/2}$
due to the large RG effects driven by the $SU(3)_C$ interactions. As a result,
the MI parameters become small for $M_{1/2}\gsim m_0$ and the gluino contribution
proportional to $\IM [(\delta^d_{LL})_{13}(\delta^d_{RR})_{31}]$ is more suppressed
than the chargino and charged Higgs contributions that are proportional to
$\IM [V^*_{31}(\delta^d_{RR})_{31}]$.

\begin{figure}[tp]
\begin{center}
\begin{tabular}{cc}
\includegraphics[scale=1.2]{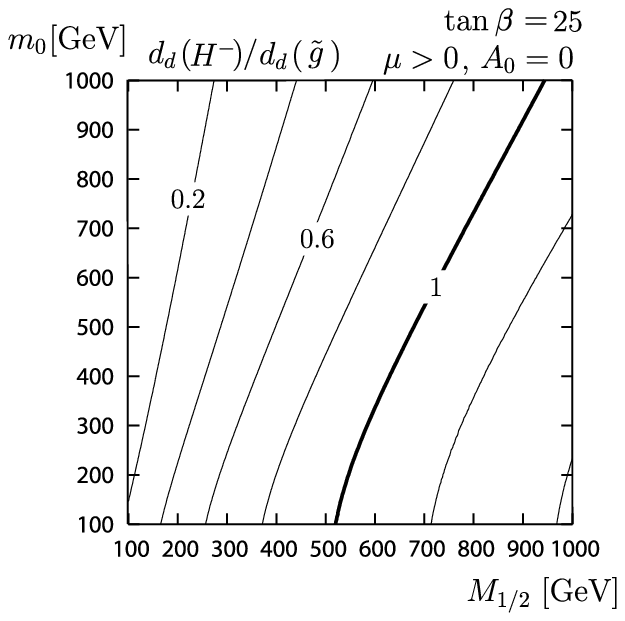} &
\includegraphics[scale=1.2]{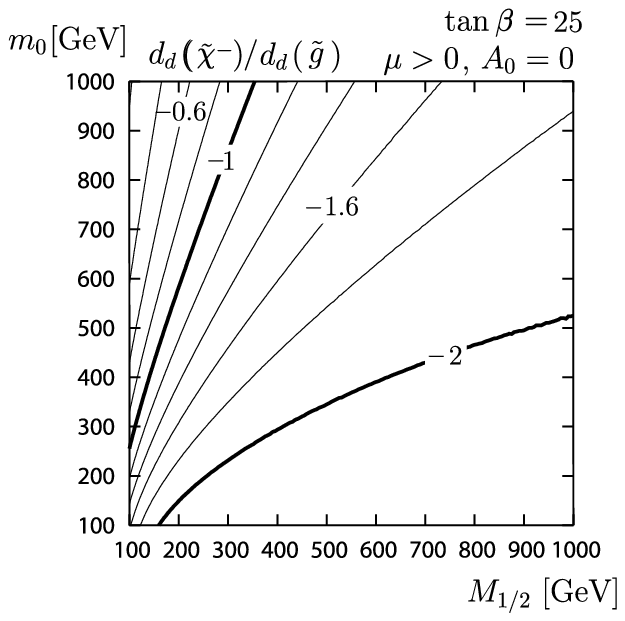} \\
\end{tabular}
\caption{\label{Fig:m0_mhalf_ratio} Left(Right): Ratio of the charged Higgs-mediated
(chargino-mediated) contributions over the gluino-mediated contributions in SUSY $SU(5)$
GUT model with right-handed neutrinos. The input parameters are the same as in Fig.~\ref{Fig:dep_m0_mhalf}(b).}
\end{center}
\end{figure}

This feature is clearly shown in Fig.~\ref{Fig:m0_mhalf_ratio} which describes the
ratio of the chargino and charged Higgs contributions over the gluino ones. In
the region where $M_{1/2} \gg m_0$, $d_d(\chi^\pm)/d_d({\tilde g})$ approximately 
approaches to a constant for a fixed value of $m_0$. This is because, as $m_0/M_{1/2}$ 
becomes small enough to be neglected, all the SUSY-breaking terms, even the off-diagonal
ones, are mostly determined by $M_{1/2}$, and all the contributions to the EDM behave as
$d_d \propto 1/M_{1/2}^2$. On the contrary, the charged Higgs contribution decouples more
slowly due to the large logarithm in the loop function, thus it provides the dominant
effects in the regions with heavy SUSY particle. This parameter dependence explains
the features of Fig.~\ref{Fig:dep_m0_mhalf}(b): for small $M_{1/2}$, the gluino is
relatively light and it provides the most important contribution to the (C)EDMs while,
as $M_{1/2}$ increases, the charged Higgs and specially the chargino contributions start 
dominating. For very large $M_{1/2}$ values, strong cancellations among amplitudes do not
appear because of the dominance of the charged Higgs contribution.

For very large $\tan\beta~(\sim 50)$, the BLO chargino mediated contributions tend to
become the dominant one in large regions of the parameter space (see Fig.~\ref{Fig:dep_m0_mhalf}(d)) due to their $\tan^2\beta$ dependence. In the case of
$\mu<0$ (see in Fig.~\ref{Fig:dep_m0_mhalf}(d)), the down quark (C)EDMs reach values
significantly larger than those for $\mu>0$. In fact, when $\mu<0$, gluino, chargino and
charged Higgs contributions have a common sign, thus, they constructively interfere with
each other; moreover the threshold corrections to the tree-level Yukawa couplings provide
enhancement contributions, in contrast to the $\mu>0$ case.

In summary, the inclusion of BLO effects for the (C)EDMs is mandatory
as long as they are generated by flavor violating terms in the MSSM
soft sector. In addition, when large cancellations among various
contributions do not occur, the predicted values for the down quark
EDMs can be within the reach of the future experimental sensitivities
such as for the neutron EDM ($d_n\sim 10^{-28}~e{\rm cm}$) and
deuteron EDM ($d_d\sim 10^{-29}~e{\rm cm}$). So, the observation of
hadronic EDMs at these future experiments could allow us to explore
the flavor-violating soft sector of the MSSM. In this respect, an
accurate theoretical prediction for the EDMs, by means of the
inclusion of the BLO effects discussed in this work, plays a
crucial role~
\footnote{
The hadronic EDMs could be also generated by
flavor-conserving but CP-violating parameters, such as the $\mu$ term
and the $A$ terms. Actually, if non-vanishing EDMs will be observed in
some experiments, powerful tools in shedding light on the source of
CP violation would be  {\it i)} correlated analyses of the EDMs for many
kind of nuclei and atoms, and also {\it ii)}  studies of the correlations among EDMs
and flavor violating transitions~\cite{ABP}. Obviously, a precise
calculation of the EDMs is essential for this purpose, too.
}.

\begin{figure}[tp]
\begin{center}
\begin{tabular}{cc}
\includegraphics[scale=1.2]{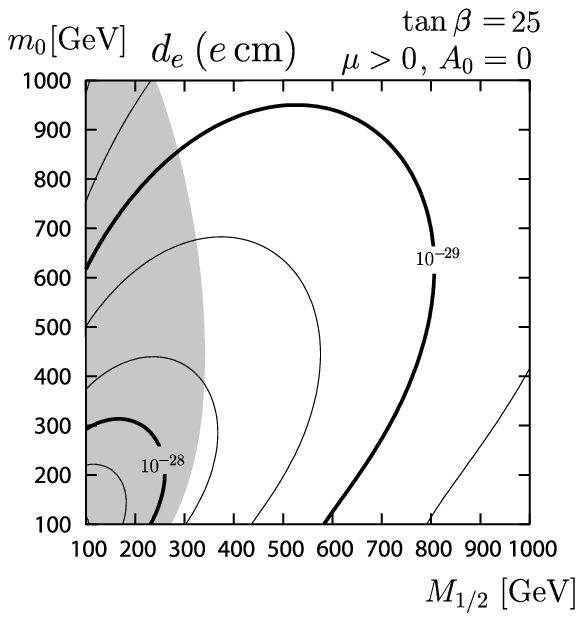} &
\includegraphics[scale=1.2]{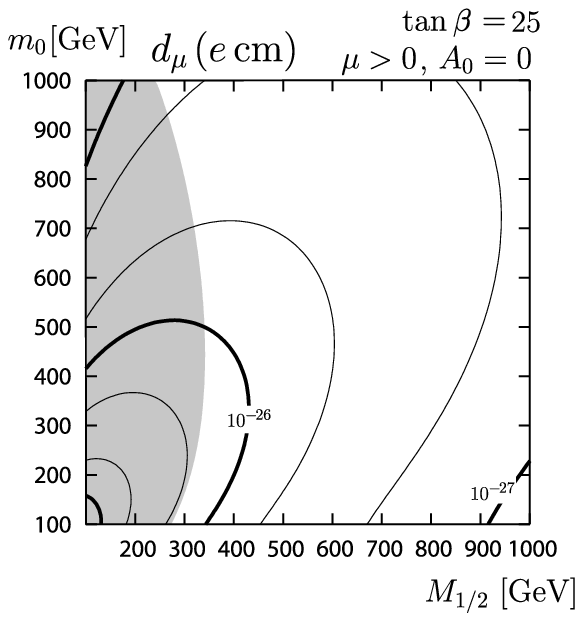} \\
(c) & (d)
\end{tabular}
\caption{\label{Fig:dep_m0_mhalf_lep} Predictions for the electron (muon) EDM $d_e$
 ($d_{\mu}$) in SUSY $SU(5)$ with right-handed neutrinos in the ($m_0$, $M_{1/2}$)
 plane. The values attained by $d_e$ and $d_{\mu}$ scale with $\tan\beta$ as
 ($\tan\beta$/25). See the caption of Fig.~\ref{Fig:dep_m0_mhalf} for additional details.
}
\end{center}
\end{figure}
For completeness, we also discuss the predictions for the electron (muon) EDM
$d_e$ ($d_{\mu}$), within the SUSY $SU(5)$ model with right-handed neutrinos.
In Fig.~\ref{Fig:dep_m0_mhalf_lep}, we show the values attained by $d_e$ and 
$d_{\mu}$ in the ($m_0$, $M_{1/2}$) plane setting $\tan\beta = 25$ (for different 
values of $\tan\beta$, $d_e$ and $d_{\mu}$ approximately scale as ($\tan\beta$/25)).
The grey region is excluded by the current experimental bound
$Br(\mu \rightarrow e \gamma)<1.2\times 10^{-11}$. As we can see, $d_e$ and $d_{\mu}$
can reach values at the level of $d_e\lesssim 10^{-28}~e{\rm cm}$ and $d_{\mu}\lesssim 10^{-26}~e{\rm cm}$, well within the expected future experimental sensitivities on
$d_e$, at least.


\section{Conclusions} \label{chap:Conclusion}

The hadronic and leptonic electric dipole moments (EDMs) represent an
ideal ground where to look for New Physics effects. In fact, on the
one side, the Standard Model predictions for the EDMs are well far
from the present experimental resolutions and, on the other side, the
EDMs are among the most sensitive observables to New Physics effects
in a broad class of theories beyond the Standard Model, included
supersymmetry.

However, there are two main obstacles to fully exploit the New Physics
sensitivity of the EDMs:
\begin{itemize}
\item  Experimentally, one measures the EDMs of composite systems, as
heavy atoms, molecules or the neutron EDM while the theoretical
predictions are relative to the EDMs of constituent particles,
{\it i.e.} quarks and leptons, thus a matching between quarks and leptons
EDMs into physical EDMs is necessary and this induces sizable
uncertainties related to QCD and nuclear and atomic interactions.

\item From a theoretical side, the supersymmetric evaluation of the
quark and lepton EDMs has been performed only at the leading order,
thus, potentially large effects from BLO contributions have been
disregarded so far.
\end{itemize}
In this paper, we have reported a detailed analysis, within supersymmetric theories,
of the hadronic and leptonic {\it flavored} EDMs, {\it i.e.} EDMs generated by
flavor-violating soft terms, at the BLO contributions, providing an answer to the
second question.

As shown by the present analysis, the impact of BLO corrections to the {\it flavored}
EDMs, when arising from ($\tan\beta$-enhanced) non-holomorphic contributions, can be
very important. We have pointed out the existence of new contributions to the EDMs that
can arise only at the two-loop level, thanks to the inclusion of ($\tan\beta$-enhanced)
non-holomorphic corrections. In fact, at one-loop level, certain contributions to the
EDMs are absent (as it happens for the charged-Higgs mediated EDMs) or completely negligible
(as for the chargino mediated EDMs). In the above cases, if new sources of flavor violation
are present, two-loop effects do not represent just sub-leading corrections to the
corresponding one-loop effects but, on the contrary, they provide the largely dominant
effects.

Moreover, the most important result of our work is that these two-loop
effects (from charged-Higgs and chargino exchanges, respectively),
previously neglected in the literature, are systematically comparable to
and often larger than the one-loop effects from gluino exchanges in a
broad region of the SUSY parameter space. This result is in contrast
to what typically happens in flavor-changing neutral current transitions.
For instance, in the $b \rightarrow s \gamma$ transition, the leading
contributions from charged-Higgs and charginos arise already at the one-loop
level and BLO effects can only provide a sub-leading correction to the
leading order result.
Moreover, altough BLO effects to the EDMs arise from ($\tan\beta$-enhanced)
non-holomorphic contributions, they turn out to be very important for the
entire allowed range of $\tan\beta$.

As a result, the evaluation of the BLO effects to the (C)EDMs presented in this work 
is an essential step forward towards a correct prediction for the (C)EDMs in supersymmetric
theories with flavor-violating soft terms. This point has been explicitly demonstrated 
in Section~\ref{Sec:GUT} for the case of a SUSY $SU(5)$ theory with right-handed neutrinos.


\section*{Acknowledgment}
The work of JH was supported by the World Premier International Center Initiative (WPI Program),
MEXT, Japan and in part by the Grant-in-Aid for Science Research, Japan Society for the Promotion
of Science (No.~20244037 and No.~2054252).
The work of MN and PP was supported in part by the Cluster of Excellence ``Origin and Structure
of the Universe'' and by the German Bundesministerium f{\"u}r Bildung und Forschung under contract 05HT6WOA.

\newpage
\appendix

\section{Effective Interactions beyond the Leading Order}
\label{Sec:effective_vertex}
The impact of non-holomorphic ($\tan\beta$-enhanced) Yukawa couplings has been 
already discussed in various processes~\cite{D'Ambrosio:2002ex,Buras:2002vd,Hall:1993gn,Hamzaoui:1998nu}.
However, in spite of their importance, a systematic analysis including general
flavor structures and CP-violating phases has not been accomplished yet. In this
section we derive various effective couplings induced by the $\tan\beta$-enhanced
radiative corrections with explicit CP phase dependence.

The derivation of the effective couplings is performed by means of the following procedure~\cite{D'Ambrosio:2002ex,Buras:2002vd}: {\it i)} evaluation of the
one-loop induced coefficients for the dimension-four operators, which modify the
Yukawa couplings, in the MI approximations, {\it ii)} determination of the
unitary matrices that diagonalize perturbatively the quark and lepton mass matrices,
treating the flavor-violating one-loop corrections as small parameters. At this
step, we specify the relation between the tree-level ({\it bare}) couplings, that appear
in the tree-level Lagrangian, and the observed ({\it physical}) parameters, {\it iii)}
derivation of the relevant effective couplings in the basis where the quarks and leptons
are in the mass eigenstates while the squarks and sleptons are in the flavor eigenstates.

In the next subsection we address the points {\it i)} and {\it ii)}
while in the last two subsections we focus on the point {\it iii)}.

\subsection{Non-holomorphic Yukawa Couplings}
In the $SU(2)_L\times U(1)_Y$ symmetric limit, the interaction Lagrangian for the
Higgs and matter fields is expressed by
\begin{eqnarray} \label{Eq:lag_NH}
 {\cal L} 
 \!\!\!&=&\!\!\!
  {{\overline u}_R^0}_i
  \left[
   {\yb_u}_i \Vb_{ij} H_2
   -
   (\eps^u \Vb)_{ij} H^\dagger_1
  \right]
  {q^0_L}_j
  +
  {{\overline d}_R^0}_i
  \left[
   {\yb_d}_i \delta_{ij} H_1
   -
   \eps^d_{ij} H^\dagger_2
  \right]
  {q_L^0}_j
  \nonumber \\
  && \!\!\!
  +~
  {{\overline e}_R^0}_i
  \left[
   {\yb_e}_i \delta_{ij} H_1
   -
   \eps^e_{ij} H^\dagger_2
 \right]
  {l_L^0}_j
  + {\rm h.c.}
  ,
\end{eqnarray}
where $\Vb$ and $\yb_f$ are CKM matrix and Yukawa couplings, respectively,
in the {\it bare} CKM basis and $\eps^f_{ij}~(f=u,d,e)$ are the non-holomorphic
radiative terms, which are generated when heavy SUSY particles are integrated
out from the effective theory. Here, the conjugate fields are defined as
\begin{eqnarray}
 {H_2^\dagger}_a \equiv \eps_{ab}({H_2}_b)^*,~
 {H_1^\dagger}_a \equiv \eps_{ab}({H_1}_b)^*\,.
\end{eqnarray}
The non-holomorphic radiative corrections are generated at one-loop level and
they are reported in Appendix~\ref{App:eff_vertex}. Notice that Eq.~(\ref{Eq:lag_NH})
contains both the corrections for the fermion mass matrices and the vertex
corrections for Higgs couplings.

After the electroweak symmetry breaking, we obtain the fermion mass matrices
which are diagonalized by the unitary matrices through
\begin{eqnarray} \label{Eq:diagonalize_mass}
 \mb_{u_i} \delta_{ij} + (\delta m_u)_{ij}
 \!\!\! &=& \!\!\!
  \frac{v \sinb}{\sqrt{2}} {\yb_u}_i
  \left( \delta_{ij} + t^{-1}_{\beta}\eps^u_{ij} \right)
  \equiv
 (U_{u_R}^\dagger \hat{m}_u U_{u_L})_{ij}\,,
 \nonumber\\
 \mb_{d_i} \delta_{ij} + (\delta m_d)_{ij}
 \!\!\! &=& \!\!\!
  -
  \frac{v \cosb}{\sqrt{2}} {\yb_d}_i
  \left( \delta_{ij} - t_{\beta}\eps^d_{ij} \right)
  \equiv
 (U_{d_R}^\dagger \hat{m}_d U_{d_L})_{ij}\,,
 \nonumber\\
 \mb_{e_i} \delta_{ij} + (\delta m_e)_{ij}
 \!\!\! &=& \!\!\! 
  -
  \frac{v \cosb}{\sqrt{2}} {\yb_e}_i
  \left( \delta_{ij} - t_{\beta}\eps^e_{ij} \right)
  \equiv
 (U_{e_R}^\dagger \hat{m}_e U_{e_L})_{ij}\,,
\end{eqnarray}
where $\sinb=\sin\beta$, $\cosb=\cos\beta$, $t_{\beta}=\sin\beta/\cos\beta$ and
$v= 2m_W/g_2$ while $\hat{m}_{f_i}$ stands for the mass eigenvalues with $U_f$
(with $f=u_L,u_R,d_L,d_R,e_L,e_R$) being the unitary matrices that diagonalize
the quark and lepton mass matrices. In the above equations, hat and bar symbols
refer to diagonal matrices and {\it bare} quantities, respectively. 
The {\it physical} SCKM basis is obtained by the unitary transformations
\begin{eqnarray} \label{Eq:unitary_transformation}
 u_L = U_{u_L} \Vb~u^0_L, \quad
 d_L = U_{d_L} d^0_L, \quad
 e_L = U_{e_L} e^0_L, \quad
 f_R = U_{f_R} f^0_R ~~(f=u,d,e)\,,
\end{eqnarray}
and the CKM matrix is expressed by
\begin{eqnarray} \label{Eq:CKM_matching}
 V = U_{u_L} \Vb U_{d_L}^\dagger\,.
\end{eqnarray}
In order to include the above threshold corrections to the Yukawa couplings, we need to derive,
for a given SUSY spectrum, the {\it bare} parameters (according to Eqs.~(\ref{Eq:diagonalize_mass})
and (\ref{Eq:CKM_matching})) that correctly reproduce physical quantities.

After that, the effective couplings among quarks and leptons (in the mass eigenstates) with the
Higgses are found performing the unitary transformations of Eq.~(\ref{Eq:unitary_transformation})
on Eq.~(\ref{Eq:lag_NH}).

The procedure for the numerical calculation is summarized in Appendix~\ref{App:eff_vertex}.
In the following, we derive the analytical expressions for $\eps^f_{ij}~(f=u,d,e)$ and
$U_{f}(f=u_L,u_R,d_L,d_R,$ $e_L,e_R)$ using the mass insertion (MI) approximation
to explicitly show their parameter dependences on the flavor structures of the MSSM.
A comparison between the numerical results obtained by means of the MI approximation
and the full computation shows that the two approaches are fully consistent provided the
MI parameters are small enough, as expected.

The calculation of the loop induced non-holomorphic Yukawa couplings requires to specify
the SUSY spectrum, which depends on the {\it bare} Yukawa couplings through the left-right mixing
terms in the sfermion mass matrices.
Therefore, it is useful to perform the calculation using the MI parameters defined in the SCKM
basis for {\it bare} Yukawa couplings ({\it bare} SCKM basis). Clearly, physical observables do not
depend on the choice of the basis and we express the MI parameters and the effective vertices
for squarks in the {\it bare} SCKM basis, just for convenience.

With an appropriate redefinition of the quark fields and the {\it bare} CKM matrix, the unitary
matrices can take the following form,
\begin{eqnarray} \label{Eq:unitary_matrices}
 U_{d_R} = e^{-i \hat{\theta}_d} V_{d_R}, \quad
 U_{d_L} = V_{d_L},  \quad
 U_{u_R} = e^{-i \hat{\theta}_u} V_{u_R}, \quad
 U_{u_L} = V_{u_L},  \quad
\end{eqnarray}
where the diagonal components of $V_f(f=d_R,d_L,u_R,u_L)$ are real. In the following, we refer to
this basis as the {\it bare} SCKM basis. The formulae given in Section~\ref{Sec:FEDM} are valid for
the {\it bare} MI parameters defined in this basis. The MI parameters defined in the {\it bare} SCKM
basis can be related to those defined in the {\it physical} SCKM basis by the following unitary 
transformation
\begin{equation}
(\delta^f_{XX})^{\rm phys}_{ij} = 
\bigg(U_{f_X} (\delta^f_{XX})^{\rm bare} U_{f_X}^\dagger\bigg)_{ij}\,,
\end{equation}
where $X=L,R$, $f=d,u$ and $i,j$ are flavor indeces. In practice, their difference can be
safely neglected as long as the off-diagonal components of the unitary matrices $U_{f_X}$
and their phase factors are small enough.


In the SM, the quark and lepton mass matrices show an highly hierarchical structure and
only the Yukawa couplings of the third generations induce large radiative corrections.
To pick up them, we define a projection operator $\Delta = {\rm diag}(0,~0,~1)$.
Under the approximations mentioned above, the corrections for the quark and lepton masses
are parameterized as
\begin{eqnarray} \label{Eq:mass_correction}
 (\delta m_u)_{ij}
 \!\!\! &=& \!\!\!
  \mb_{u_i} \epsgp_i \tgb^{-1}\delta_{ij} 
   +
   \frac{\epsLRp_{ij}\tgb^{-1}}{6} \overline{m}_t 
   (\delta^u_{RR}\Delta\delta^u_{LL})_{ij}
   -
   \frac{{\epsRp}_{ji}\tgb^{-1}}{3} ( \delta^u_{RR} )_{ij} 
   \mb_{u_j}
 \nonumber\\
 && \!\!\!
   -
   \overline{m}_{u_i} 
   \left[
    \frac{\epsLp_{ij}\tgb^{-1}}{3} ( \delta^u_{LL} )_{ij}
    -
    \epsYp \tgb^{-1} (\Vb \Delta \Vb^\dagger)_{ij}
    +
    \frac{\epsLYp_k \tgb^{-1}}{3} \Vb_{ik} (\delta^d_{LL} \Delta \Vb^\dagger)_{kj}
   \right],
  \nonumber\\
 (\delta m_d)_{ij}
 \!\!\! &=& \!\!\!
   \mb_{d_i} \epsg_i \tgb\delta_{ij} 
   +
   \frac{\epsLR_{ij}\tgb}{6} \overline{m}_b 
   (\delta^d_{RR}\Delta\delta^d_{LL})_{ij}
   -
   \frac{{\epsR}_{ji}\tgb}{3} ( \delta^d_{RR} )_{ij} 
   \mb_{d_j}
 \nonumber\\
 && \!\!\!
   -
   \overline{m}_{d_i} 
   \left[
    \frac{\epsL_{ij}\tgb}{3} ( \delta^d_{LL} )_{ij}
    -
    \epsY\tgb (\Vb^\dagger \Delta \Vb)_{ij}
    +
    \frac{\epsLY_k\tgb}{3} (\Vb^\dagger)_{ik} (\delta^u_{LL} \Delta \Vb)_{kj}
   \right]\,,
  \nonumber\\
 (\delta m_e)_{ij}
 \!\!\! &=& \!\!\!
   \mb_{e_i} {\eps_g^e}_i\tgb \delta_{ij} 
   +
   \frac{{\eps_{LR}^e}_{ij}\tgb}{6} \overline{m}_\tau 
   (\delta^e_{RR}\Delta\delta^e_{LL})_{ij}
 \nonumber\\
 &&  \!\!\!
   -
   \frac{{\eps_R^e}_{ji}\tgb}{3} ( \delta^e_{RR} )_{ij} 
   \mb_{e_j}
   -
   \overline{m}_{e_i} 
   \frac{{\eps_L^e}_{ij}\tgb}{3} ( \delta^e_{LL} )_{ij},
\end{eqnarray}
where the loop factors are given in Appendix~\ref{App:loop_factor} and a summation over the
index $k$ is assumed in Eq.~(\ref{Eq:mass_correction}). Assuming equal SUSY masses, it turns out that
$\epsLR\!=\!\epsL\!=\!\epsR=\epsg=\epsLRp=\epsLp=\epsRp=\epsgp={\rm sign}(\mu M_3)\times\alpha_s/3\pi$,
$\epsY=\epsLY=-{\rm sign}(\mu A_t)\times y_t^2/32\pi^2$ and $\epsYp=\epsLYp=-{\rm sign}(\mu A_b)\times y_b^2/32\pi^2$ for quarks. For leptons, $\eps^e_g=\eps^e_L=-{\rm sign}(\mu M_2)\times 3\alpha_2/16\pi$,~$\eps^e_{LR}={\rm sign}(\mu M_1)\times\alpha_Y/8\pi$ and $\eps^e_R=0$.

Next, we derive the approximate expressions for the unitary matrices $U_{d_R}$ and $U_{d_L}$ that
diagonalize the down-type quark mass matrix. As stated before, it is convenient to choose the basis 
where
\begin{eqnarray}
 U_{d_R} = e^{-i \hat{\theta}_d} V_{d_R}, \quad
 U_{d_L} = V_{d_L},
\end{eqnarray}
where the phase rotation $e^{-i \hat{\theta}_d}$ has been introduced to make the quark masses real.
From Eq.~(\ref{Eq:diagonalize_mass}), $U_{d_R}$, $U_{d_L}$ and $e^{-i \hat{\theta}_d}$ are determined
by the following matching condition,
\begin{eqnarray}
   V_{d_R} (\hat{\mb}_d + \delta m_d) V_{d_L}^\dagger 
   = \hat{m}_{d}~e^{i \hat{\theta}_d}\,.
\end{eqnarray}
Expanding $V_{d_{R,L}}\simeq 1 + \delta V_{d_{R,L}}$, we find
\begin{eqnarray}
 (\delta V_{d_L})_{ij} 
 \!\!\! &\simeq& \!\!\!
  ~~\frac{1}{{m_d}_i^2 - {m_d}_j^2}
  \left[ 
   ({\mb_d}_i+(\delta m_d^\dagger)_{ii})(\delta{m}_d)_{ij}
   +
   (\delta{m}_d^\dagger)_{ij}({\mb_d}_j+(\delta m_d)_{jj})
  \right]\,,
 \nonumber\\
 (\delta V_{d_R})_{ij} 
 \!\!\! &\simeq& \!\!\!
    \frac{1}{{m_d}_i^2 - {m_d}_j^2}
  \left[ 
   ({\mb_d}_i+({\delta m_d})_{ii})(\delta{m}_d^\dagger)_{ij}
   +
   (\delta{m}_d)_{ij}({\mb_d}_j+({\delta m_d^\dagger})_{jj})
  \right]\,,
\end{eqnarray}
and
\begin{eqnarray}\label{Eq:mass_matching}
 {m_d}_i e^{i{\theta_d}_i}
 \!\!\! &=& \!\!\!
  {\mb_d}_i(1 + \eps_{i}\tgb + \eps^{(2)}_i \ttgb)\,,
\end{eqnarray}
where
\begin{eqnarray}
 \eps_i \tgb
 &\simeq&
  \frac{(\delta m_d)_{ii}}{{\mb_d}_i} \quad (i=1,2,3)\,,
 \nonumber\\
 \eps_i^{(2)} \ttgb
 &\simeq&
   - \frac{(\delta m_d)_{i3}(\delta m_d)_{3i}}{(1+\eps_3 \tgb)\ \mb_{d_i} \mb_b} \quad (i=1,2)\,,
 \nonumber\\
 \eps_3^{(2)}\ttgb
 &\simeq&
   - 2\sum_{i=1,2}\frac{(\delta m_d)_{i3}(\delta m_d)_{3i}}{(1+\eps_{3}\tgb)\mb_b^2}\,.
\end{eqnarray}
In the expression of $\eps_i^{(2)}\ttgb$, we have systematically neglected subdominant terms of
order ${\cal O}(|(\delta m_d)_{i3}|^2/m_b^2)$ because always negligible. Moreover, we have also
neglected the second-order terms in the expansion of the unitary matrices $V_{d_L}$ and $V_{d_R}$;
this approximation is valid as long as the MI parameters are significantly lesser than one.

Notice that Eq.~(\ref{Eq:mass_matching}) determines the phase factor $\theta_{d_i}$ and relates
{\it bare} and {\it physical} quark masses in the following way,
\begin{eqnarray}
{\mb_d}_i = \frac{{m_d}_i}{|1 + \eps_{i}\tgb + \eps^{(2)}_i \ttgb|}\,, \quad
 \theta_{d_i} = {\rm arg}(1 + \eps_{i}\tgb + \eps^{(2)}_i \ttgb)\,.
\end{eqnarray}
Using the parameterization of Eq.~(\ref{Eq:mass_correction}), we derive
\begin{eqnarray} \label{Eq:Unitary_components}
 (\delta V_{d_L})_{3i}
 \!\!\! &\simeq& \!\!\!
  - \frac{\epsL_{3i}\tgb}{3(1+\eps_3\tgb)}(\delta_{LL}^d)_{3i}
  + \frac{\epsY\tgb}{1+\eps_3\tgb}\Vb_{3i}
  - \frac{{\mb_d}_i}{\mb_b}
    \frac{{\overline{z}_R}_i \epsR_{3i}^*\tgb}{3(1+\eps_3^*\tgb)}(\delta_{RR}^d)_{3i}\,,
 \nonumber\\
 (\delta V_{d_R})_{i3}
 \!\!\! &\simeq& \!\!\!
  \frac{\epsR_{3i}\tgb}{3(1+\eps_3\tgb)}(\delta_{RR}^d)_{i3}
  - \frac{{\mb_d}_i}{\mb_b}
    \frac{{z_Y}_i \epsY^*\tgb}{1+\eps_3^*\tgb}\Vb_{3i}^*
  + \frac{{\mb_d}_i}{\mb_b}
    \frac{{\overline{z}_L}_i \epsL_{3i}^*\tgb}{3(1+\eps_3^*\tgb)}(\delta_{LL}^d)_{i3}\,,
\end{eqnarray}
and
\begin{eqnarray} \label{Eq:epsilon}
 {\eps}_i
 \!\!\!&=&\!\!\! 
   \epsg_i 
   + 
   \frac{\epsLR_{ii}}{6}
   \frac{\overline{m}_b}{\overline{m}_{d_i}}
   (\delta^d_{RR}\Delta\delta^d_{LL})_{ii}\,,
   -
   \frac{\epsLY_i}{3}(\delta^d_{LL} \Delta \Vb)_{ii}\,,
 \nonumber\\
 {\eps}^{(2)}_i
 \!\!\!&=&\!\!\! 
   -
   \frac{\epsL_{3i}\epsR_{3i}}{9(1+\eps_3\tgb)}
   \frac{\overline{m}_b}{\overline{m}_{d_i}}
   (\delta^d_{RR}\Delta\delta^d_{LL})_{ii}
   +
   \frac{\epsY\epsR_{3i}}{3(1+\eps_3\tgb)}
   \frac{\overline{m}_b}{\overline{m}_{d_i}}
   (\delta^d_{RR}\Delta \Vb)_{ii}
 \nonumber\\
 && \!\!\!
   +
    \frac{\epsY (\epsL_{i3}-\epsL_{3i}+\epsLY_i)}{3(1+\eps_3\tgb)}
   (\delta^d_{LL}\Delta \Vb)_{ii}\,,
\end{eqnarray}
for $i=1,2$ and
\begin{eqnarray}
 \eps_3
 \!\!\!&=&\!\!\!
  \epsg_3+\epsY - \sum_{i=1,2} \frac{\epsLY_i}{3}(\Vb^\dagger\Delta \delta^d_{LL})_{ii}\,,
 \nonumber\\
 {\eps}^{(2)}_3
 \!\!\!&=&\!\!\!
   2\sum_{i=1,2}\bigg(-\frac{\epsL_{3i}\epsR_{3i}}{9(1+\eps_3\tgb)}
   (\delta^d_{RR}\Delta\delta^d_{LL})_{ii}
   +
   \frac{\epsY\epsR_{3i}}{3(1+\eps_3\tgb)}
   (\delta^d_{RR}\Delta \Vb)_{ii}\bigg)\,.
\end{eqnarray}
The parameters ${z_Y}, {\overline{z}_R}$ and ${\overline{z}_L}$ appearing in the above expressions 
are defined in Appendix~\ref{App:loop_factor}.
Although terms proportional to these parameters are suppressed by the light quark masses  (and thus negligible in many cases), they are still relevant for the calculation of the (C)EDMs when these arise
from $J^{(q)}_{LL}$.

The components of the {\it bare} CKM matrix can be expressed as a combination of {\it physical} CKM
matrix elements and mass insertion parameters. In Eq.~(\ref{Eq:CKM_matching}), the contribution
from $U_{u_L}$ is negligible due to the suppression of $\tgb^{-1}$ as will be seen later. Thus,
the components of the {\it bare} CKM matrix are given as
\begin{eqnarray} \label{Eq:CKMcomponents_matching2}
 \Vb_{3i}
 \!\!\! &\simeq& \!\!\!
  \frac{1+\eps_3\tgb}{1+\epsb_3\tgb}
  V_{3i}  -
  \frac{\epsL_{3i}\tgb}{3(1+\epsb_3\tgb)}
  (\delta^d_{LL})_{3i}
  \nonumber\\
  &&
  -
  \frac{\mb_{d_i}}{\mb_b}
  \frac{\overline{z}_{Ri}(1+\eps_3\tgb)\epsR_{3i}^*\tgb}
  {3(1+\epsb_3\tgb)(1+\eps^*_3\tgb)} (\delta^d_{RR})_{3i}\,,
  \nonumber\\
 \Vb_{i3}
 \!\!\! &\simeq& \!\!\!
  \frac{1+\eps_3^*\tgb}{1+\epsb_3^*\tgb}
  (1-p_i)V_{i3}
  +
  \frac{\epsL^*_{3i}\tgb}{3(1+\epsb_3^*\tgb)}
  (\delta^d_{LL})_{i3}
  \nonumber\\
  &&
  +
  \frac{\mb_{d_i}}{\mb_b}
  \frac{\overline{z}^*_{Ri}(1+\eps^*_3\tgb)\epsR_{3i}\tgb}
  {3(1+\epsb^*_3\tgb)(1+\eps_3\tgb)} (\delta^d_{RR})_{i3}
  -V_{ij}(\delta V_{d_L}^*)_{3j}\,,
\end{eqnarray}
where $\epsb_3=\eps_3-\epsY$ and
\begin{eqnarray}
 p_i 
 \!\!\! &=& \!\!\!
  \frac{\epsY^*\tgb}{1+\eps_3^*\tgb}
  \left( 1+\frac{V_{3i}^*}{V_{i3}} \right)\,,
\end{eqnarray}
where $i \neq j$ with $i,j= 1,2$. Although $p_1 \simeq {\cal O}(\eps\tgb)$ is comparable
to other terms in Eq.~(\ref{Eq:CKMcomponents_matching2}), we set hereafter $p_i=0$, for
simplicity. In terms of the {\it physical} CKM matrix, Eqs.~(\ref{Eq:Unitary_components}) and
(\ref{Eq:epsilon}) can be rewritten as
\begin{eqnarray}
 (\delta V_{d_L})_{3i}
 \!\!\! &\simeq& \!\!\!
  - \frac{\epsL_{3i}\tgb}{3(1+\epsb_3\tgb)}(\delta_{LL}^d)_{3i}
  + \frac{\epsY\tgb}{1+\epsb_3\tgb}V_{3i}
  - \frac{{\mb_d}_i}{\mb_b}
    \frac{{z_R}_i \epsR_{3i}^*\tgb}{3(1+\eps_3^*\tgb)}(\delta_{RR}^d)_{3i},
 \nonumber\\
 (\delta V_{d_R})_{i3}
 \!\!\! &\simeq& \!\!\!
  \frac{\epsR_{3i}\tgb}{3(1+\eps_3\tgb)}(\delta_{RR}^d)_{i3}
  - \frac{{\mb_d}_i}{\mb_b}
    \frac{{z_Y}_i \epsY^*\tgb}{1+\epsb_3^*\tgb}V_{3i}^*
  + \frac{{\mb_d}_i}{\mb_b}
    \frac{{z_L}_i \epsL_{3i}^*\tgb}{3(1+\epsb_3^*\tgb)}(\delta_{LL}^d)_{i3},
\end{eqnarray}
where ${z_L}$ and $z_R$ are defined in Appendix~\ref{App:loop_factor} and
\begin{eqnarray}
 {\eps}_i
 \!\!\!&=&\!\!\! 
   \epsg_i 
   + 
   \frac{\epsLR_{ii}}{6}
   \frac{\overline{m}_b}{\overline{m}_{d_i}}
   (\delta^d_{RR}\Delta\delta^d_{LL})_{ii}
   - 
   \frac{\epsLY_i}{3}
   (\delta^d_{LL}\Delta V)_{ii}
   ,
 \nonumber\\
 {\eps}^{(2)}_i
 \!\!\!&=&\!\!\! 
   -
   \frac{\epsL_{3i}\epsR_{3i}}{9(1+\epsb_3\tgb)}
   \frac{\overline{m}_b}{\overline{m}_{d_i}}
   (\delta^d_{RR}\Delta\delta^d_{LL})_{ii}
   +
   \frac{\epsY\epsR_{3i}}{3(1+\epsb_3\tgb)}
   \frac{\overline{m}_b}{\overline{m}_{d_i}}
   (\delta^d_{RR}\Delta V)_{ii}\,,
\end{eqnarray}
for $i=1,2$ and
\begin{eqnarray}
 \eps_3
 \!\!\!&=&\!\!\!
  \epsg_3+\epsY - \sum_{i=1,2} \frac{\epsLY_i(1+\eps^*_3\tgb)}{3(1+\epsb^*_3\tgb)}
  (V^\dagger\Delta \delta^d_{LL})_{ii}\,,
 \nonumber\\
 {\eps}^{(2)}_3
 \!\!\!&=&\!\!\!
   2\sum_{i=1,2}
   \left(
   -\frac{\epsL_{3i}\epsR_{3i}}{9(1+\epsb_3\tgb)}
   (\delta^d_{RR}\Delta\delta^d_{LL})_{ii}
   +
   \frac{\epsY\epsR_{3i}}{3(1+\epsb_3\tgb)}
   (\delta^d_{RR}\Delta V)_{ii}
   \right).
\end{eqnarray}
In the above expression for ${\eps}^{(2)}_i$, we have omitted terms proportional
to $\epsY\epsL_{i3}$ and $\epsY^*\epsL^*_{3i}$ since they vanish for degenerate 
squarks and vanishing flavor-conserving CP phases.


Turning to the up-type quarks, a similar calculation can be implemented and the
related expressions are obtained by the corresponding expressions for the down-type
quarks by exchanging $\Vb$ ($\tgb$) with $\Vb^\dagger$ ($\cotb$).
We define unitary matrices as
\begin{eqnarray}
 U_{u_R} = e^{-i \hat{\theta}_u} V_{u_R}, \quad
 U_{u_L} = V_{u_L} ,
\end{eqnarray}
and we match the {\it bare} Yukawa couplings with {\it physical} masses. In contrast to the 
case of down-type quarks, the corrections for up-type quark masses are suppressed by $\tan\beta$,
\begin{eqnarray} \label{Eq:mu_relation}
{\mb_u}_i = \frac{{m_u}_i}{|1 + \epsp_{i}\tgb^{-1}|}, \quad
 \theta_{u_i} = {\rm arg}(1 + \epsp_i\cotb),
\end{eqnarray}
where
\begin{eqnarray}
 \epsp_i
 \!\!\!&=&\!\!\! 
   \epsgp_i 
   + 
   \frac{\epsLRp_{ii}}{6}
   \frac{\overline{m}_t}{\overline{m}_{u_i}}
   (\delta^u_{RR}\Delta\delta^u_{LL})_{ii}
   - 
   \frac{\epsLYp_{i}}{3}(\delta^d_{LL}\Delta\Vb^\dagger)_{ii} \quad (i=1,2)\,,
 \nonumber\\
 \epsp_3
 \!\!\!&=&\!\!\! 
   \epsgp_3  + \epsYp
   -
   \sum_{i=1,2}
   \frac{\epsLYp_i}{3}(\Vb\Delta \delta^d_{LL})_{ii}\,,
\end{eqnarray}
and $\epsbp_3=\epsp_3-\epsYp$. In the above expression, while terms proportional to $\tgb^{-2}$
can be safely neglected, those proportional to $\tgb^{-1}$ should be retained. In fact, the $\epsp_{i}$
parameter in Eq.~(\ref{Eq:mu_relation}) is related to the vertex corrections for the up quark couplings
with Higgs, leading to $\tan\beta$-enhanced corrections, as we will discuss in the next subsection.
Concerning the unitary matrices, by means of the parameterization of Eq.~(\ref{Eq:mass_correction}),
we find
\begin{eqnarray}
 (V_{u_L})_{3i}
 \!\!\! &\simeq& \!\!\!
  - \frac{1}{3}\epsLp_{3i}\cotb(\delta_{LL}^u)_{3i}
  + 
  \epsYp\cotb 
  \left[
   \frac{1+\eps_3\tgb}{1+\epsb_3\tgb}
   V^*_{i3}
   +
   \frac{\epsL_{3i}\tgb}{3(1+\epsb_3\tgb)}
   (\delta^d_{LL})_{3i}
  \right]
  \nonumber\\ 
  && 
  - \frac{{\mb_u}_i}{\mb_t}
    \frac{{z'_R}_i}{3} \epsR_{3i}^*\cotb(\delta_{RR}^u)_{3i}\,,
 \nonumber\\
 (V_{u_R})_{i3}
  \!\!\! &\simeq & \!\!\!
  \frac{1}{3}\epsRp_{3i}\cotb(\delta_{RR}^u)_{i3}
  - \frac{{\mb_u}_i}{\mb_t}
    \frac{{z'_Y}_i}{3} \cotb \epsYp^*
    \left[
     \frac{1+\eps_3^*\tgb}{1+\epsb_3^*\tgb}
     V_{i3}
     +
     \frac{\epsL^*_{3i}\tgb}{3(1+\epsb^*_3\tgb)}
     (\delta^d_{LL})_{i3}
    \right]
  \nonumber\\
  && \!\!\!
  + \frac{{\mb_u}_i}{\mb_t}
    \frac{{z'_L}_i}{3} \epsLp_{3i}^* \cotb (\delta_{LL}^u)_{i3}\,,
\end{eqnarray}
where terms proportional to $\tgb^{-2}$ are safely neglected and ${z'_Y}, {z'_L}$ and ${z'_R}$
are defined in Appendix~\ref{App:loop_factor}.


The same formalism used for the quarks applies to the leptons, too. In particular,
the unitary matrices that diagonalize the lepton mass matrix can be defined as
\begin{eqnarray}
 U_{e_R} = e^{-i \hat{\theta}_e} V_{e_R}, \quad
 U_{e_L} = V_{e_L}\,,
\end{eqnarray}
and the phase factors, which make the lepton masses real, and the corrections to the 
lepton masses can be parameterized as
\begin{eqnarray}
{\mb_e}_i = \frac{{m_e}_i}{|1 + \eps^e_{i}\tgb + {\eps}^{e(2)}_i \ttgb|}, \quad
 \theta_{e_i} = {\rm arg}(1 + \eps^e_{i}\tgb + \eps^{e(2)}_i \ttgb)\,.
\end{eqnarray}
With the parameters in Eq.~(\ref{Eq:mass_correction}), we find
\begin{eqnarray} 
 (\delta V_{e_L})_{3i}
 \!\!\! &\simeq& \!\!\!
  - \frac{{\eps^e_L}_{3i}\tgb}{3(1+\eps^e_3\tgb)}(\delta_{LL}^e)_{3i}\,,
 \nonumber\\
 (\delta V_{e_R})_{i3}
 \!\!\! &\simeq& \!\!\!
  \frac{{\eps^e_R}_{3i}\tgb}{3(1+\eps^e_3\tgb)}(\delta_{RR}^e)_{i3}\,,
\end{eqnarray}
and
\begin{eqnarray}
 {\eps}^{e}_i
 \!\!\!&=&\!\!\! 
   {\eps^e_g}_i 
   + 
   \frac{{\eps^e_{LR}}_{ii}}{6}
   \frac{\overline{m}_\tau}{\overline{m}_{e_i}}
   (\delta^e_{RR}\Delta\delta^e_{LL})_{ii} \quad (i=1,2)\,,
 \nonumber\\
 {\eps}^{e(2)}_i
 \!\!\!&=&\!\!\! 
   -
   \frac{{\eps^e_L}_{3i}{\eps^e_R}_{3i}}{9(1+\eps^e_3\tgb)}
   \frac{\overline{m}_\tau}{\overline{m}_{e_i}}
   (\delta^e_{RR}\Delta\delta^e_{LL})_{ii} \quad(i=1,2)\,,
 \nonumber\\
 {\eps}^{e(2)}_3
 \!\!\!&=&\!\!\! 
   -2 \sum_{i=1,2}
   \frac{{\eps^e_L}_{3i}{\eps^e_R}_{3i}}{9(1+\eps^e_3\tgb)}
   (\delta^e_{RR}\Delta\delta^e_{LL})_{ii} \,,
\end{eqnarray}
and ${\eps}^{e}_3={\eps^e_g}_3$.

\subsection{Effective Higgs Interactions}
\label{enhc}

We are now ready to derive the effective interactions of matter fields at the BLO.
In this subsection, we consider the interactions of fermions with Higgs fields.
The effective couplings in the {\it physical} CKM basis are obtained by performing
the unitary transformations of Eq.~(\ref{Eq:unitary_transformation}) to the Higgs
couplings with quarks in the {\it bare} SCKM basis, which is described by Eq.~(\ref{Eq:lag_NH}).

For charged Higgs interactions, we obtain
\begin{eqnarray}
 {\cal L}_{\rm eff}
 \!\!\! &=& \!\!\!
  \frac{g_2 m_{u_i}}{\sqrt{2}m_W} \cotb \left( C_{d_L}^{H^\pm} \right)_{ij}
  \uRb_i \dL_j H^+
  +
  \frac{g_2 m_{d_j}}{\sqrt{2}m_W} \tgb \left( C_{d_R}^{H^\pm} \right)_{ij}
  \uLb_i \dR_j H^+
  \nonumber\\
 && \!\!\!
  +
  \frac{g_2 m_{e_j}}{\sqrt{2}m_W} \tgb \left( C_{e_R}^{H^\pm} \right)_{ij}
  \nLb_i \eR_j H^+
  +
  {\rm h.c.}\,,
\end{eqnarray}
where
\begin{eqnarray}
 ( C_\dL^{H^\pm} )_{ij}
 \!\!\! &=& \!\!\!
  \frac{1}{{m_u}_i}
  (U_\uR \mb_u U_\uL^\dagger V )_{ij}(1+\tgb^2)
  -
  V_{ij} \tgb^2,
\label{Eq:C_dL_H}
  \\
 ( C_\dR^{H^\pm} )_{ij}
 \!\!\! &=& \!\!\!
  \frac{1}{{m_d}_j}
  (V U_\dL \mb_d U_\dR^\dagger)_{ij}(1+\tgb^{-2})
  -
  V_{ij} \tgb^{-2}\,,
\label{Eq:C_dR_H}
  \\
 ( C_\eR^{H^\pm} )_{ij}
 \!\!\! &=& \!\!\!
  \frac{1}{{m_e}_j}
  ( U_\eL \mb_e U_{e_R}^\dagger)_{ij}(1+\tgb^{-2})
  -
  \delta_{ij} \tgb^{-2}\,.
\label{Eq:C_eR_H}
\end{eqnarray}
In the above expressions, the non-holomorphic Yukawa terms $\eps^{d,u,e}_{ij}$ are expressed
by the Yukawa couplings and the unitary matrices through Eq.~(\ref{Eq:diagonalize_mass}),
then terms proportional to $\ttgb$ or $\tgb^{-2}$ in Eqs.~(\ref{Eq:C_dL_H}-\ref{Eq:C_eR_H})
come from the one-loop vertex corrections for charged Higgs couplings while the remaining 
terms are generated by the mass term corrections.
Notice that, in the limit of $U_f=1~(f=d_L,d_R,u_L,u_R,e_L,e_R)$ and $\mb_f=m_f~(f=u,d,e)$,
the vertex corrections are vanishing, as it should be. This means that the vertex corrections
are encoded in the difference between {\it bare} and {\it physical} Yukawa couplings and in the
deviation of unitary matrices from the identity matrix. Especially, we need to keep
$\tan\beta$-suppressed terms in $U_{u_L}$, $U_{u_R}$ and $\mb_u$ for $C^{H^\pm}_{d_L}$.

Neutral Higgs interactions with quarks and leptons are obtained in a similar way by
\begin{eqnarray}
 {\cal L}_{\rm eff}
 \!\!\! &=& \!\!\!
  -\frac{g {m_f}_i}{2\sinb\cosb m_W} (C_{f_L}^{H_a})_{ij}
  {\overline{f}_R}_i {f_L}_j H_a
  ~+~
  {\rm h.c.}\,,
\end{eqnarray}
where
\begin{eqnarray}
 (C_\uL^{H_a})_{ij}
 \!\!\! &=& \!\!\!
  ~
  \frac{1}{{m_u}_i}
  (U_\uR \mb_u U_{u_L}^\dagger)_{ij} {a^a}^*
  +
  \delta_{ij} \sinb {b_1^a}^*\,,
  \\
 (C_\dL^{H_a})_{ij}
 \!\!\! &=& \!\!\!
  -\frac{1}{{m_d}_i}
  (U_\dR \mb_d U_{d_L}^\dagger)_{ij} a^a
  +
  \delta_{ij} \cosb {b_2^a}^*\,,
  \\
 (C_\eL^{H_a})_{ij}
 \!\!\! &=& \!\!\!
  -\frac{1}{{m_e}_i}
  (U_\eR \mb_e U_{e_L}^\dagger)_{ij} a^a
  +
  \delta_{ij} \cosb {b_2^a}^*\,,
\end{eqnarray}
with $a^a \equiv \cosb {b_2^a}^* - \sinb b_1^a$. The Higgs fields $H_1^0$ and $H_2^0$ are
expanded as linear combinations of mass eigenstates $H_a=h,H,A,$ and the coefficients $b_1^a$
and $b_2^a$ are defined as
\begin{eqnarray}
 H_1^0 = \frac{1}{\sqrt{2}} (v\cosb + b_1^a H_a),\quad 
 H_2^0 = \frac{1}{\sqrt{2}} (v\sinb + b_2^a H_a)\,.
\end{eqnarray}
If there is no CP violation in the Higgs sector, $b_1^a=\{-s_\alpha,~c_\alpha,~i s_\beta\}$,
$b_2^a=\{c_\alpha,~s_\alpha ,~i c_\beta \}$, $a^a=\{c_{\alpha-\beta},~s_{\alpha-\beta},~-i \}$
and then we can identify $h$ and $H$ as the CP-even Higgs bosons and $A$ as a the CP-even Higgs
boson. In the presence of CP violation, mass eigenstates can not be assigned with CP charges,
and all of the above coefficients become complex numbers as
\begin{eqnarray}
 b_1^a = {\cal O}_{1a} + i\sinb {\cal O}_{3a}, \quad
 b_2^a = {\cal O}_{2a} + i\cosb {\cal O}_{3a},
\end{eqnarray}
where the derivation of the matrix ${\cal O}_{ia}$ can be found
in Refs.~\cite{Carena:2000yi,Pilaftsis:1999qt}.

To calculate the physical observables keeping their dependence on the flavor- and CP-violating
parameters in the MSSM, it is useful to derive analytical approximate formulae for these vertices.
Since the unitary matrices and tree-level Yukawa couplings are already derived by means of MI
parameters and loop factors, we can easily derive the expressions for the desired vertices. Here,
we present only the vertices that are relevant for the calculations of the {\it flavored} EDMs.

In the calculation of quark (C)EDMs, flavor-violating quark couplings with the charged Higgs have a
very important role, as shown in Ref.~\cite{Hisano:2006mj}. The corrections to the Higgs couplings
with quarks, induced by the non-holomorphic Yukawa couplings, are given as
\begin{eqnarray}
 (C_{d_R}^{H^\pm})_{3i} 
 \!\!\! &\simeq& \!\!\! 
  \frac{{\mb_d}_i}{{m_d}_i} e^{i{\theta_d}_i} \Vb_{3i}
  +
  \frac{{\mb_b}}{{m_d}_i} e^{i{\theta_d}_i} (V_{d_R})^*_{i3}
 \nonumber\\
 \!\!\! &\simeq& \!\!\! 
  \frac{ e^{i{\theta_d}_i} }{ |1+\eps_i\tgb+\eps_i^{(2)}\ttgb| } 
  \left[
   \frac{1+\eps_3\tgb-{z_Y}_i^*\epsY\tgb}{1+\epsb_3\tgb}V_{3i}
   -
   \frac{(1-{z_L}_i^*) \epsL_{3i}\tgb}{3(1+\epsb_3\tgb)}(\delta_{LL}^d)_{3i}
  \right]
  \nonumber \\
  && \!\!\!
  +~
  \frac{\epsR_{3i}^*\tgb}{3(1+\eps_3^*\tgb)}
  \frac{e^{i\theta_{d_i}} }{| 1+\eps_3^*\tgb |} \frac{m_b}{m_{d_i}}
  (\delta_{RR}^d)_{3i}\,,
 \label{Eq:C_CH_dR}
 \\
 (C_\dL^{H^\pm})_{3i} 
 \!\!\! &\simeq& \!\!\! 
  (1-\epsp_3\tgb) V_{3i} 
  - \ttgb \sum_{j\neq 3} (V_\uL)_{3j} V_{ji}
  \nonumber\\
 \!\!\! &\simeq& \!\!\! 
  \left[
   1 - \epsbp_3\tgb + \frac{\epsY\epsYp\ttgb}{1+\epsb_3\tgb}
  \right] V_{3i} 
  + 
  \left[
   \frac{1}{3}\epsLp_{3i}\tgb 
   -
   \frac{\epsYp\epsL_{3j}\ttgb}{3(1+\epsb_3\tgb)}
  \right]
  (\delta^d_{LL})_{3i}\,,
 \label{Eq:C_CH_dL}
\end{eqnarray}
for $i=1,2$. Here terms proportional to $z'_X~(X=R,L,Y)$ are neglected since they are
suppressed by $m_{u_i}/m_t$, thus, irrelevant.
The last term in Eq.~(\ref{Eq:C_CH_dR}) is significantly enhanced by the bottom Yukawa
coupling (through the mass corrections for down-type quarks); the enhancement factor
$(m_b/m_{d_i})$ and the appearance of an extra $\tgb$ factor partially compensate
the suppression arising from the loop factor and the MI parameters. As a result, these
flavor-violating terms can become comparable in size to the tree-level couplings and
with a different phase.

For light up-type quarks, the couplings with the charged Higgs are given by
\begin{eqnarray}
 (C_\dR^{H^\pm})_{i3} 
 \!\!\! &\simeq& \!\!\! 
  \frac{\mb_b}{m_b} e^{i \theta_b } \Vb_{i3}
 \nonumber\\
 \!\!\! &\simeq& \!\!\! 
  \frac{ e^{i \theta_b } }{ |1+\eps_3\tgb| }
  \left[
   \frac{1+\eps_3^*\tgb}{1+\epsb_3^*\tgb}
   V_{i3}
   +
   \frac{\epsL^*_{3i}\tgb}{3(1+\epsb_3^*\tgb)}
   (\delta^d_{LL})_{i3}
  \right]\,,
 \label{Eq:C_CH_uL}
 \\
 (C_\dL^{H^\pm})_{i3} 
 \!\!\! &\simeq& \!\!\! 
  (1-\epsp_i\tgb) V_{i3} 
  + \ttgb \sum_{j\neq i} (V_{u_L}^\dagger)_{ij} V_{j3}
  + \frac{\mb_t}{{m_u}_i}(U_{u_R})_{i3} \ttgb
  \nonumber\\
 \!\!\! &\simeq& \!\!\! 
  \left[
   1 - \epsp_i\tgb + \epsYp^* \tgb
     + \frac{\epsY^*\epsYp^* \ttgb}{1+\epsb_3^*\tgb}
  \right] V_{i3}
  \nonumber\\
  && \!\!\!
  -
  \left[
   \frac{1}{3}\epsLp^*_{i3}\tgb 
   -
   \frac{\epsYp^* \epsL^*_{3j}\ttgb}{3(1+\epsb^*_3\tgb)}
  \right]
  (\delta^d_{LL})_{i3}
  +
  \frac{1}{3}\epsRp_{3i} \tgb 
  \frac{m_t}{m_{u_i}} (\delta^u_{RR})_{i3}\,,
 \label{Eq:C_CH_uR}
\end{eqnarray}
for $i=1,2$. In the above derivation, we have assumed $(\delta_{LL}^u V)_{3i}\simeq(\delta_{LL}^d V)_{3i}
\simeq (\delta_{LL}^d)_{3i}$ and terms proportional to $z'_X~(X=R,L,Y)$ have been neglected.
The $\tan\beta$-enhanced corrections appear from the vertex corrections in contrast to the case of
down-quarks, where the corrections come from the mass corrections, and especially the last term of Eq.~(\ref{Eq:C_CH_uR}) is enhanced by the large top quark mass.

For the neutral Higgs interactions, flavor-violating interactions are expressed in a similar way.
Moreover, flavor-conserving CP-violating interactions appear at the BLO and they are important
for the calculation of Higgs mediated CP-violating four-Fermi interactions. In the limit of large
$\tan\beta$ and $m_A$, these couplings are expressed as
\begin{eqnarray} 
\label{Eq:NHiggs_coupling_MI}
 (C_\dL^{H})_{ii} 
  &\simeq&
  \frac{e^{-i \theta_{d_i}}}{|1+\eps_i\tgb+\eps_i^{(2)}\ttgb|}
  \left(
   1 + \frac{\mb_b}{\mb_{d_i}} (V_{d_R}\Delta V_{d_L}^\dagger)_{ii}
  \right)
  \nonumber\\
  &\simeq& 
  \frac{e^{-i \theta_{d_i}}}{|1+\eps_i\tgb+\eps_i^{(2)}\ttgb|}
  \left(
   1 - \frac{\eps_i^{(2)}\ttgb}{1+\eps_3\tgb}
  \right)\qquad (i=1,2)\,,
  \\
 (C_\dL^{A})_{ii} &\simeq& i (C_\dL^{H})_{ii},
 \\
 (C_\uL^{H})_{ii} 
  &\simeq&
   -\cosb^2~(1 - \epsp_i \tgb),
 \\
 (C_\uL^{A})_{ii} &\simeq& i\cosb^2~(1 + \epsp_i \tgb)\,,
\end{eqnarray}
and $(C_\dL^{h})_{ii} \simeq (C_\uL^{h})_{ii}\simeq \sinb\cosb$ since
the lightest Higgs boson behaves as the SM Higgs boson in this
limit. As we can see from Eq.~(\ref{Eq:NHiggs_coupling_MI}), CP-violating
interactions between down-type quarks and the Higgses $H$ and $A$ can
arise from {\it i)} diagonal corrections to the quark mass matrix
($e^{-i\theta_i}$), an effect which has been already discussed in the
literature, or {\it ii)} from the combination ($(V_{d_R}\Delta
V_{d_L}^\dagger)_{ii}$), {\it i.e.} from pure flavor-violating effects to
the quark mass matrix.

Although these last effects are suppressed by small flavor-violating parameters, they are highly
enhanced by the heaviest quark Yukawas thus, they can provide potentially large effects.

\subsection{Effective Fermion-Sfermion Interactions}

The effects of non-holomorphic corrections appear not only in the
interactions with Higgs bosons but also in those with the Higgsino and
gauginos.  Here we discussed the effective vertices including the BLO
$\tan\beta$-enhanced corrections.

First of all, we need to specify the basis in which we work. In our convention,
while quarks are expressed in the mass eigenstates, we adopt the {\it bare} SCKM basis
for squarks and flavor eigenstates for gauginos and Higgsinos to keep the dependence
of flavor and CP-violating parameters. Thus, the left-right mixing terms proportional
to the $\mu$ term in the squark mass matrices become flavor blind (see Eqs.~(\ref{Eq:msu_RL})
and (\ref{Eq:msd_RL})) while BLO corrections induce flavor-violating couplings even
for neutral SUSY particles such as the gluino and neutralinos. If we use {\it physical}
SCKM basis, these couplings with neutral particles become flavor blind but there appear
non-trivial flavor violations in the left-right squark mixing terms. Obviously, physical
observables do not depend on the choice of the squark basis. In Appendix~\ref{App:eff_vertex},
we also provide the relevant vertices for all particles in mass eigenstates.

First, gluino interactions are modified by the BLO corrections as
\begin{eqnarray}
 -{\cal L}^{\tilde{g}}_{\rm int}
 \!\!\! &=& \!\!\!
  ( C^{\tilde g}_{q_L} )_{ij}~
  {\tilde q}^*_{L_i} \overline{{\tilde g}}_R^a t_a 
  {q_L}_j
  +
  ( C^{\tilde g}_{q_R} )_{ij}~
  {\tilde q}^*_{R_i} \overline{{\tilde g}}_L^a t_a 
  {q_R}_j
  ~+~ {\rm h.c.},
\end{eqnarray}
where
\begin{eqnarray}
 &&
 ( C^{\tilde g}_{q_L} )_{ij}
 =
  \sqrt{2} g_s(U^\dagger_{q_L})_{ij},
  \nonumber\\
 &&
 ( C^{\tilde g}_{q_R} )_{ij}
 =
  -\sqrt{2} g_s (U^\dagger_{q_R})_{ij}
\end{eqnarray}
with $q =u,d$. The flavor-mixing terms are generated by the mismatch between the {\it bare} and
{\it physical} SCKM bases. It is worth mentioning that even the flavor-conserving couplings can have
CP-violating phases at the BLO due to the phases $\theta_q$ in the unitary matrix $U_{q_R}$.

The charginos are the mass eigenstates of charged Higgsino and wino, which are mixed each
other after the electroweak symmetry breaking.
Including the BLO corrections, the vertices of these charged Higgsino and wino are given by
\begin{eqnarray}
 -{\cal L}^{\tilde{\chi}^\pm}_{\rm int}
 \!\!\! &=& \!\!\!
   (C_\dL^{{\tilde W}^\pm} )_{ij} \tilde{u}_{Li}^* \tWRmb \dL_j
  +(C_\dL^{{\tilde H}^\pm} )_{ij} \tilde{u}_{Ri}^* \tHuRmb \dL_j
  +(C_\dR^{{\tilde H}^\pm} )_{ij} \tilde{u}_{Li}^* \tHdLmb \dR_j
 \nonumber\\
  && \!\!\!
  +(C_\uL^{{\tilde W}^\pm} )_{ij} \tilde{d}_{Li}^* \tWRpb \uL_j
  +(C_\uL^{{\tilde H}^\pm} )_{ij} \tilde{d}_{Ri}^* \tHdRpb \uL_j
  +(C_\uR^{{\tilde H}^\pm} )_{ij} \tilde{d}_{Li}^* \tHuLpb \uR_j
  \nonumber\\
  && \!\!\!
  +(C_\eL^{{\tilde W}^\pm} )_{ij} \tilde{\nu}_{Li}^* \tWRmb \eL_j
  +(C_\eR^{{\tilde H}^\pm} )_{ij} \tilde{\nu}_{Li}^* \tHdLmb \eR_j
 \nonumber\\
  && \!\!\!
  +(C_\nL^{{\tilde W}^\pm} )_{ij} \tilde{e}_{Li}^* \tWRpb \nL_j
  +(C_\nL^{{\tilde H}^\pm} )_{ij} \tilde{e}_{Ri}^* \tHdRpb \nL_j
  \nonumber\\
  && \!\!\!
  +~{\rm h.c.}\,,
\end{eqnarray}
where
\begin{eqnarray}
 \label{Eq:chargino_coupling_first}
 (C_\dL^{\tilde{W}^\pm})_{ij}
 \!\!\! &=& \!\!\!
  g_2 (U_{u_L}^\dagger V)_{ij}\,,
  \nonumber\\
 (C_{d_L}^{\tilde{H}^\pm})_{ij}
 \!\!\! &=& \!\!\!
  -(\hat{\yb}_u U_{u_L}^\dagger V)_{ij}\,,
  \nonumber\\
 (C_{d_R}^{\tilde{H}^\pm})_{ij}
 \!\!\! &=& \!\!\!
  (U_{u_L}^\dagger V U_{d_L} \hat{\yb}_d U_{d_R}^\dagger)_{ij}\,,
  \nonumber\\
 (C_{u_L}^{\tilde{W}^\pm})_{ij}
 \!\!\! &=& \!\!\!
  g_2 (U_{d_L}^\dagger V^\dagger)_{ij}\,,
  \nonumber\\
 (C_{u_L}^{\tilde{H}^\pm})_{ij}
 \!\!\! &=& \!\!\!
  (\hat{\yb}_d U_{d_L}^\dagger V^\dagger)_{ij}\,,
  \nonumber\\
 (C_{u_R}^{\tilde{H}^\pm})_{ij}
 \!\!\! &=& \!\!\!
  -(U_{d_L}^\dagger V^\dagger U_{u_L} \hat{\yb}_u U_{u_R}^\dagger)_{ij}\,,
  \nonumber\\
 (C_\eL^{\tilde{W}^\pm})_{ij}
 \!\!\! &=& \!\!\!
  g_2 (U_{e_L}^\dagger)_{ij}\,,
  \nonumber\\
 (C_{e_R}^{\tilde{H}^\pm})_{ij}
 \!\!\! &=& \!\!\!
  (\hat{\yb}_e U_{e_R}^\dagger)_{ij}\,,
  \nonumber\\
 (C_{\nu_L}^{\tilde{W}^\pm})_{ij}
 \!\!\! &=& \!\!\!
  g_2 \delta_{ij},
  \nonumber\\
 (C_{\nu_L}^{\tilde{H}^\pm})_{ij}
 \!\!\! &=& \!\!\!
  (\hat{\yb}_e)_{ij}\,.
 \label{Eq:chargino_coupling_last}
\end{eqnarray}
The couplings of neutralinos, which are composed by bino, wino and neutral Higgsinos, are given by
\begin{eqnarray}
 -{\cal L}^{\tilde{\chi}^0}_{\rm int}
 =
  \!\!\! && \!\!\!
   (C_\dL^{\tilde{W}^0} )_{ij} \tilde{d}_{Li}^* \tWRzb \dL_j
  +(C_\dL^{\tilde{B}  } )_{ij} \tilde{d}_{Li}^* \tBRb \dL_j
  +(C_\dR^{\tilde{B}  } )_{ij} \tilde{d}_{Ri}^* \tBLb \dR_j
 \nonumber\\
  && \!\!\!
  +(C_\dL^{\tilde{H}^0} )_{ij} \tilde{d}_{Ri}^* \tHuRzb \dL_j
  +(C_\dR^{\tilde{H}^0} )_{ij} \tilde{d}_{Li}^* \tHdLzb \dR_j
 \nonumber\\
  && \!\!\!
  +(C_\uL^{\tilde{W}^0} )_{ij} \tilde{u}_{Li}^* \tWRzb \uL_j
  +(C_\uL^{\tilde{B}  } )_{ij} \tilde{u}_{Li}^* \tBRb \uL_j
  +(C_\uR^{\tilde{B}  } )_{ij} \tilde{u}_{Ri}^* \tBLb \uR_j
 \nonumber\\
  && \!\!\!
  +(C_\uL^{\tilde{H}^0} )_{ij} \tilde{u}_{Ri}^* \tHdRzb \uL_j
  +(C_\uR^{\tilde{H}^0} )_{ij} \tilde{u}_{Li}^* \tHuLzb \uR_j
 \nonumber\\
  && \!\!\!
  +(C_\eL^{\tilde{W}^0} )_{ij} \tilde{e}_{Li}^* \tWRzb \eL_j
  +(C_\eL^{\tilde{B}  } )_{ij} \tilde{e}_{Li}^* \tBRb \eL_j
  +(C_\eR^{\tilde{B}  } )_{ij} \tilde{e}_{Ri}^* \tBLb \eR_j
 \nonumber\\
  && \!\!\!
  +(C_\eL^{\tilde{H}^0} )_{ij} \tilde{e}_{Ri}^* \tHuRzb \eL_j
  +(C_\eR^{\tilde{H}^0} )_{ij} \tilde{e}_{Li}^* \tHdLzb \eR_j
 \nonumber\\
  && \!\!\!
  +(C_\nL^{\tilde{W}^0} )_{ij} \tilde{\nu}_{Li}^* \tWRzb \nL_j
  +(C_\nL^{\tilde{B}  } )_{ij} \tilde{\nu}_{Li}^* \tBRb \nL_j
  ~+~ {\rm h.c.}\,,
\end{eqnarray}
where
\begin{eqnarray}
 \label{Eq:neutralino_coupling_first}
 (C_\dL^{\tilde{W}^0})_{ij}
 \!\!\! &=& \!\!\!
  -\frac{\sqrt{2} g_2}{2} (U_{d_L}^\dagger)_{ij}\,,
  \nonumber\\
 (C_\dL^{\tilde{B}  })_{ij}
 \!\!\! &=& \!\!\!
  \sqrt{2} g_Y Y_{q_L} (U_{d_L}^\dagger)_{ij}\,,
  \nonumber\\
 (C_\dR^{\tilde{B}  })_{ij}
 \!\!\! &=& \!\!\!
  \sqrt{2} g_Y Y_{d_R} (U_{d_R}^\dagger)_{ij}\,,
  \nonumber\\
 (C_{d_L}^{\tilde{H}^0})_{ij}
 \!\!\! &=& \!\!\!
  -(\hat{\yb}_d U_{d_L}^\dagger)_{ij}\,,
  \nonumber\\
 (C_{d_R}^{\tilde{H}^0})_{ij}
 \!\!\! &=& \!\!\!
  -(\hat{\yb}_d U_{d_R}^\dagger)_{ij}\,,
  \nonumber\\
 (C_\uL^{\tilde{W}^0})_{ij}
 \!\!\! &=& \!\!\!
  \frac{\sqrt{2} g_2}{2} (U_{u_L}^\dagger)_{ij}\,,
  \nonumber\\
 (C_\uL^{\tilde{B}  })_{ij}
 \!\!\! &=& \!\!\!
  \sqrt{2} g_Y Y_{q_L} (U_{u_L}^\dagger)_{ij}\,,
  \nonumber\\
 (C_\uR^{\tilde{B}  })_{ij}
 \!\!\! &=& \!\!\!
  \sqrt{2} g_Y Y_{u_R} (U_{u_R}^\dagger)_{ij}\,,
  \nonumber\\
 (C_{u_L}^{\tilde{H}^0})_{ij}
 \!\!\! &=& \!\!\!
  (\hat{\yb}_u U_{u_L}^\dagger)_{ij}\,,
  \nonumber\\
 (C_{u_R}^{\tilde{H}^0})_{ij}
 \!\!\! &=& \!\!\!
  (\hat{\yb}_u U_{u_R}^\dagger)_{ij}
  \nonumber\\
 (C_\eL^{\tilde{W}^0})_{ij}
 \!\!\! &=& \!\!\!
  -\frac{\sqrt{2} g_2}{2} (U_{e_L}^\dagger)_{ij}\,,
  \nonumber\\
 (C_\eL^{\tilde{B}  })_{ij}
 \!\!\! &=& \!\!\!
  \sqrt{2} g_Y Y_{l_L} (U_{e_L}^\dagger)_{ij}\,,
  \nonumber\\
 (C_\eR^{\tilde{B}  })_{ij}
 \!\!\! &=& \!\!\!
  \sqrt{2} g_Y Y_{e_R} (U_{e_R}^\dagger)_{ij}\,,
  \nonumber\\
 (C_{e_L}^{\tilde{H}^0})_{ij}
 \!\!\! &=& \!\!\!
  -(\hat{\yb}_e U_{e_L}^\dagger)_{ij}\,,
  \nonumber\\
 (C_{e_R}^{\tilde{H}^0})_{ij}
 \!\!\! &=& \!\!\!
  -(\hat{\yb}_e U_{e_R}^\dagger)_{ij}\,,
  \nonumber\\
 (C_\nL^{\tilde{W}^0})_{ij}
 \!\!\! &=& \!\!\!
  \frac{\sqrt{2} g_2}{2} \delta_{ij}\,,
  \nonumber\\
 (C_\nL^{\tilde{B}  })_{ij}
 \!\!\! &=& \!\!\!
  \sqrt{2} g_Y Y_{l_L} \delta_{ij}
 \label{Eq:neutralino_coupling_last}\,,
\end{eqnarray}
where $Y_{q_L}=1/6,~Y_{d_R}=1/3,~Y_{u_R}=-2/3,~Y_{l_L}=-1/2$ and $Y_{e_R}=1$.

It is easy to see the parameter dependence of these effective couplings just considering
the approximate expressions for the unitary matrices. For example, flavor-violating gluino
interactions with down-type quarks can be written as
\begin{eqnarray}
 ( C^{\tilde g}_{d_L} )_{3i}
 \!\!\! &\simeq & \!\!\!
  \sqrt{2} g_s
  \left[
    \frac{\epsL_{3i}\tgb}{3(1+\epsb_3\tgb)}(\delta_{LL}^d)_{3i}
  - \frac{\epsY\tgb}{1+\epsb_3\tgb}V_{3i}
  + \frac{{\mb_d}_i}{\mb_b}
    \frac{{z_R}_i \epsR_{3i}^*\tgb}{3(1+\eps_3^*\tgb)}(\delta_{RR}^d)_{3i}
  \right]\,,
  \\
 ( C^{\tilde g}_{d_R} )_{3i}
 \!\!\! &\simeq & \!\!\!
    -\sqrt{2} g_s~e^{i\theta_{d_i}}
  \bigg[
    \frac{\epsR^{*}_{3i}\tgb}{3(1+\eps^{*}_{3}\tgb)}(\delta_{RR}^d)_{3i}
  \nonumber\\
  & &\qquad
  - \frac{{\mb_d}_i}{\mb_b}
    \frac{{z_Y}^*_i \epsY\tgb}{3(1+\epsb_3\tgb)}V_{3i}
  + \frac{{\mb_d}_i}{\mb_b}
    \frac{{z_L}^*_i \epsL_{3i}\tgb}{3(1+\epsb_3\tgb)}(\delta_{LL}^d)_{3i}
  \bigg]\,,
\end{eqnarray}
for $i=1,2,$ and the flavor conserving couplings of the right-handed down-type quarks have CP
phases as $( C^{\tilde g}_{d_R} )_{ii}=-\sqrt{2} g_s~e^{i \theta_{d_i}}$, in our convention.


To discuss the down-type quark (C)EDMs, it may be convenient to take the basis of
left-handed up squarks same as that of left-handed down squarks. In that case,
effective couplings $C_\dL^{\tilde{W}^\pm}$ and $C_{d_R}^{\tilde{H}^\pm}$ should
be replaced by $C_\dL^{\tilde{W}^0}$ and $C_{d_R}^{\tilde{H}^0}$, respectively,
and also the left-right mixing mass term of up squarks is given by
\begin{eqnarray}
 -{\cal L} = (m^2_{RL})_{ij}\, {\tilde u}_{Ri}^* {\tilde u}_{Lj} + {\rm h.c.}\,,
\end{eqnarray}
where
\begin{eqnarray}
 (m^2_{RL})_{ij}
 \simeq
  - {\mb_u}_i (A_u + \mu^*\cot\beta) (U_{u_L} V U_{d_L}^\dagger)_{ij}\,.
\end{eqnarray}
Here, the $A$ terms are assumed to be proportional to the Yukawa couplings as
$(a_u)_{ij}\simeq A_u{\yb_u}_i$ in the {\it bare} SCKM basis.


\section{Loop Factors}  
\label{App:loop_factor}
In Appendix~\ref{Sec:effective_vertex}, non-holomorphic Yukawa corrections are
expressed by the combination of mass insertion parameters and loop factors.
Here, we present the expression for the loop factors.

For down quarks, the loop factors are given by
\begin{eqnarray}
 \epsg_i 
 \!\!\!&=&\!\!\! 
  \frac{\alpha_s}{3\pi} \mu^* M_3^* ~
  I_3(|M_3|^2, m^2_{\tilde{d}_{Li}},m^2_{\tilde{d}_{Ri}})
  -
  \frac{3\alpha_2}{16\pi} \mu^* M_2^*~
  I_3(|M_2|^2, |\mu|^2, m^2_{\tilde{d}_{Li}})
  \nonumber\\
  && \!\!\!
  -
  \frac{\alpha_Y}{144\pi} \mu^* M_1^*
  \Big[
   2
   I_3(|M_1|^2, m^2_{\tilde{d}_{Li}}, m^2_{\tilde{d}_{Ri}})
   +
   3
   I_3(|M_1|^2, |\mu|^2, m^2_{\tilde{d}_{Li}})
   +
   6
   I_3(|M_1|^2, |\mu|^2, m^2_{\tilde{d}_{Ri}})
  \Big]\,,
 \nonumber\\
 \epsL_{ij}
 \!\!\!&=&\!\!\! 
  \frac{\alpha_s}{3\pi} \mu^* M_3^* m^2_{\tilde q}~
  I_4(|M_|^2, m^2_{\tilde{d}_{Ri}}, 
      m^2_{\tilde{d}_{Li}}, m^2_{\tilde{d}_{Lj}})
  -
  \frac{3\alpha_2}{16\pi} \mu^* M_2^* m^2_{\tilde q} ~
  I_4(|M_2|^2, |\mu|^2, m^2_{\tilde{d}_{Li}}, m^2_{\tilde{d}_{Lj}})
  \nonumber\\
  && \!\!\!
  -
  \frac{\alpha_Y}{72\pi} \mu^* M_1^* m^2_{\tilde q}
  \left[
   I_4(|M_1|^2,m^2_{\tilde{d}_{Ri}},m^2_{\tilde{d}_{Li}}, m^2_{\tilde{d}_{Lj}})
   +
   3
   I_4(|M_1|^2, |\mu|^2, m^2_{\tilde{d}_{Li}}, m^2_{\tilde{d}_{Lj}})
  \right]\,,
 \nonumber
\end{eqnarray}
\begin{eqnarray}
 \epsR_{ij}
 \!\!\!&=&\!\!\! 
  \frac{\alpha_s}{3\pi} \mu^* M_3^* m^2_{\tilde d} ~
  I_4(|M_3|^2, m^2_{\tilde{d}_{Li}}, 
      m^2_{\tilde{d}_{Ri}}, m^2_{\tilde{d}_{Rj}})
  \nonumber\\
  && \!\!\!
  -
  \frac{\alpha_Y}{72\pi} \mu^* M_1^* m^2_{\tilde d}
  \left[
   I_4(|M_1|^2,m^2_{\tilde{d}_{Li}},m^2_{\tilde{d}_{Ri}}, m^2_{\tilde{d}_{Rj}})
   +
   3
   I_4(|M_1|^2, |\mu|^2, m^2_{\tilde{d}_{Ri}}, m^2_{\tilde{d}_{Rj}})
  \right]\,,
 \nonumber\\
 \epsLR_{ij}
 \!\!\!&=&\!\!\! 
  \frac{\alpha_s}{3\pi} \mu^* M_3^* m^{2}_{\tilde d} m^{2}_{\tilde q} ~
  I_5(|M_3|^2, m^2_{\tilde{d}_{Li}}, m^2_{\tilde{b}_{L}}\,,
      m^2_{\tilde{d}_{R_i}}, m^2_{\tilde{b}_{R}})
  \nonumber\\
  && \!\!\!
  -
  \frac{\alpha_Y}{72\pi}
  \mu^* M_1^* m^{2}_{\tilde d} m^{2}_{\tilde q} ~
  I_5(|M_1|^2, m^2_{\tilde{d}_{Li}}, m^2_{\tilde{b}_{L}}\,,
      m^2_{\tilde{d}_{Ri}}, m^2_{\tilde{b}_{R}})\,,
 \nonumber\\
 \epsY
 \!\!\!&=&\!\!\! 
  -
  \frac{\alpha_2}{16\pi}
  \frac{m_t^2}{m_W^2} A_t^* \mu^*
  \left( 1 + \frac{1}{\ttgb} \right)~
  I_3(|\mu|^2, m^2_{\tilde{t}_{R}}, m^2_{\tilde{t}_{L}})\,,
 \nonumber\\
 \epsLY_i
 \!\!\!&=&\!\!\! 
  -
  \frac{\alpha_2}{16\pi}
  \frac{m_t^2}{m_W^2} A_t^* \mu^*
  \left( 1 + \frac{1}{\ttgb} \right) m^2_{\tilde q}~
  I_4(|\mu|^2, m^2_{\tilde{t}_{R}}, m^2_{\tilde{t}_{L}}, m^2_{\tilde{u}_{L_i}})\,.
\end{eqnarray}
The functions $I_3,\ I_4$ and $I_5$ are defined in Appendix~\ref{App:loop_func}.

For the up quarks, the loop factors are
\begin{eqnarray}
 \epsgp_i
 \!\!\!&=&\!\!\! 
  \frac{\alpha_s}{3\pi} \mu^* M_3^* ~
  I_3(|M_3|^2, m^2_{\tilde{u}_{Li}},m^2_{\tilde{u}_{Ri}})
  -
  \frac{3\alpha_2}{16\pi} \mu^* M_2^*~
  I_3(|M_2|^2, |\mu|^2, m^2_{\tilde{u}_{Li}})
  \nonumber\\
  && \!\!\!
  +
  \frac{\alpha_Y}{144\pi} \mu^* M_1^*
  \Big[
   4
   I_3(|M_1|^2, m^2_{\tilde{d}_{Li}}, m^2_{\tilde{u}_{Ri}})
   +
   3
   I_3(|M_2|^2, |\mu|^2, m^2_{\tilde{u}_{Li}})
   -
   12
   I_3(|M_2|^2, |\mu|^2, m^2_{\tilde{u}_{Ri}})
  \Big]\,,
 \nonumber\\
 \epsLp_{ij}
 \!\!\!&=&\!\!\! 
  \frac{\alpha_s}{3\pi} \mu^* M_3^* m^2_{\tilde q} ~
  I_4(|M_3|^2, m^2_{\tilde{u}_{Ri}}, 
      m^2_{\tilde{u}_{Li}}, m^2_{\tilde{u}_{Lj}})
  -
  \frac{3\alpha_2}{16\pi} \mu^* M_2^* m^2_{\tilde q} ~
  I_4(|M_2|^2, |\mu|^2, m^2_{\tilde{u}_{Li}}, m^2_{\tilde{u}_{Lj}})
  \nonumber\\
  && \!\!\!
  +
  \frac{\alpha_Y}{144\pi} \mu^* M_1^* m^2_{\tilde q} 
  \left[
   4
   I_4(|M_1|^2,m^2_{\tilde{u}_{Ri}},m^2_{\tilde{u}_{Li}}, m^2_{\tilde{u}_{Lj}})
   +
   3
   I_3(|M_2|^2, |\mu|^2, m^2_{\tilde{u}_{Li}}, m^2_{\tilde{u}_{Lj}})
  \right]\,,
 \nonumber\\
 \epsRp_{ij}
 \!\!\!&=&\!\!\! 
  \frac{\alpha_s}{3\pi} \mu^* M_3^* m^2_{\tilde u}~
  I_4(|M_3|^2, m^2_{\tilde{u}_{Li}}, 
      m^2_{\tilde{u}_{Ri}}, m^2_{\tilde{u}_{Rj}})
  \nonumber\\
  && \!\!\!
  +
  \frac{\alpha_Y}{36\pi} \mu^* M_1^* m^2_{\tilde u}
  \left[
   I_4(|M_1|^2,m^2_{\tilde{u}_{Li}},m^2_{\tilde{u}_{Ri}}, m^2_{\tilde{u}_{Rj}})
   +
   3
   I_4(|M_2|^2, |\mu|^2, m^2_{\tilde{u}_{Ri}}, m^2_{\tilde{u}_{Rj}})
  \right]\,,
 \nonumber\\
 \epsLRp_{ij}
 \!\!\!&=&\!\!\! 
  \frac{\alpha_s}{3\pi} \mu^* M_3^* m^2_{\tilde u} m^2_{\tilde q}~
  I_5(|M_3|^2, m^2_{\tilde{u}_{Li}}, m^2_{\tilde{t}_{L}}\,,
      m^2_{\tilde{u}_{R_i}}, m^2_{\tilde{t}_{R}})
  \nonumber\\
  && \!\!\!
  +
  \frac{\alpha_Y}{36\pi}
  \mu^* M_1^*  m^2_{\tilde u} m^2_{\tilde q}~
  I_5(|M_1|^2, m^2_{\tilde{u}_{Li}}, m^2_{\tilde{t}_{L}}\,,
      m^2_{\tilde{u}_{Ri}}, m^2_{\tilde{t}_{R}}),
 \nonumber\\
 \epsYp
 \!\!\!&=&\!\!\! 
  -
  \frac{\alpha_2}{16\pi}
  \frac{m_b^2}{m_W^2}
  \left( 1 + \ttgb \right)~
  I_3(|\mu|^2, m^2_{\tilde{b}_{R}}, m^2_{\tilde{b}_{L}})\,,
 \nonumber\\
 \epsLYp
 \!\!\!&=&\!\!\! 
  -
  \frac{\alpha_2}{16\pi}
  \frac{m_b^2}{m_W^2}
  \left( 1 + \ttgb \right) m^2_{\tilde q}~
  I_4(|\mu|^2, m^2_{\tilde{b}_{R}}, m^2_{\tilde{b}_{L}}, m^2_{\tilde{d}_{Li}})\,.
\end{eqnarray}
Concerning the expression for the off-diagonal components of the unitary matrix that diagonalizes
the quark mass matrices, it is useful to introduce some combinations of the $\epsilon$ parameters.
For down quarks, we define
\begin{eqnarray}
 {z_Y}_i
 \!\!\!&=&\!\!\!
  \frac{1+\eps_i\tgb}{1+\eps_3\tgb}
  + \frac{\epsY}{\epsY^*}\frac{1+\eps_3^*\tgb}{1+\eps_3\tgb},
 \nonumber\\
 {z_R}_i
 \!\!\!&=&\!\!\!
  \frac{1+\eps_i\tgb}{1+\epsb_3\tgb}
  + \frac{\epsR_{i3}}{\epsR_{3i}^*}\frac{1+\eps_3^*\tgb}{1+\epsb_3\tgb}\,,
 \nonumber\\
 {z_L}_i
 \!\!\!&=&\!\!\!
  \frac{1+\eps_i\tgb}{1+\eps_3\tgb}
  + \frac{\epsL_{i3}+\epsLY_i}{\epsL_{3i}^*}\frac{1+\epsb_3^*\tgb}{1+\eps_3\tgb}
  + \frac{\epsY\tgb}{1+\eps_3\tgb}\,,
  \nonumber\\
 {\overline{z}_R}_i
 \!\!\!&=&\!\!\!
  \frac{1+\eps_i\tgb}{1+\eps_3\tgb}
  + \frac{\epsR_{i3}}{\epsR_{3i}^*}\frac{1+\eps_3^*\tgb}{1+\eps_3\tgb}\,,
 \nonumber\\
 {\overline{z}_L}_i
 \!\!\!&=&\!\!\!
  \frac{1+\eps_i\tgb}{1+\eps_3\tgb}
  + \frac{\epsL_{i3}+\epsLY_i}{\epsL_{3i}^*}\frac{1+\eps_3^*\tgb}{1+\eps_3\tgb}\,,
\end{eqnarray}
and for up quarks we define
\begin{eqnarray}
 {z'_Y}_i
 \simeq
  1 + \frac{\epsYp}{\epsYp^*}, \quad
 {z'_R}_i
 \simeq
  1 + \frac{\epsRp_{i3}}{\epsRp_{3i}^*}, \quad
 {z'_L}_i
 \simeq
  1
  + \frac{\epsLp_{i3}}{\epsLp_{3i}^*}\,.
\end{eqnarray}

The loop factors for charged leptons are calculated as
\begin{eqnarray}
 {\eps_g^e}_i
 \!\!\!&=&\!\!\! 
  -\frac{3\alpha_2}{16\pi} \mu^* M_2^*~
  I_3(|M_2|^2, |\mu|^2, m^2_{\tilde{e}_{Li}})
  +
  \frac{\alpha_Y}{16\pi} \mu^* M_1^*
  \Big[
   2
   I_3(|M_1|^2, m^2_{\tilde{e}_{Li}}, m^2_{\tilde{e}_{Ri}})
  \nonumber\\
  && \!\!\!
   + 
   I_3(|M_1|^2, |\mu|^2, m^2_{\tilde{e}_{Li}})
   -
   2
   I_3(|M_1|^2, |\mu|^2, m^2_{\tilde{e}_{Ri}})
  \Big],
 \nonumber\\
 {\eps_L^e}_{ij}
 \!\!\!&=&\!\!\! 
  -\frac{3\alpha_2}{16\pi} \mu^* M_2^* m^2_{\tilde e} ~
  I_4(|M_2|^2, |\mu|^2, m^2_{\tilde{e}_{Li}}, m^2_{\tilde{e}_{Lj}})
  \nonumber\\
  && \!\!\!
  +
  \frac{\alpha_Y}{16\pi} \mu^* M_1^* m^2_{\tilde e}
  \left[
   2
   I_4(|M_1|^2,m^2_{\tilde{e}_{Ri}},m^2_{\tilde{e}_{Li}}, m^2_{\tilde{e}_{Lj}})
   +
   I_4(|M_1|^2, |\mu|^2, m^2_{\tilde{e}_{Li}}, m^2_{\tilde{e}_{Lj}})
  \right]\,,
 \nonumber\\
 {\eps^e_R}_{ij}
 \!\!\!&=&\!\!\! 
  \frac{\alpha_Y}{8\pi} \mu^* M_1^* m^2_{\tilde e}
  \left[
   I_4(|M_1|^2,m^2_{\tilde{e}_{Li}},m^2_{\tilde{e}_{Ri}}, m^2_{\tilde{e}_{Rj}})
   -
   I_4(|M_1|^2, |\mu|^2, m^2_{\tilde{e}_{Ri}}, m^2_{\tilde{e}_{Rj}})
  \right]\,,
 \nonumber\\
 {\eps^e_{LR}}_{ij}
 \!\!\!&=&\!\!\! 
  \frac{\alpha_Y}{8\pi}
  \mu^* M_1^* m^4_{\tilde e} ~
  I_5(|M_1|^2, m^2_{\tilde{e}_{Li}}, m^2_{\tilde{\tau}_{L}}\,,
      m^2_{\tilde{e}_{Ri}}, m^2_{\tilde{\tau}_{R}})\,,
\end{eqnarray}
where the functions $I_3,\ I_4$ and $I_5$ are defined in Appendix~\ref{App:loop_func}.

\section{Effective Vertices for Mass Eigenstates}
\label{App:eff_vertex}

In this Appendix, we present the numerical procedure to derive the effective vertices
for the scalar mass eigenstates, which are used in our numerical calculation of
Section~\ref{Sec:GUT}.

First, we need to derive the non-holomorphic Yukawa corrections for a given set of tree-level
Yukawa couplings and soft {SUSY-breaking} parameters. To this end, we have to diagonalize the
sfermion mass matrices by means of the unitary matrix $U_{\tilde f}$ as
\begin{eqnarray}
  U_{\tilde f}~m^2_{\tilde f}~U_{\tilde f}^\dagger = \hat{m}^2_{\tilde f}
  \qquad (f=q,u,d,l,e)\,.
\end{eqnarray}
Using the unitary matrices and tree-level Yukawa couplings ${\yb_u}\!=\!\hat{\yb}_u \Vb$,
$\yb_d\!=\!\hat{\yb}_d$ and $\yb_e\!=\!\hat{\yb}_e$, the non-holomorphic corrections defined
in Eq.~(\ref{Eq:lag_NH}) are described as follow
\begin{eqnarray}
(\epsilon^u \Vb)_{ij}
 &=&
  \frac{\alpha_3}{3\pi} \mu^* M_3^* \,
  (U_{\tilde u}^\dagger)_{ik}
  (U_{\tilde u}\yb_u U_{\tilde q}^\dagger)_{kl} (U_{\tilde q})_{lj}\,
  I_3\left(|M_3|^2,{m_{\tilde u}^2}_k,{m_{\tilde q}^2}_l\right)
  \nonumber\\
 &&
  -
  \frac{1}{32\pi^2}\mu^* \,
  (\yb_u U_{\tilde q}^\dagger)_{ik}
  (U_{\tilde q} a_d^\dagger U_{\tilde d}^\dagger)_{kl} (U_{\tilde d})_{lj}\,
  I_3\left(|\mu|^2,{m_{\tilde q}^2}_k,{m_{\tilde d}^2}_l\right)
  \nonumber\\
 &&
  -
  \frac{3\alpha_2}{16\pi} \mu^* M_2^* \,
  (\yb_u U_{\tilde q}^\dagger)_{ik} (U_{\tilde q})_{kj}\,
  I_3\left(|M_2|^2,|\mu|^2,{m_{\tilde q}^2}_k\right)
  \nonumber\\
 &&
  +
  \frac{\alpha_Y}{48\pi} \mu^* M_1^* \,
  (\yb_u U_{\tilde q}^\dagger)_{ik} (U_{\tilde q})_{kj}\,
  I_3\left(|M_1|^2,|\mu|^2,{m_{\tilde q}^2}_k\right)
  \nonumber\\
 &&
  -
  \frac{\alpha_Y}{12\pi} \mu^* M_1^* \,
  (U_{\tilde u}^\dagger)_{ik}
  (U_{\tilde u}\yb_u)_{kj}
  I_3\left(|M_1|^2,|\mu|^2,{m_{\tilde q}^2}_k\right)
  \nonumber\\
 &&
  +
  \frac{\alpha_Y}{36\pi} \mu^* M_1^* \,
  (U_{\tilde u}^\dagger)_{ik}
  (U_{\tilde u}\yb_u U_{\tilde q}^\dagger)_{kl} (U_{\tilde q})_{lj}\,
  I_3\left(|M_1|^2,{m_{\tilde u}^2}_k,{m_{\tilde q}^2}_l\right)\,,
\end{eqnarray}
\begin{eqnarray}
 \epsilon^d_{ij}
 &=&
  -
  \frac{\alpha_3}{3\pi} \mu^* M_3^* \,
  (U_{\tilde d}^\dagger)_{ik}
  (U_{\tilde d}\yb_d U_{\tilde q}^\dagger)_{kl} (U_{\tilde q})_{lj}\,
  I_3\left(|M_3|^2,{m_{\tilde d}^2}_k,{m_{\tilde q}^2}_l\right)
  \nonumber\\
 &&
  +
  \frac{1}{32\pi^2}\mu^* \,
  (\yb_d U_{\tilde q}^\dagger)_{ik}
  (U_{\tilde q} a_u^\dagger U_{\tilde u}^\dagger)_{kl} (U_{\tilde u})_{lj}\,
  I_3\left(|\mu|^2,{m_{\tilde q}^2}_k,{m_{\tilde u}^2}_l\right)
  \nonumber\\
 &&
  +
  \frac{3\alpha_2}{16\pi} \mu^* M_2^* \,
  (\yb_d U_{\tilde q}^\dagger)_{ik} (U_{\tilde q})_{kj}\,
  I_3\left(|M_2|^2,|\mu|^2,{m_{\tilde q}^2}_k\right)
  \nonumber\\
 &&
  +
  \frac{\alpha_Y}{48\pi} \mu^* M_1^* \,
  (\yb_d U_{\tilde q}^\dagger)_{ik} (U_{\tilde q})_{kj}\,
  I_3\left(|M_1|^2,|\mu|^2,{m_{\tilde q}^2}_k\right)
  \nonumber\\
 &&
  +
  \frac{\alpha_Y}{24\pi} \mu^* M_1^* \,
  (U_{\tilde d}^\dagger)_{ik}
  (U_{\tilde d}\yb_d)_{kj}
  I_3\left(|M_1|^2,|\mu|^2,{m_{\tilde q}^2}_k\right)
  \nonumber\\
 &&
  +
  \frac{\alpha_Y}{72\pi} \mu^* M_1^* \,
  (U_{\tilde d}^\dagger)_{ik}
  (U_{\tilde d}\yb_d U_{\tilde q}^\dagger)_{kl} (U_{\tilde q})_{lj}\,
  I_3\left(|M_1|^2,{m_{\tilde d}^2}_k,{m_{\tilde q}^2}_l\right)\,,
\end{eqnarray}
\begin{eqnarray}
\epsilon^e_{ij}
 &=&
  \frac{3\alpha_2}{16\pi} \mu^* M_2^* \,
  (\yb_e U_{\tilde l}^\dagger)_{ik} (U_{\tilde l})_{kj}\,
  I_3\left(|M_2|^2,|\mu|^2,{m_{\tilde l}^2}_k\right)
  \nonumber\\
 &&
  -
  \frac{\alpha_Y}{16\pi} \mu^* M_1^* \,
  (\yb_e U_{\tilde l}^\dagger)_{ik} (U_{\tilde l})_{kj}\,
  I_3\left(|M_1|^2,|\mu|^2,{m_{\tilde l}^2}_k\right)
  \nonumber\\
 &&
  +
  \frac{\alpha_Y}{8\pi} \mu^* M_1^* \,
  (U_{\tilde e}^\dagger)_{ik}
  (U_{\tilde e}\yb_e)_{kj}
  I_3\left(|M_1|^2,|\mu|^2,{m_{\tilde l}^2}_k\right)
  \nonumber\\
 &&
  -
  \frac{\alpha_Y}{8\pi} \mu^* M_1^* \,
  (U_{\tilde e}^\dagger)_{ik}
  (U_{\tilde e}\yb_e U_{\tilde l}^\dagger)_{kl} (U_{\tilde l})_{lj}\,
  I_3\left(|M_1|^2,{m_{\tilde e}^2}_k,{m_{\tilde l}^2}_l\right)\,.
\end{eqnarray}
The loop function $I_3(x,y,z)$ is defined in the Appendix~\ref{App:loop_func}.

Mass eigenstates for quarks and leptons are determined by Eq.~(\ref{Eq:diagonalize_mass})
and we can numerically derive the unitary matrices $U_f\ (f\!=\!d_L,d_R,u_L,u_R,e_L,e_R)$
which relate the mass eigenstates and flavor eigenstates. In the procedure, tree-level Yukawa
couplings must be iteratively adjusted to reproduce experimentally-observed values of the CKM
matrix and mass eigenvalues. After the achievement of the adjustment with the desired accuracy,
tree-level Yukawa couplings are determined, which fix the SUSY spectrum, and finally unitary
matrices $U_f$ are obtained.

Although the effective vertices among quarks/leptons and SUSY particles have already been
derived in terms of $U_f$ and $\bar{y}_f$ in Appendix~\ref{Sec:effective_vertex}, 
we derive here these vertices also in the mass eigenstates, for numerical convenience.

Firstly, unitary matrices that diagonalize the mass matrices of SUSY particles must be defined.
For the sfermions, the unitary matrices $U_{\tilde f}$ are given as
\begin{eqnarray}
 U_{\tilde f}
 \left(
 \begin{array}{cc}
 m^2_{LL} & m_{RL}^{2\dagger} \\
 m^2_{RL} & m^2_{RR}
 \end{array}
 \right)
 U_{\tilde f}^\dagger = \hat{m}^{2}_{\tilde{f}}\,,
\end{eqnarray}
with $f=d,u,e$ and mass eigenstates ${\tilde f}_I$ and flavor eigenstates in the {\it bare}
SCKM basis ${\tilde f}_i=(\tilde{f}_L,~\tilde{f}_R)$ are related by
\begin{eqnarray}
 {\tilde f}_I = ({U_{\tilde f}})_{Ii} {\tilde f}_i \quad (I=1,\cdots ,6,~i=1,\cdots ,6)\,.
\end{eqnarray}
In our convention, for up quarks
\begin{eqnarray}
 (m_{LL}^2)_{ij} 
  &=& (\Vb m_{\tilde q}^2 \Vb^\dagger)_{ij}
    + \left(\frac{1}{2}-\frac{2}{3}\sin^2\theta_W \right)
      m_Z^2 \cos 2\beta~\delta_{ij}\,,\\
 (m_{RR}^2)_{ij}
  &=& (m_{\tilde u}^2)_{ij}
    +\frac{2}{3}\sin^2\theta_W~m_Z^2 \cos 2\beta~\delta_{ij},\\
 (m_{RL}^2)_{ij}
  &=&
    - \frac{v\sin\beta}{\sqrt{2}}\left((a_u\Vb^\dagger)_{ij}
    + \yb_{u_i} \delta_{ij} \mu^* \cot\beta \right)\,,
\label{Eq:msu_RL}
\end{eqnarray}
and for down quarks 
\begin{eqnarray}
 (m_{LL}^2)_{ij} 
  &=& (m_{\tilde q}^2)_{ij}
    + \left(-\frac{1}{2}+\frac{1}{3}\sin^2\theta_W \right)
      m_Z^2 \cos 2\beta~\delta_{ij}\,,\\
 (m_{RR}^2)_{ij}
  &=& (m_{\tilde d}^2)_{ij}
    -\frac{1}{3}\sin^2\theta_W~m_Z^2 \cos 2\beta~\delta_{ij}\,,\\
 (m_{RL}^2)_{ij}
  &=&
    - \frac{v\cos\beta}{\sqrt{2}}\left( {a_d}_{ij}
    + \yb_{d_i} \delta_{ij} \mu^* \tgb \right)\,,
\label{Eq:msd_RL}
\end{eqnarray}
in the {\it bare} SCKM basis. The mass matrix for charged sleptons can be read as
\begin{eqnarray}
 (m_{LL}^2)_{ij} 
  &=& (m_{\tilde l}^2)_{ij}
    + \left(-\frac{1}{2}+\sin^2\theta_W \right)
      m_Z^2 \cos 2\beta~\delta_{ij},\\ 
 (m_{RR}^2)_{ij}
  &=& (m_{\tilde e}^2)_{ij}
    -\sin^2\theta_W~m_Z^2 \cos 2\beta~\delta_{ij},\\
 (m_{RL}^2)_{ij}
  &=&
    - \frac{v\cos\beta}{\sqrt{2}}\left( {a_e}_{ij}
    + \yb_{e_i} \delta_{ij} \mu^* \tgb \right)\,,
\label{Eq:mse_RL}
\end{eqnarray}
while that for left-handed sneutrinos is defined as
\begin{eqnarray}
 (m_{LL}^2)_{ij} 
  &=& (m_{\tilde l}^2)_{ij}
    + \frac{1}{2} m_Z^2 \cos 2\beta~\delta_{ij}\,,
\label{Eq:msn_LL}
\end{eqnarray}
and it should be diagonalized by a 3-by-3 unitary matrix.

The chargino mass matrix is diagonalized by
\begin{eqnarray}
 O_R
 \left(
 \begin{array}{cc}
  M_2 & \sqrt{2}m_W\cos\beta \\
  \sqrt{2}m_W\sin\beta & \mu 
 \end{array}
 \right) O_L^\dagger =  
 \left(
 \begin{array}{cc}
  M_{C_1} &  \\
  & M_{C_2}
 \end{array}
 \right)\,.
\end{eqnarray}
Charginos are mixed states of charged wino and Higgsino defined as
\begin{eqnarray}
 \left[
 \begin{array}{c}
  {\tilde \chi}^-_{1L} \\
  {\tilde \chi}^-_{2L}
 \end{array}
 \right]
 \!\!\! &=& \!\!\!
  O_{L}
  \left[
  \begin{array}{c}
   \tWLm \\
   \tHdLm
  \end{array}
  \right]\,,
  \nonumber\\
 \left[
 \begin{array}{c}
  {\tilde \chi}^-_{1R} \\
  {\tilde \chi}^-_{2R}
 \end{array}
 \right]
 \!\!\! &=& \!\!\!
  O_{R}
  \left[
  \begin{array}{c}
   {\tilde W}_R^- \\
   {\tilde H}_{2R}^-
  \end{array}
  \right]\,,
\end{eqnarray}
and the charge conjugate fields are defined by $\psi_{L/R}^+ = ( \psi_{R/L}^- )^c$ 
and $\psi^c = C \bar{\psi}^T$.

The neutralino mass matrix is diagonalized by
\begin{eqnarray} 
\label{N_mass_diagonalization}
 U_N^*
 \left(
 \begin{array}{cccc}
  M_1 & 0 & -m_Z s_W \cosb & m_Z s_W \sinb \\
  0 & M_2 & m_Z c_W \cosb & -m_Z c_W \sinb \\
  -m_Z s_W \cosb & m_Z c_W \cosb & 0 & -\mu \\
  m_Z s_W \sinb & -m_Z c_W \sinb & -\mu & 0
 \end{array}
 \right)
 U_{N}^\dagger 
 = 
 \left(
 \begin{array}{cccc}
  M_{N_1} &  &  & \\
  & M_{N_2} & &  \\
  & & M_{N_3} & \\
  & & & M_{N_4}
 \end{array}
 \right)
\end{eqnarray}
This unitary matrix is easily obtained by considering the square of both sides 
of Eq.~(\ref{N_mass_diagonalization}).
Neutralinos are mixed states of bino, neutral wino and Higgsinos defined by
\begin{eqnarray}
 {\tilde \chi}^0_{iL}
 =
  \sum_{j=1}^4
  (U_{N})_{ij} 
  ({\tilde B}_L,~{\tilde W}^0_L,~{\tilde H}^0_{1L},~{\tilde H}^0_{2L})_j\,,
\end{eqnarray}
and $\tilde{\chi}_{iR}^0 = (\tilde{\chi}_{iL}^0)^c$.

Using the above unitary matrices and the effective vertices in the flavor basis given in Appendix.~\ref{Sec:effective_vertex}, which are expressed by tree-level Yukawa couplings
and unitary matrices $U_{f}$, one can find the effective vertices in the mass eigenstates.

The couplings of the gluino to the quarks are given by
\begin{eqnarray}
 -{\cal L}^{\tilde{g}}_{\rm int}
 \!\!\! &=& \!\!\!
  {\tilde q}^*_I \overline{{\tilde g}}^a t_a 
  \left[
   ( \Gamma^{\tilde g}_{q_L} )_{Ij} P_L
   +
   ( \Gamma^{\tilde g}_{q_R} )_{Ij} P_R
  \right]
  q_j
  ~+~ {\rm h.c.}\,,
\end{eqnarray}
where
\begin{eqnarray}
 ( \Gamma^{\tilde g}_{q_L} )_{Ij}
 \!\!\! &=& \!\!\!
  \sqrt{2} g_s~
  \sum_{k=1}^3
  (U_{\tilde q})_{Ik}(U^\dagger_{q_L})_{kj}\,,
  \\
 ( \Gamma^{\tilde g}_{q_R} )_{Ij}
 \!\!\! &=& \!\!\!
  -\sqrt{2} g_s~
  \sum_{k=1}^3
  (U_{\tilde q})_{Ik+3}(U^\dagger_{q_R})_{kj}\,.
\end{eqnarray}

The couplings of the chargino to the quarks and leptons are given by
\begin{eqnarray}
 -{\cal L}^{\tilde{\chi}^\pm}_{\rm int}
 =
  \!\!\! && \!\!\!
   (\Gamma_\dL^{\tilde{\chi}^\pm} )^k_{Ij} \tilde{u}_I^* \tchiRmb \dL_j
  +(\Gamma_\dR^{\tilde{\chi}^\pm} )^k_{Ij} \tilde{u}_I^* \tchiLmb \dR_j
  +(\Gamma_\uL^{\tilde{\chi}^\pm} )^k_{Ij} \tilde{d}_I^* \tchiRpb \uL_j
  +(\Gamma_\uR^{\tilde{\chi}^\pm} )^k_{Ij} \tilde{d}_I^* \tchiLpb \uR_j
 \nonumber
 \\ && \!\!\!
  +(\Gamma_\eL^{\tilde{\chi}^\pm} )^k_{Ij} \tilde{\nu}_I^* \tchiRmb \eL_j
  +(\Gamma_\eR^{\tilde{\chi}^\pm} )^k_{Ij} \tilde{\nu}_I^* \tchiLmb \eR_j
  +(\Gamma_\nL^{\tilde{\chi}^\pm} )^k_{Ij} \tilde{e}_I^* \tchiRpb \nL_j
  ~+~ {\rm h.c.}\,,
\end{eqnarray}
where
\begin{eqnarray}
 ( \Gamma^{\tilde{\chi}^\pm}_{d_L} )^k_{Ij}
 \!\!\! &=& \!\!\!
  \sum_{i=1}^3
  \Big[
  (O_R)_{k1} (U_{\tilde{u}})_{Ii} (C_{d_L}^{\tilde{W}^\pm})_{ij}
  + 
  (O_R)_{k2} (U_{\tilde{u}})_{Ii+3} (C_{d_L}^{\tilde{H}^\pm})_{ij}
  \Big]\,,
  \nonumber
  \\
 ( \Gamma^{\tilde{\chi}^\pm}_{d_R} )^k_{Ij}
 \!\!\! &=& \!\!\!
  \sum_{i=1}^3
  (O_L)_{k2} (U_{\tilde{u}})_{Ii} (C_{d_R}^{\tilde{H}^\pm})_{ij}\,,
  \nonumber
  \\
 ( \Gamma^{\tilde{\chi}^\pm}_{u_L} )^k_{Ij}
 \!\!\! &=& \!\!\!
  \sum_{i=1}^3
  \Big[
  (O_L^*)_{k1} (U_{\tilde{d}})_{Ii} (C_{u_L}^{\tilde{W}^\pm})_{ij}
  + 
  (O_L^*)_{k2} (U_{\tilde{d}})_{Ii+3} (C_{u_L}^{\tilde{H}^\pm})_{ij}
  \Big]\,,
  \nonumber
  \\
 ( \Gamma^{\tilde{\chi}^\pm}_{u_R} )^k_{Ij}
 \!\!\! &=& \!\!\!
  \sum_{i=1}^3
  (O_R^*)_{k2} (U_{\tilde{d}})_{Ii} (C_{u_R}^{\tilde{H}^\pm})_{ij}\,,
\end{eqnarray}
for quarks and
\begin{eqnarray}
 ( \Gamma^{\tilde{\chi}^\pm}_{e_L} )^k_{Ij}
 \!\!\! &=& \!\!\!
  \sum_{i=1}^3
  (O_R)_{k1} (U_{\tilde{\nu}})_{Ii} (C_{e_L}^{\tilde{W}^\pm})_{ij}\,,
  \nonumber
  \\
 ( \Gamma^{\tilde{\chi}^\pm}_{e_R} )^k_{Ij}
 \!\!\! &=& \!\!\!
  \sum_{i=1}^3
  (O_L)_{k2} (U_{\tilde{\nu}})_{Ii} (C_{e_R}^{\tilde{H}^\pm})_{ij}\,,
  \nonumber
  \\
 ( \Gamma^{\tilde{\chi}^\pm}_{\nu_L} )^k_{Ij}
 \!\!\! &=& \!\!\!
  \sum_{i=1}^3
  \Big[
  (O_L^*)_{k1} (U_{\tilde{e}})_{Ii} (C_{\nu_L}^{\tilde{W}^\pm})_{ij}
  + 
  (O_L^*)_{k2} (U_{\tilde{e}})_{Ii+3} (C_{\nu_L}^{\tilde{H}^\pm})_{ij}
  \Big]\,,
\end{eqnarray}
for leptons. The couplings in the flavor eigenstates are defined in
Eq.~(\ref{Eq:chargino_coupling_last}).

The couplings of the neutralino to the quarks are given by
\begin{eqnarray}
 -{\cal L}^{\tilde{\chi}^0}_{\rm int}
 =
 \!\!\! && \!\!\!
   (\Gamma_\dL^{\tilde{\chi}^0} )^k_{Ij} \tilde{d}_I^* \tchiRzb \dL_j
  +(\Gamma_\dR^{\tilde{\chi}^0} )^k_{Ij} \tilde{d}_I^* \tchiLzb \dR_j
  +(\Gamma_\uL^{\tilde{\chi}^0} )^k_{Ij} \tilde{u}_I^* \tchiRzb \uL_j
  +(\Gamma_\uR^{\tilde{\chi}^0} )^k_{Ij} \tilde{u}_I^* \tchiLzb \uR_j
 \nonumber\\
 && \!\!\!
  +(\Gamma_\eL^{\tilde{\chi}^0} )^k_{Ij} \tilde{e}_I^* \tchiRzb \eL_j
  +(\Gamma_\eR^{\tilde{\chi}^0} )^k_{Ij} \tilde{e}_I^* \tchiLzb \eR_j
  +(\Gamma_\nL^{\tilde{\chi}^0} )^k_{Ij} \tilde{\nu}_I^* \tchiRzb \nL_j
  + {\rm h.c.}\,,
\end{eqnarray}
where
\begin{eqnarray}
 ( \Gamma^{\tilde{\chi}^0}_{d_L} )^k_{Ij}
 \!\!\! &=& \!\!\!
  \sum_{i=1}^3
  \Big[
  (U_N)_{k1} (U_{\tilde{d}})_{Ii} (C_{d_L}^{\tilde{B}})_{ij}
  + 
  (U_N)_{k2} (U_{\tilde{d}})_{Ii} (C_{d_L}^{\tilde{W}^0})_{ij}
  \nonumber
  + 
  (U_N)_{k3} (U_{\tilde{d}})_{Ii+3} (C_{d_L}^{\tilde{H}^0})_{ij}
  \Big]\,,
  \nonumber
  \\
 ( \Gamma^{\tilde{\chi}^0}_{d_R} )^k_{Ij}
 \!\!\! &=& \!\!\!
  \sum_{i=1}^3
  \Big[
  (U_N^*)_{k1} (U_{\tilde{d}})_{Ii+3} (C_{d_R}^{\tilde{B}})_{ij}
  + 
  (U_N^*)_{k3} (U_{\tilde{d}})_{Ii} (C_{d_R}^{\tilde{H}^0})_{ij}
  \Big]\,,
  \nonumber
  \\
 ( \Gamma^{\tilde{\chi}^0}_{u_L} )^k_{Ij}
 \!\!\! &=& \!\!\!
  \sum_{i=1}^3
  \Big[
  (U_N)_{k1} (U_{\tilde{u}})_{Ii} (C_{u_L}^{\tilde{B}})_{ij}
  + 
  (U_N)_{k2} (U_{\tilde{u}})_{Ii} (C_{u_L}^{\tilde{W}^0})_{ij} 
  \nonumber
  + 
  (U_N)_{k4} (U_{\tilde{u}})_{Ii+3} (C_{u_L}^{\tilde{H}^0})_{ij}
  \Big],
  \nonumber
  \\
 ( \Gamma^{\tilde{\chi}^0}_{u_R} )^k_{Ij}
 \!\!\! &=& \!\!\!
  \sum_{i=1}^3
  \Big[
  (U_N^*)_{k1} (U_{\tilde{u}})_{Ii+3} (C_{u_R}^{\tilde{B}})_{ij}
  + 
  (U_N^*)_{k4} (U_{\tilde u})_{Ii} (C_{u_R}^{\tilde{H}^0})_{ij}
  \Big]\,,
\end{eqnarray}
for quarks and
\begin{eqnarray}
 ( \Gamma^{\tilde{\chi}^0}_{e_L} )^k_{Ij}
 \!\!\! &=& \!\!\!
  \sum_{i=1}^3
  \Big[
  (U_N)_{k1} (U_{\tilde{e}})_{Ii} (C_{e_L}^{\tilde{B}})_{ij}
  + 
  (U_N)_{k2} (U_{\tilde{e}})_{Ii} (C_{e_L}^{\tilde{W}^0})_{ij}
  \nonumber
  + 
  (U_N)_{k3} (U_{\tilde{e}})_{Ii+3} (C_{e_L}^{\tilde{H}^0})_{ij}
  \Big]\,,
  \nonumber
  \\
 ( \Gamma^{\tilde{\chi}^0}_{e_R} )^k_{Ij}
 \!\!\! &=& \!\!\!
  \sum_{i=1}^3
  \Big[
  (U_N^*)_{k1} (U_{\tilde{e}})_{Ii+3} (C_{e_R}^{\tilde{B}})_{ij}
  + 
  (U_N^*)_{k3} (U_{\tilde{e}})_{Ii} (C_{e_R}^{\tilde{H}^0})_{ij}
  \Big]\,,
  \nonumber
  \\
 ( \Gamma^{\tilde{\chi}^0}_{\nu_L} )^k_{Ij}
 \!\!\! &=& \!\!\!
  \Big[
  \sum_{i=1}^3
  (U_N)_{k1} (U_{\tilde{\nu}})_{Ii} (C_{\nu_L}^{\tilde{B}})_{ij}
  + 
  (U_N)_{k2} (U_{\tilde{\nu}})_{Ii} (C_{\nu_L}^{\tilde{W}^0})_{ij} 
  \Big]\,,
\end{eqnarray}
for leptons. The couplings in the flavor eigenstates are defined in
Eq.~(\ref{Eq:neutralino_coupling_last}).

\section{(C)EDMs Formulae for Mass Eigenstates}
\label{App:edm_formulae}
Here, we present the full expressions for the quark (C)EDMs and lepton EDMs
in the mass eigenstates basis using, in particular, the effective vertices 
derived in the previous section.

The gluino-mediated contributions read
\begin{eqnarray}
 \left\{
 d_{d_i} ,~
 d^c_{d_i}
 \right\}_{\tilde g} 
 = 
 - \frac{1}{32\pi^2}
 \sum_{j=1}^6 
  \frac{1}{{m^2_{\tilde d}}_j}
  {\rm Im} \left[ M_{3}\, {(\Gamma^{\tilde g}_{d_L})_{ji}}^*
  (\Gamma^{\tilde g}_{d_R})_{ji} 
  \right]
  \left\{
  f^{(0)}_3\left( x_j \right) 
  ,\,
  f^{(0)}_2\left( x_j \right)
  \right\}\,,
\end{eqnarray}
where $x_j={|M_3|^2}/{{m^2_{\tilde d}}_j}$. The chargino-mediated contributions read
\begin{eqnarray}
 \left\{
 d_{d_i},~
 d^c_{d_i}
 \right\}_{{\tilde \chi}^\pm} 
 = 
 - \frac{1}{32\pi^2}
 \sum_{j=1}^6 
 \sum_{k=1}^2 
  \frac{{{M_C}_k}}{{m^2_{\tilde u}}_j}
  {\rm Im} \left[ {(\Gamma^{{\tilde \chi}^\pm}_{d_L})^{k}_{ji}}^*
  (\Gamma^{{\tilde \chi}^\pm}_{d_R})^{k}_{ji} 
  \right]
  \left\{
  f^{(0)}_5\left( x^k_j \right) 
  ,\, 
  f^{(0)}_0\left( x^k_j \right)
  \right\}\,,
\end{eqnarray}
where $x^k_j={{{M^2_C}_k}}/{{m^2_{\tilde u}}_j}$. The neutralino-mediated contributions read
\begin{eqnarray}
 \left\{
 d_{d_i},~
 d^c_{d_i}
 \right\}_{{\tilde \chi}^0} 
 = 
 - \frac{1}{32\pi^2}
 \sum_{j=1}^6 
 \sum_{k=1}^4 
  \frac{{{M_N}_k}}{{m^2_{\tilde d}}_j}
  {\rm Im} \left[ {(\Gamma^{{\tilde \chi}^0}_{d_L})^{k}_{ji}}^*
  (\Gamma^{{\tilde \chi}^0}_{d_R})^{k}_{ji} 
  \right]
  \left\{
  f^{(0)}_7\left( x^k_j \right) 
  ,\, 
  f^{(0)}_0\left( x^k_j \right)
  \right\}\,,
\end{eqnarray}
where $x^k_j={{{M^2_N}_k}}/{{m^2_{\tilde d}}_j}$. Finally, the Higgs-mediated contributions
have the following expression,
\begin{eqnarray}
 \left\{
 d_{d_i},~
 d^c_{d_i}
 \right\}_{H^\pm} 
 = 
 - \frac{g_2^2}{64\pi^2}
 \sum_{j=1}^3 
  \frac{m^2_{u_j}}{m_W^2}
  \frac{m_{d_i}}{M_{H^\pm}^2}
  {\rm Im} 
  \left[
  {(C^{H^\pm}_{d_L})_{ji}}^*
  {(C^{H^\pm}_{d_R})_{ji}} 
  \right]
  \left\{ 
  f^{(0)}_9\left( x_j \right) 
  ,\, 
  f^{(0)}_1\left( x_j \right)
  \right\}\,,
\end{eqnarray}
where $x_j={m^2_{u_j}}/{M_{H^\pm}^2}$.

Passing to the up quark (C)EDMs, the gluino-mediated contribution is given by
\begin{eqnarray}
 \left\{
 d_{u_i},~
 d^c_{u_i}
 \right\}_{\tilde g} 
 = 
 - \frac{1}{32\pi^2}
 \sum_{j=1}^6 
  \frac{1}{{m^2_{\tilde u}}_j}
  {\rm Im} \left[ M_{3}\, {(\Gamma^{\tilde g}_{u_L})_{ji}}^*
  (\Gamma^{\tilde g}_{u_R})_{ji} 
  \right]
  \left\{
  f^{(0)}_4\left( x_j \right) 
  ,\,
  f^{(0)}_2\left( x_j \right) 
  \right\}\,,
\end{eqnarray}
where $x_j=|M_3|^2/{m^2_{\tilde u}}_j$. The chargino-mediated contribution reads
\begin{eqnarray}
 \left\{
 d_{u_i},~
 d^c_{u_i}
 \right\}_{{\tilde \chi}^\pm} 
 = 
 - \frac{1}{32\pi^2}
 \sum_{j=1}^6 
 \sum_{k=1}^2 
  \frac{{{M_C}_k}}{{m^2_{\tilde d}}_j}
  {\rm Im} \left[ {(\Gamma^{{\tilde \chi}^\pm}_{u_L})^{k}_{ji}}^*
  (\Gamma^{{\tilde \chi}^\pm}_{u_R})^{k}_{ji} 
  \right]
  \left\{
  f^{(0)}_6\left( x^k_j \right) 
  ,\, 
  f^{(0)}_0\left( x^k_j \right)
  \right\}\,,
\end{eqnarray}
where $x^k_j={{{M^2_C}_k}}/{{m^2_{\tilde d}}_j}$. The neutralino-mediated contribution reads
\begin{eqnarray}
 \left\{
 d_{u_i},~
 d^c_{u_i}
 \right\}_{{\tilde \chi}^0} 
 = 
 - \frac{1}{32\pi^2}
 \sum_{j=1}^6 
 \sum_{k=1}^4 
  \frac{{{M_N}_k}}{{m^2_{\tilde u}}_j}
  {\rm Im} \left[ {(\Gamma^{{\tilde \chi}^0}_{u_L})^{k}_{ji}}^*
  (\Gamma^{{\tilde \chi}^0}_{u_R})^{k}_{ji} 
  \right]
  \left\{
  f^{(8)}_0\left( x^k_j \right) 
  ,\, 
  f^{(0)}_0\left( x^k_j \right)
  \right\}\,,
\end{eqnarray}
where $x^k_j={{{M^2_N}_k}}/{{m^2_{\tilde u}}_j}$. Finally, the Higgs-mediated contribution
is given by
\begin{eqnarray}
 \left\{
 d_{u_i},~
 d^c_{u_i}
 \right\}_{H^\pm} 
 = 
 - \frac{g_2^2}{64\pi^2}
 \sum_{j=1}^3 
  \frac{m^2_{d_j}}{m_W^2}
  \frac{m_{u_i}}{M_{H^\pm}^2}
  {\rm Im} 
  \left[
  {(C^{H^\pm}_{d_R})_{ij}}
  {(C^{H^\pm}_{d_L})_{ij}}^* 
  \right]
  \left\{ 
  f^{(0)}_{10}\left( x_j \right) 
  ,\, 
  f^{(0)}_1\left( x_j \right)
  \right\}\,,
\end{eqnarray}
where $x_j={m^2_{d_j}}/{M_{H^\pm}^2}$.

The lepton EDMs receive contributions by the only chargino and neutralino sectors.
The chargino-mediated contribution reads
\begin{eqnarray}
 \left\{ d_{e_i} \right\}_{{\tilde \chi}^\pm}
 =
 \frac{1}{32\pi^2} 
 \sum_{j=1}^3 
 \sum_{k=1}^2 
  \frac{{{M_C}_k}}{{m^2_{\tilde n}}_j}
  {\rm Im} \left[ {(\Gamma^{{\tilde \chi}^\pm}_{e_L})^{k}_{ji}}^*
  (\Gamma^{{\tilde \chi}^\pm}_{e_R})^{k}_{ji} 
  \right]
  f^{(0)}_1\left( x^k_j \right)\,,
\end{eqnarray}
where $x^k_j={{{M^2_C}_k}}/{{m^2_{\tilde n}}_j}$. The neutralino-mediated contribution
is given by
\begin{eqnarray}
 \left\{ d_{e_i} \right\}_{{\tilde \chi}^0}
 = 
 \frac{1}{32\pi^2} 
 \sum_{j=1}^6 
 \sum_{k=1}^4 
  \frac{{{M_N}_k}}{{m^2_{\tilde e}}_j}
  {\rm Im} \left[ {(\Gamma^{{\tilde \chi}^0}_{e_L})^{k}_{ji}}^*
  (\Gamma^{{\tilde \chi}^0}_{e_R})^{k}_{ji} 
  \right]
  f^{(0)}_0\left( x^k_j \right)\,,
\end{eqnarray}
where $x^k_j={{{M^2_N}_k}}/{{m^2_{\tilde e}}_j}$.


\section{Loop functions for EDM Formulae} \label{App:loop_func}

The loop functions that appear in Sections~\ref{Sec:Jarlskog} and \ref{Sec:FEDM} are defined as
\begin{eqnarray}
 f_0^{(0)}(x)
 \!\!\! &=& \!\!\!
  \frac{1-x^2+2x\log x}{(1-x)^3}\,,
 \nonumber\\
 f_1^{(0)}(x)
 \!\!\! &=& \!\!\!
  \frac{3-4x+x^2+2\log x}{(1-x)^3}\,,
 \nonumber\\
 f_0^{(1)}(x)
 \!\!\! &=& \!\!\!
  \frac{-1-4x+5x^2-2x(x+2)\log x}{(1-x)^4}\,,
 \nonumber
\end{eqnarray}
\begin{eqnarray}
 f_1^{(1)}(x)
 \!\!\! &=& \!\!\!
  \frac{-5+4x+x^2-2(1+2x)\log x}{(1-x)^4}\,,
 \nonumber\\
 f_0^{(2)}(x)
 \!\!\! &=& \!\!\!
  \frac{1+9x-9x^2-x^3+6x(x+1)\log x}{(1-x)^5}\,,
 \nonumber\\
 f_1^{(2)}(x)
 \!\!\! &=& \!\!\!
  \frac{2(3-3x^2+(1+4x+x^2)\log x)}{(1-x)^5}\,,
 \nonumber\\
 f_0^{(3)}(x)
 \!\!\! &=& \!\!\!
  \frac{-3-44x+36x^2+12x^3-x^4-12x(3x+2)\log x}{3(1-x)^6}\,,
 \nonumber\\
 f_1^{(3)}(x)
 \!\!\! &=& \!\!\!
  2~\frac{-10-9x+18x^2+x^3-3(1+6x+3x^2)\log x}{3(1-x)^6}\,,
\end{eqnarray}
and
\begin{eqnarray}
 g_k^{(i)}(x,y)
 \!\!\! &=& \!\!\!
  \frac{f_k^{(i)}(x)-f_k^{(i)}(y)}{x-y}
 \qquad
 (i,k=0,1).
\end{eqnarray}
In the limit of degenerate arguments, the functions are given as
\begin{eqnarray}
 &&
 f_0^{(i)}(1)
 =
  \left\{{1 \over 3},~-{1\over 6},~{1\over 10},~-{1\over 15} \right\}
 \qquad (i=0,1,2,3)\,,
 \nonumber\\
 &&
 f_1^{(i)}(1)
 =
  \left\{-{2\over 3},~{1\over 6},~-{1\over 15},~{1\over 30} \right\}
 \qquad (i=0,1,2,3)\,,
 \nonumber\\
 &&
 g_0^{(j)}(1,1)
 =
  \left\{-{1 \over 6},~{2\over 15} \right\}
 \qquad \qquad (j=0,1)\,,
 \nonumber\\
 &&
 g_1^{(j)}(1,1)
 =
 \left\{{1\over 2},~-{1\over 5} \right\}
 \qquad \qquad (j=0,1)\,.
\end{eqnarray}
The following combinations of functions are also used
\begin{eqnarray}
f^{(i)}_2(x) 
 \!\!\! &=& \!\!\!
 -\frac{1}{6}f^{(i)}_0(x) + \frac{3}{2}f^{(i)}_1(x)\,, 
\nonumber\\
f^{(i)}_3(x) 
 \!\!\! &=& \!\!\!
 -\frac{4}{9}f^{(i)}_0(x)\,,  
\nonumber\\
f^{(i)}_4(x) 
 \!\!\! &=& \!\!\!
  \frac{8}{9}f^{(i)}_0(x)\,,   
\nonumber\\
f^{(i)}_5(x) 
 \!\!\! &=& \!\!\!
  \frac{2}{3}f^{(i)}_0(x) - f^{(i)}_1(x)\,, 
\nonumber\\
f^{(i)}_6(x) 
 \!\!\! &=& \!\!\!
 -\frac{1}{3}f^{(i)}_0(x) + f^{(i)}_1(x)\,, 
\nonumber\\
f^{(i)}_7(x) 
 \!\!\! &=& \!\!\!
 -\frac{1}{3}f^{(i)}_0(x)\,, 
\nonumber\\
f^{(i)}_8(x) 
 \!\!\! &=& \!\!\!
  \frac{2}{3}f^{(i)}_0(x)\,, 
\nonumber
\end{eqnarray}
\begin{eqnarray}
f^{(i)}_9(x) 
 \!\!\! &=& \!\!\!
 -f^{(i)}_0(x) + \frac{2}{3}f^{(i)}_1(x)\,, 
\nonumber\\
f^{(i)}_{10}(x) 
 \!\!\! &=& \!\!\!
 f^{(i)}_0(x) - \frac{1}{3}f^{(i)}_1(x)\,. 
\end{eqnarray}


\section{Loop functions for Yukawa coupling
  corrections} 

\label{App:loop_func2}

The loop functions used in the derivation of non-holomorphic Yukawa couplings are given by
\begin{eqnarray}
 I_3(a,b,c)
 \!\!\!&=&\!\!\!
  -2
  \left(
   \frac{a \log a}{(c-a)(a-b)}
   +\frac{b \log b}{(a-b)(b-c)}
   +\frac{c \log c}{(b-c)(c-a)}
  \right)\,,
 \nonumber\\
 I_4(a,b,c,d)
 \!\!\!&=&\!\!\!
  -6
  \left(
   \frac{a \log a}{(a-b) (a-c) (a-d)}
   +
   \mbox{3 cyclic permutations}
  \right)\,,
 \nonumber\\
 I_5(a,b,c,d,e)
 \!\!\!&=&\!\!\!
  12
  \left(
   \frac{a \log a}{(a-b) (a-c) (a-d) (a-e)}
   +
   \mbox{4 cyclic permutations}
  \right)\,.
\end{eqnarray}
These functions are normalized as $~I_3(1,1,1)=I_4(1,1,1,1)=I_5(1,1,1,1,1)=1$.

\newpage
%
%
\newcommand{\Journal}[4]{{\sl #1} {\bf #2} {(#3)} {#4}}
\newcommand{\APJ}{Ap. J.}
\newcommand{\CJP}{Can. J. Phys.}
\newcommand{\MPL}{Mod. Phys. Lett.}
\newcommand{\NC}{Nuovo Cimento}
\newcommand{\NP}{Nucl. Phys.}
\newcommand{\PL}{Phys. Lett.}
\newcommand{\PR}{Phys. Rev.}
\newcommand{\PRep}{Phys. Rep.}
\newcommand{\PRL}{Phys. Rev. Lett.}
\newcommand{\PTP}{Prog. Theor. Phys.}
\newcommand{\SJNP}{Sov. J. Nucl. Phys.}
\newcommand{\ZP}{Z. Phys.}

\end{document}